\documentclass[reprint,aps,prd,twocolumn,superscriptaddress,showpacs,preprintnumbers,amsmath,amssymb,floatfix,nofootinbib]{revtex4-1}%

\pdfoutput=1

\usepackage{graphicx}
\usepackage{epstopdf}
\usepackage[mathscr]{euscript}
\usepackage{mathtools} 
\usepackage{textcomp} 
\usepackage{dcolumn}
\usepackage[caption=false]{subfig}
\usepackage{bigstrut}
\usepackage[english]{babel}
\usepackage{bm}
\usepackage[mathlines]{lineno}
\usepackage{xspace} 
\usepackage[colorlinks,citecolor=blue]{hyperref}

\newcommand{\FIG}{Fig.~}
\newcommand{\FIGS}{Figs.~}
\newcommand{\SEC}{Sec.~}
\newcommand{\SECS}{Secs.~}
\newcommand{\TAB}{Table~}

\newcommand{\EQ}{Eq.~}
\newcommand{\EQS}{Eqs.~}
\newcommand{\REF}{Ref.~}
\newcommand{\REFS}{Refs.~}
\newcommand{\wrt}{{with respect to}\xspace}
\newcommand{\ie}{\textit{i.e.}\xspace}
\newcommand{\eg}{\textit{e.g.}\xspace}
\newcommand{\vs}{\textit{vs.}\xspace}
\newcommand{\etal}{\textit{et al.}\xspace}

\newcommand{\dm}{dark matter\xspace}
\newcommand{\w}{WIMP\xspace}
\newcommand{\ws}{WIMPs\xspace}
\newcommand{\lws}{low-mass WIMPs\xspace}
\newcommand{\eh}{electron-hole\xspace}
\newcommand{\lf}{low-frequency\xspace}
\newcommand{\dXLF}{$\Delta\chi^2_{\text{LF}}$\xspace}
\newcommand{\dXLFeq}{\Delta\chi^2_{\text{LF}}\xspace}
\newcommand{\Ereeeq}{E_{\text{r,ee}}\xspace} 
\newcommand{\Eree}{$E_{\text{r,ee}}$\xspace} 
\newcommand{\Ernreq}{E_{\text{r,nr}}\xspace} 
\newcommand{\Ernr}{$E_{\text{r,nr}}$\xspace} 
\newcommand{\Ereq}{E_{\text{r}}\xspace} 
\newcommand{\Er}{$E_{\text{r}}$\xspace} 
\newcommand{\Eteq}{E_{\text{t}}\xspace} 
\newcommand{\Et}{$E_{\text{t}}$\xspace}
\newcommand{\ELeq}{E_{\text{NTL}}\xspace}
\newcommand{\EL}{$E_{\text{NTL}}$\xspace}
\newcommand{\Eqeq}{E_{\text{Q}}\xspace}
\newcommand{\Eq}{$E_{\text{Q}}$\xspace}
\newcommand{\Vbeq}{V_{\text{b}}\xspace}
\newcommand{\Vb}{$V_{\text{b}}$\xspace}
\newcommand{\dVeq}{\Delta V\xspace}
\newcommand{\dV}{$\Delta V$\xspace}
\newcommand{\Neheq}{N_{\text{e/h}}\xspace}
\newcommand{\Neh}{$N_{\text{e/h}}$\xspace}
\def\evt{eV$_\text{t}$\xspace}

\def\evee{eV$_\text{ee}$\xspace}
\def\evnr{eV$_\text{nr}$\xspace}
\def\kevt{keV$_\text{t}$\xspace}

\def\kevee{keV$_\text{ee}$\xspace}
\def\kevnr{keV$_\text{nr}$\xspace}

\def\eveeeq{\text{eV}_\text{ee}\xspace}

\def\kevteq{\text{keV}_\text{t}\xspace}
\def\keveeeq{\text{keV}_\text{ee}\xspace}
\def\kevnreq{\text{keV}_\text{nr}\xspace}
\def\epgeq{\varepsilon_{\gamma}}
\def\cpdee{counts\,$\left[\keveeeq\,\text{kg\,d}\right]^{-1}$\xspace}

\def\cm2{cm$^\text{2}$\xspace}
\def\mev{MeV/$c^2$\xspace}
\def\gev{GeV/$c^2$\xspace}
\def\tev{TeV/$c^2$\xspace}

\def\cf{$^{252}$Cf\xspace}
\newcommand{\Om}[1]{$\mathcal{O}\left( #1 \right)$\xspace} 
\newcommand{\runOne}{Run~1\xspace}
\newcommand{\runTwo}{Run~2\xspace}
\newcommand{\perOne}{Period~1\xspace}
\newcommand{\perTwo}{Period~2\xspace}
\newcommand{\cdmsII}{CDMS\,II\xspace}
\newcommand{\geant}{G\textsc{eant}4}
\renewcommand{\d}[1]{\ensuremath{\operatorname{d}\!{#1}}}
\newcolumntype{d}[1]{D{,}{.}{#1}}
\newcolumntype{p}[1]{D{,}{\,\pm\,}{#1}}
\newcolumntype{h}[1]{D{,}{\text{--}}{#1}}
\begin{document}
\title{Low-mass dark matter search with CDMSlite}

\affiliation{Division of Physics, Mathematics, \& Astronomy, California Institute of Technology, Pasadena, CA 91125, USA}
\affiliation{Institute for Particle Physics Phenomenology, Department of Physics, Durham University, Durham  DH1 3LE, UK}
\affiliation{Fermi National Accelerator Laboratory, Batavia, IL 60510, USA}
\affiliation{Lawrence Berkeley National Laboratory, Berkeley, CA 94720, USA}
\affiliation{Department of Physics, Massachusetts Institute of Technology, Cambridge, MA 02139, USA}
\affiliation{School of Physical Sciences, National Institute of Science Education and Research, HBNI, Jatni 752050, India}
\affiliation{Department of Physics \& Astronomy, Northwestern University, Evanston, IL 60208-3112, USA}
\affiliation{Pacific Northwest National Laboratory, Richland, WA 99352, USA}
\affiliation{Department of Physics, Queen's University, Kingston, ON K7L 3N6, Canada}
\affiliation{Department of Physics, Santa Clara University, Santa Clara, CA 95053, USA}
\affiliation{SLAC National Accelerator Laboratory/Kavli Institute for Particle Astrophysics and Cosmology, Menlo Park, CA 94025, USA}
\affiliation{SNOLAB, Creighton Mine \#9, 1039 Regional Road 24, Sudbury, ON P3Y 1N2, Canada}
\affiliation{Department of Physics, South Dakota School of Mines and Technology, Rapid City, SD 57701, USA}
\affiliation{Department of Physics, Southern Methodist University, Dallas, TX 75275, USA}
\affiliation{Department of Physics, Stanford University, Stanford, CA 94305, USA}
\affiliation{Department of Physics, Syracuse University, Syracuse, NY 13244, USA}
\affiliation{Department of Physics and Astronomy, and the Mitchell Institute for Fundamental Physics and Astronomy, Texas A\&M University, College Station, TX 77843, USA}
\affiliation{Instituto de F\'{\i}sica Te\'orica UAM/CSIC, Universidad Aut\'onoma de Madrid, 28049 Madrid, Spain}
\affiliation{Department of Physics \& Astronomy, University of British Columbia, Vancouver, BC V6T 1Z1, Canada}
\affiliation{Department of Physics, University of California, Berkeley, CA 94720, USA}
\affiliation{Department of Physics, University of California, Santa Barbara, CA 93106, USA}
\affiliation{Departments of Physics and Electrical Engineering, University of Colorado Denver, Denver, CO 80217, USA}
\affiliation{Department of Physics, University of Evansville, Evansville, IN 47722, USA}
\affiliation{Department of Physics, University of Florida, Gainesville, FL 32611, USA}
\affiliation{Department of Physics, University of Illinois at Urbana-Champaign, Urbana, IL 61801, USA}
\affiliation{School of Physics \& Astronomy, University of Minnesota, Minneapolis, MN 55455, USA}
\affiliation{Department of Physics, University of South Dakota, Vermillion, SD 57069, USA}
\affiliation{Department of Physics, University of Toronto, Toronto, ON M5S 1A7, Canada}
\affiliation{TRIUMF, Vancouver, BC V6T 2A3, Canada}

\author{R.~Agnese} \affiliation{Department of Physics, University of Florida, Gainesville, FL 32611, USA}
\author{A.J.~Anderson} \affiliation{Fermi National Accelerator Laboratory, Batavia, IL 60510, USA}
\author{T.~Aralis} \affiliation{Division of Physics, Mathematics, \& Astronomy, California Institute of Technology, Pasadena, CA 91125, USA}
\author{T.~Aramaki} \affiliation{SLAC National Accelerator Laboratory/Kavli Institute for Particle Astrophysics and Cosmology, Menlo Park, CA 94025, USA}
\author{I.J.~Arnquist} \affiliation{Pacific Northwest National Laboratory, Richland, WA 99352, USA}
\author{W.~Baker} \affiliation{Department of Physics and Astronomy, and the Mitchell Institute for Fundamental Physics and Astronomy, Texas A\&M University, College Station, TX 77843, USA}
\author{D.~Balakishiyeva} \affiliation{Department of Physics, Southern Methodist University, Dallas, TX 75275, USA}
\author{D.~Barker} \affiliation{School of Physics \& Astronomy, University of Minnesota, Minneapolis, MN 55455, USA}
\author{R.~Basu~Thakur} \affiliation{Fermi National Accelerator Laboratory, Batavia, IL 60510, USA}\affiliation{Department of Physics, University of Illinois at Urbana-Champaign, Urbana, IL 61801, USA}
\author{D.A.~Bauer} \affiliation{Fermi National Accelerator Laboratory, Batavia, IL 60510, USA}
% ...

\smallskip

\date{\today}

\collaboration{SuperCDMS Collaboration}
\noaffiliation

\smallskip

\begin{abstract}
The SuperCDMS experiment is designed to directly detect weakly interacting massive particles (WIMPs) that may constitute the dark matter in our Galaxy. During its operation at the Soudan Underground Laboratory, germanium detectors were run in the CDMSlite mode to gather data sets with sensitivity specifically for WIMPs with masses ${<}10$~\gev.  In this mode, a higher detector-bias voltage is applied to amplify the phonon signals produced by drifting charges. This paper presents studies of the experimental noise and its effect on the achievable energy threshold, which is demonstrated to be as low as 56~\evee (electron equivalent energy). The detector-biasing configuration is described in detail, with analysis corrections for voltage variations to the level of a few percent. Detailed studies of the electric-field geometry, and the resulting successful development of a fiducial parameter, eliminate poorly measured events, yielding an energy resolution ranging from ${\sim}9$~\evee at 0~keV to $101$~\evee at ${\sim}10$~\kevee. New results are derived for astrophysical uncertainties relevant to the WIMP-search limits, specifically examining how they are affected by variations in the most probable WIMP velocity and the Galactic escape velocity. These variations become more important for WIMP masses below $10$~\gev. Finally, new limits on spin-dependent low-mass WIMP-nucleon interactions are derived, with new parameter space excluded for WIMP masses $\lesssim$3~\gev.
\end{abstract}

\pacs{95.35.+d, 14.80.Ly, 29.40.Wk, 95.55.Vj}

\maketitle



\section{Introduction}
\label{sec:intro}

In the last few decades, astronomical observations have consistently indicated that most of the matter content of the Universe is nonluminous and nonbaryonic \dm~\cite{Olive2014,Ade2015}. There is strong evidence that \dm is distributed in large halos encompassing the visible matter in galaxies, including the Milky Way.  If this \dm is composed of particles that interact with normal matter through a nongravitational force, it may be possible to directly detect it in laboratory experiments. 

The first generation of direct detection experiments searched for dark matter in the form of weakly interacting massive particles (\ws), with particle masses spanning from a few \gev to a few \tev, and interaction strengths with normal matter less than the weak force~\cite{Jungman1996,Goodman1985}. These searches were partly motivated by supersymmetric theories in which the lightest neutral particles are WIMPs and thus natural dark matter candidates.  However, no confirmed WIMP signals have been found, and there is no evidence as yet for supersymmetry at the LHC~\cite{Aad2015,Khachatryan2015}.

Other theoretical models have been developed, motivated by possible symmetries between normal and dark matter (\eg asymmetric dark matter~\cite{Kaplan2009}) or the possibility of a parallel “dark sector” that may contain many dark matter particles~\cite{Alexander2016}. These new models predict dark matter particles with masses ${<}10$~\gev, stimulating experiments to search in this region.

\ws are expected to scatter elastically and coherently from atomic nuclei, producing nuclear recoils (NRs). Neutrons also produce nuclear recoils, but often scatter multiple times in a detector; \ws interact too weakly to scatter more than once. Residual radioactivity in the experimental apparatus predominantly interacts with atomic electrons, causing electron recoils (ERs) that are the dominant source of background. Experiments try to reduce the rate of all backgrounds using layers of radiopure shielding and through the detection of multiple types of signals to discriminate between electron and nuclear recoils.
 
The nuclear-recoil energy spectrum expected from simple \w models is featureless and quasiexponential~\cite{Jungman1996,Lewin1996}.  The differential nuclear-recoil rate is
\begin{equation}
	\frac{\d{R}}{\d{\Ereq}} = \frac{N_T m_T}{2m_{\chi}\mu_T^2}\left[\sigma_0^{\text{SI}}F_{\text{SI}}^2{\left(\Ereq\right)}+\sigma_0^{\text{SD}}F_{\text{SD}}^2{\left(\Ereq\right)}\right]\mathcal{I}_{\text{halo}},
	\label{eq:wimpRate}
\end{equation}
where $m_{\chi}$ and $m_T$ are the masses of the \w and the target nucleus, respectively, $\mu_T=m_{\chi}m_T/\left(m_{\chi}+m_T\right)$ is the reduced mass of the \w-target system, $N_T$ is the number of nuclei per target mass, and \Er is the energy of the recoiling nucleus.  The spin-independent (SI) and spin-dependent (SD) cross sections for the \w-nucleus scattering are each factored into a total zero-energy cross section $\sigma_0^{\text{SI/SD}}$ and nuclear form factor $F_{\text{SI/SD}}^2{\left(\Ereq\right)}$.

The rate's dependence on the astrophysical description of the \w halo is encompassed by the halo-model factor $\mathcal{I}_{\text{halo}}$.  This factor depends on the velocities of the \ws in the halo's frame $\bm{v}$ and the velocity of the Earth with respect to the halo $\bm{v}_E$ as
\begin{equation}
	\mathcal{I}_{\text{halo}} = \frac{\rho_0}{k}\int_{v_{\text{min}}}^{v_{\text{max}}}\frac{f{\left(\bm{v},\bm{v}_E\right)}}{v}\d{^3\bm{v}},
	\label{eq:astroIdef}
\end{equation}
where $\rho_0$ is the local dark matter mass density, $k$ is a normalization constant, and the halo's velocity distribution with respect to the Earth $f{\left(\bm{v},\bm{v}_E\right)}$ is integrated from the minimum $v_{\text{min}}$ to the maximum $v_{\text{max}}$ \w velocities that can cause a recoil of energy \Er.  The maximum velocity is related to the Galactic escape velocity $v_{\text{esc}}$, while the minimum velocity is $v_{\text{min}}=\sqrt{m_T\Ereq/2\mu_T^2}$.  Assuming the standard Maxwellian velocity distribution with a characteristic velocity $v_0$ (see \SEC\ref{sec:halo}) gives an expression for $\mathcal{I}_{\text{halo}}$ as~\cite{Donato1998}
\begin{widetext}
\begin{equation}
	\mathcal{I}_{\text{halo}} = \frac{k_0}{k}\frac{\rho_0}{2yv_0}
	\begin{cases}
		\text{erf}{\left(x+y\right)}-\text{erf}{\left(x-y\right)}-\frac{4}{\sqrt{\pi}}ye^{-z^2} 		& 0< x < z-y \\
		\text{erf}{\left(z\right)}-\text{erf}{\left(x-y\right)}-\frac{2}{\sqrt{\pi}}\left(y+z-x\right)e^{-z^2}	& z-y < x < y+z \\
		0															& y+z < x,
	\end{cases}
	\label{eq:astroIsolve}
\end{equation}
\end{widetext}
where  $x=v_{\text{min}}/v_0$, $y=v_E/v_0$, $z=v_{\text{esc}}/v_0$, $k_0=\left(\pi v_0^2\right)^{3/2}$, and $k=k_0\left[\text{erf}{\left(z\right)}-\left(2/\sqrt{\pi}\right)z\exp{\left(-z^2\right)}\right]$.  The final case in this expression is set to zero to avoid unphysical negative rates.

\begin{figure}
	\centering
	\includegraphics[width=\columnwidth]{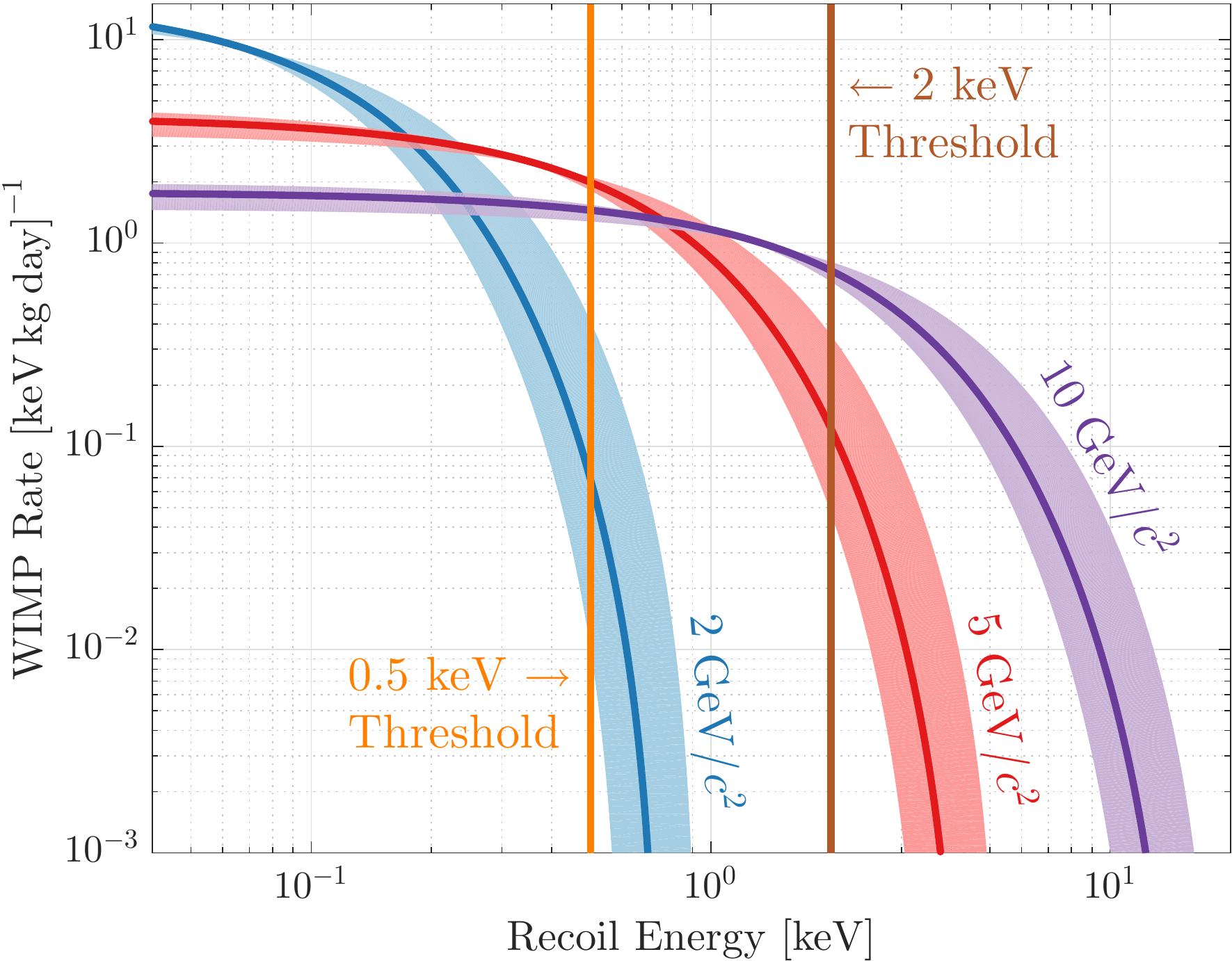}
	\caption{Differential rates for \w recoils on a germanium target as functions of recoil energy.  \ws with \w-nucleon spin-independent cross section of $10^{-41}$~\cm2 and masses of 2, 5, and 10 \gev are considered.  The bands encompassing each curve are computed by varying the astrophysical parameters of the dark matter halo within known observational uncertainties.  The vertical lines designate example nuclear-recoil thresholds of 0.5 and 2 keV, respectively.}
	\label{fig:light_wimp_1}
\end{figure}

Figure~\ref{fig:light_wimp_1} shows the predicted differential rates on a germanium target for three \lws with spin-independent \w-nucleon cross-sections of $10^{-41}$~\cm2.  Lowering the experimental energy threshold boosts the signal-to-background ratio, assuming a flat background spectrum, and reduces the dependence of the \w signal on astrophysical uncertainties.  A lower threshold thus dramatically increases an experiment's sensitivity to lower-mass WIMPs.

The Cryogenic Dark Matter Search low ionization threshold experiment (CDMSlite) uses a technique developed by the SuperCDMS Collaboration to reduce the experiment's energy threshold and increase sensitivity to \lws~\cite{Agnese2014,Agnese2016}.  This paper presents further details of the published CDMSlite analyses and some new results.  The organization of the paper is as follows.  Section~\ref{sec:technique} discusses the experimental technique, CDMSlite data sets, and data-reduction improvements.  Section~\ref{sec:noise} discusses the analysis and removal of noise.  Section~\ref{sec:threshAndRes} discusses an energy resolution model and energy thresholds.  Section~\ref{sec:bias} discusses the effects of bias instability in the analyses and the steps taken to account for those effects.  Section~\ref{sec:backfid} discusses the definition of a fiducial volume and its effect on backgrounds.  Finally, new WIMP results are given in \SEC\ref{sec:results} based on the effects of astrophysical uncertainties on the spin-independent \w-nucleon scattering limit presented in \REF\cite{Agnese2016} and new spin-dependent \w-nucleon scattering limits.

\section{Description of the experiment}
\label{sec:technique}

The SuperCDMS Soudan experiment was located at the Soudan Underground Laboratory and used the same cryogenics system, shielding, and electronics as the earlier \cdmsII experiment~\cite{Akerib2005,Agnese2015a}.  Five towers, each consisting of three germanium interleaved Z-sensitive ionization and phonon detectors (iZIPs), were operated from 2011 to 2015~\cite{Agnese2013a}.  Each iZIP was roughly cylindrical with a ${\sim}$76~mm diameter, ${\sim}$25~mm height, and ${\sim}$600~g mass. Particle interactions in these semiconductor crystals excite \eh charge pairs as well as lattice vibrations (phonons). The top and bottom circular faces of an iZIP are instrumented with electrodes for sensing the charge signal and tungsten transition edge sensors (TESs) for measuring phonons. The electrons and holes are drifted to the electrodes by applying a bias voltage across the crystal (nominally 4 V), while athermal phonons are absorbed by Al fins that are coupled to the TESs.  During data taking, the output traces from the detectors were recorded (``triggering'' the experiment) if the analog sum of any detector's raw phonon traces exceeded a user-set hardware threshold~\cite{Bauer2011}.

Measuring both the charge and phonon signals allows for discrimination between NRs and ERs through the ionization yield $Y$:
\begin{equation}
	Y{\left(E_{\text{r}}\right)}\equiv\frac{\Eqeq}{\Ereq},
    \label{eq:yield}
\end{equation}
where \Eq is the charge signal, and, for electron recoils, $\Eqeq\equiv\Ereq$.  The efficiency of producing electron-hole pairs is lower for nuclear recoils, leading to yields of $Y\sim0.3$ for $\Ereq\gtrsim10~\text{keV}$.  Below this energy, electronic noise causes the widths of the ER and NR populations to increase until they largely overlap at ${\sim}1~\text{keV}$, and complex background modeling must be used to separate the recoil types~\cite{Agnese2014a}.  This, coupled with the additional difficulty of separating low-energy events from noise, requires the typical iZIP analysis threshold to be set above the overlap region.

\subsection{CDMSlite}
\label{sec:CDMSlite}

\begin{figure}
	\centering
	\includegraphics[width=0.8\columnwidth]{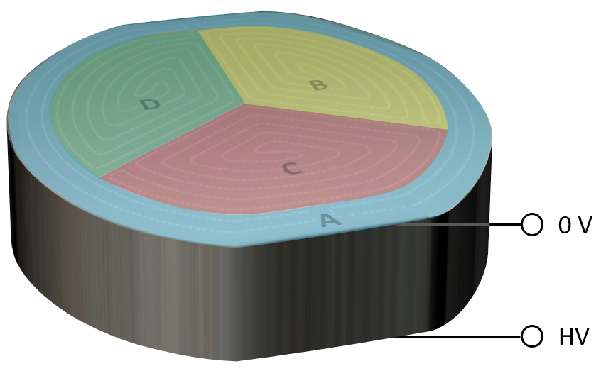}
	\caption{Schematic showing the general coverage of the four phonon read-out channels ($A$--$D$) overlaying the sensor pattern for the CDMSlite detector. The sensors on the bottom side are exclusively used for applying the high-voltage bias (HV) and are not read out.  All sensors on the top side are held at ground.}
	\label{fig:CDMSlite_Det}
\end{figure}

In 2012, SuperCDMS began running detectors in the alternate CDMSlite operating mode, where the detector potential difference was raised to 50--80~V.  The standard iZIP electronics and biasing configuration were adapted for this higher-voltage operating mode; phonon and ionization sensors on one side of the detector were set to the given bias, while all of the sensors on the opposite face were held near ground potential. Figure~\ref{fig:CDMSlite_Det} shows the phonon sensor layout and biasing scheme of the CDMSlite detectors. The sensors on the grounded side of the detector were then read out.  The limitations of the \cdmsII electronics board prohibited two-sided operation as the board could not simultaneously be floated to a potential and read out.

The CDMSlite operating mode takes advantage of phonon amplification via the Neganov-Trofimov-Luke (NTL) effect~\cite{Neganov1985,Luke1988}. Electrons and holes liberated by the initial recoil drift across the detector, driven by the applied electric potential.\footnote{This discussion of electron and hole transportation in a germanium crystal is taken from the rigorous calculations in \REF\cite{Sundqvist2012}.  See, \eg, Chaps. 2 and 4 of the reference for further details.}  During transport, they collide with Ge atoms and reach a scattering-limited drift velocity of \Om{10^6~\text{cm\,s}^{-1}} in ${\lesssim}1$~ns. When the kinetic energy of the charge carriers is ${\gtrsim}30$~meV, high rates of optical and intervalley phonon scattering limit further acceleration and cause them to reach a terminal velocity.  The additional work done in drifting these charge carriers, as they collide with the lattice (50--80~eV per \eh pair at the biases under discussion), is emitted as phonons. The residual kinetic energy of ${\sim}30$~meV per \eh pair, along with the band gap energy of 0.74~eV~\cite{Kittel2005}, is eventually released as phonons, called relaxation or recombination phonons, when the charge carriers relax to the Fermi sea near detector boundaries. The phonons emitted during charge transport are called NTL phonons, and the net energy in these phonons, \EL , is the work done by the electric field
\begin{equation}
	\ELeq = \Neheq e\dVeq.
\end{equation}
Here, \Neh is the number of \eh pairs created in the recoil, $e$ is the elementary charge, and \dV is the potential difference traversed by the pairs.  \dV is nominally the absolute value of the bias applied by the power supply \Vb.  The advantage of operating at relatively high bias potentials is an amplification of the charge signal (as observed in the phonon signal) due to increased NTL-phonon production.

The total phonon energy in the crystal is thus the sum of ionization-associated NTL phonons, primary phonons created at the initial recoil site, and relaxation phonons created near detector surfaces.  The sum of the primary and relaxation phonons is \Er and thus the total energy is
\begin{equation}
	\Eteq = \Ereq + \ELeq = \Ereq + \Neheq e\dVeq.
    \label{eq:Etot}
\end{equation}
The number of \eh pairs created by a recoil depends on the recoil type.  For electron recoils in germanium, the average (photoexcitation) energy required to generate a single \eh pair is taken to be $\epgeq=3~\text{eV}$~\cite{Emery1965,*Pehl1968}.  This gives $\Neheq=\Eqeq/\epgeq=Y{\left(\Ereq\right)}\Ereq/\epgeq$, where \EQ\ref{eq:yield} is used for the second equality.  Substituting this last expression into \EQ\ref{eq:Etot} gives
\begin{equation}
	\Eteq = \Ereq\left(1+Y{\left(\Ereq\right)}\frac{e\dVeq}{\epgeq}\right).
    \label{eq:totAmp}
\end{equation}
As only one of two faces of an iZIP are read out in CDMSlite mode, the energy absorbed by the operable phonon sensors is half that of \EQ\ref{eq:totAmp}.

The calibration of the measured phonon signal proceeds in three steps, with three corresponding energy scales, using \EQ\ref{eq:totAmp} assuming $\dVeq=\Vbeq$.  The first step is to convert the raw output to the ``total phonon energy scale,'' with units of \kevt, using calibration data taken at the standard operating bias of 4~V and the expectation from \EQ\ref{eq:totAmp} (see \SEC\ref{sec:totEscale}).  Converting the calibrated \Et to the interaction's \Er requires knowledge of the yield. Because CDMSlite only measures phonons, the yield cannot be constructed on an event-by-event basis and a model for $Y{\left(\Ereq\right)}$ is required.  Two further energy scales are defined corresponding to the assumed ER/NR recoil type.  The ER scale is stretched considerably compared to the NR scale with its smaller \eh production efficiency; this further increases the signal-to-background ratio for CDMSlite.

The recoil energies are next calibrated assuming all events are ERs, \ie, $Y{\left(\Ereq\right)}=1$, called ``electron-equivalent'' energy in units of \kevee and denoted by \Eree.  This scale is useful for characterization of the backgrounds, which are primarily ERs.  An ER calibration is available from electron-capture decays of $^{71}$Ge.  Thermal neutron capture on $^{70}$Ge (20.6\,\% natural abundance) creates $^{71}$Ge, which then decays by electron-capture with a half-life of 11.43 days~\cite{Hampel1985}.  The $K$-, $L$-, and $M$-shell binding energies of the resulting $^{71}$Ga are 10.37, 1.30, and 0.16~keV, respectively~\cite{Bearden1967}.  In the experiment, $^{71}$Ge was created in the detector by exposing it to a $^{252}$Cf source two to five times per CDMSlite data set.  The $K$-shell peak, clearly visible in the data following such an activation, is used to calibrate the energy scale to \kevee and to correct for any changes in the energy scale with time (see \SEC\ref{sec:bias}).

WIMP scatters are expected to be NRs; so a nuclear-recoil energy is ultimately constructed, called ``nuclear-recoil equivalent'' energy in units of \kevnr and denoted by \Ernr.  The calibration to \kevnr is performed by comparing \EQ\ref{eq:totAmp}, assuming the detector sees the full \Vb bias, for an ER and NR with the same \Et, and solving for \Ernr,
\begin{equation}
	\Ernreq = \Ereeeq\left(\frac{1+e\Vbeq/\epgeq}{1+Y{\left(\Ernreq\right)}e\Vbeq/\epgeq}\right),
	\label{eq:ee2nr}
\end{equation}
where $Y{\left(\Ernreq\right)}$ is the yield as a function of nuclear-recoil energy, for which a model is needed.  The model used is that of Lindhard~\cite{Lindhard1963,*Lindhard1963a,*Lindhard1968}
\begin{equation}
	Y{\left(\Ernreq\right)} = \frac{k\cdot g{\left(\varepsilon\right)}}{1+k\cdot g{\left(\varepsilon\right)}},
    \label{eq:lindhard}
\end{equation}
where $g{\left(\varepsilon\right)}=3\varepsilon^{0.15}+0.7\varepsilon^{0.6}+\varepsilon$, $\varepsilon=11.5\Ernreq{\left(\kevnreq\right)}Z^{-7/3}$, and $Z$ is the atomic number of the material.  For germanium, $k=0.157$.  The Lindhard model has been shown to roughly agree with measurements in germanium down to ${\sim}$250~\evnr~\cite{Barker2012,Soma2016}, although measurements in this energy range are difficult, and relatively few exist~\cite{Jones1975,Barbeau2007,Scholz2016}. The SuperCDMS Collaboration has a campaign planned to directly measure the nuclear-recoil energy scale for germanium (and silicon) down to very low energies, since this will be required for the upcoming SuperCDMS SNOLAB experiment.

\subsection{Data Sets and Previous Results}
\label{sec:runs}

A single detector was operated in CDMSlite mode during two operational periods, \runOne in 2012 and \runTwo in 2014.\footnote{Only a single detector was operated for each run due to limitations of the Soudan electronics and to preserve the live time for the standard iZIP data taken concurrently.}  The initial analyses of these data sets, published in \REFS\cite{Agnese2014,Agnese2016}, respectively, applied various selection criteria (cuts) to the data sets and used the remaining events to compute upper limits on the SI WIMP-nucleon interaction.  These limits were computed using the optimum interval method~\cite{Yellin2002,*Yellin2007}, the nuclear form factor of Helm~\cite{Helm1956,Lewin1996}, and assuming that the SI interaction is isoscalar.  Under this last assumption, the WIMP-nucleon cross section $\sigma_N^{\text{SI}}$ is related to $\sigma_0^{\text{SI}}$ in \EQ\ref{eq:wimpRate} as $\sigma_0^{\text{SI}}=\left(A\mu_T/\mu_N\right)^2\sigma_N^{\text{SI}}$, where $\mu_N$ is the reduced mass of the WIMP-nucleon system.

\begin{figure}
	\centering
	\includegraphics[width=\columnwidth]{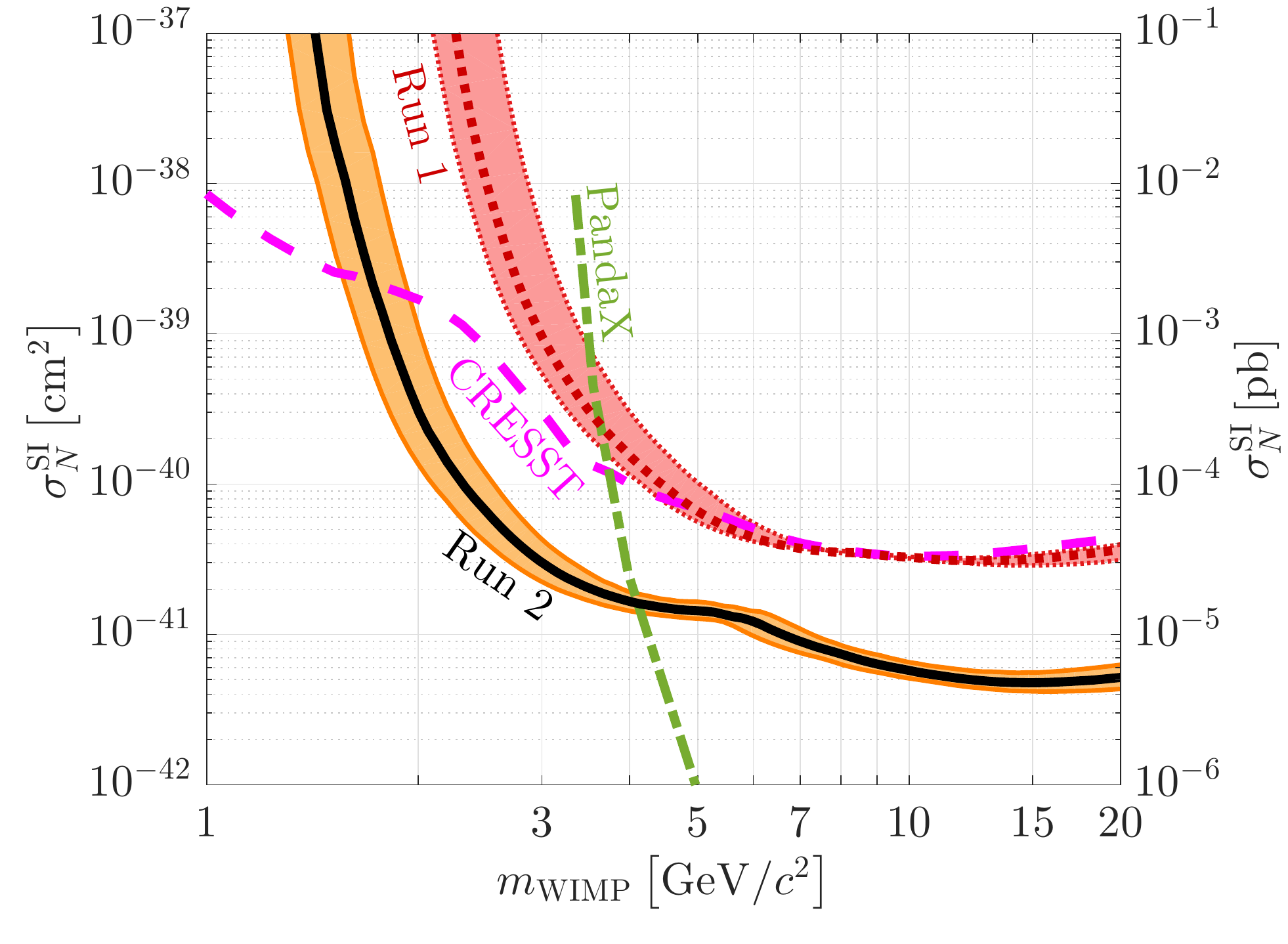}
	\caption{Spin-independent WIMP-nucleon cross section 90\,\% upper limits from CDMSlite \runOne (red dotted curve with red uncertainty band)~\cite{Agnese2014} and \runTwo (black solid curve with orange uncertainty band)~\cite{Agnese2016} compared to the other (more recent) most sensitive results in this mass region: CRESST-II (magenta dashed curve)~\cite{Angloher2016}, which is more sensitive than CDMSlite \runTwo for $m_{\text{WIMP}}\lesssim 1.7$~\gev, and PandaX-II (green dot-dashed curve)~\cite{Tan2016}, which is more sensitive than CDMSlite \runTwo for $m_{\text{WIMP}}\gtrsim4$~\gev.  The \runOne uncertainty band gives the conservative bounding values due to the systematic uncertainty in the nuclear-recoil energy scale.  The \runTwo band additionally accounts for the uncertainty on the analysis efficiency and gives the 95\,\% uncertainty on the limit.}
	\label{fig:limits_si}
\end{figure}

CDMSlite \runOne was a proof of principle and the first time WIMP-search data were taken in CDMSlite mode.  For \runOne, the detector was operated at a nominal bias of $-69$~V and an analysis threshold of 170~\evee was achieved.  In an exposure of just 6.25 kg\,d (9.56 kg\,d raw), the experiment reached the SI sensitivity shown in \FIG\ref{fig:limits_si} (labeled ``\runOne''), which was world leading for \ws lighter than 6~\gev at the time of publication~\cite{Agnese2014}.

\begin{figure}
	\centering
	\includegraphics[width=\columnwidth]{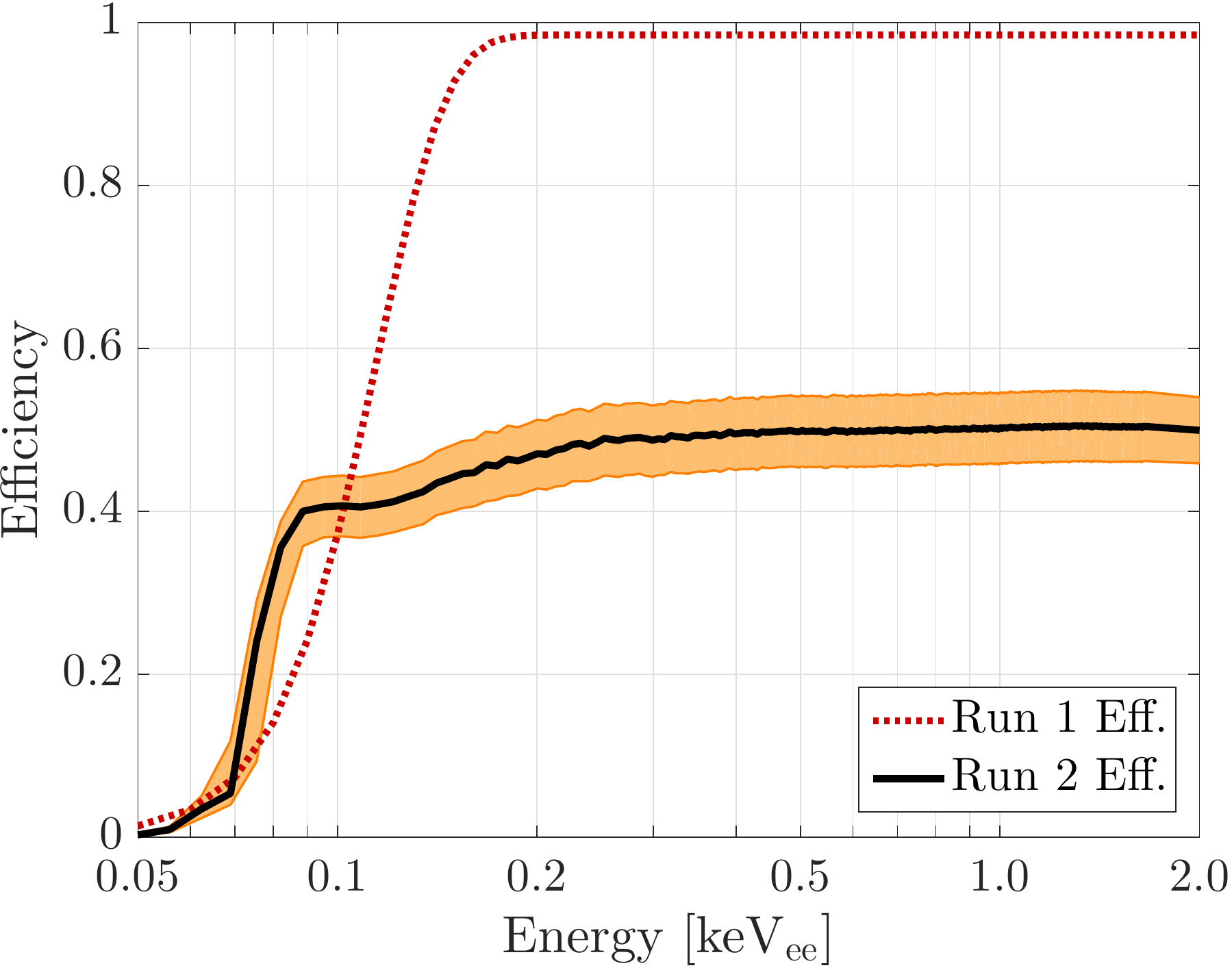}
	\caption{Total combined trigger and analysis efficiencies for \runOne (red dotted curve) and \runTwo (black solid curve with orange 68\,\% uncertainty band).  The implementation of a fiducial-volume cut is primarily responsible for the reduction in efficiency at high recoil energies between the two analyses.}
	\label{fig:effs}
\end{figure}

\begin{figure}
	\centering
	\includegraphics[width=\columnwidth]{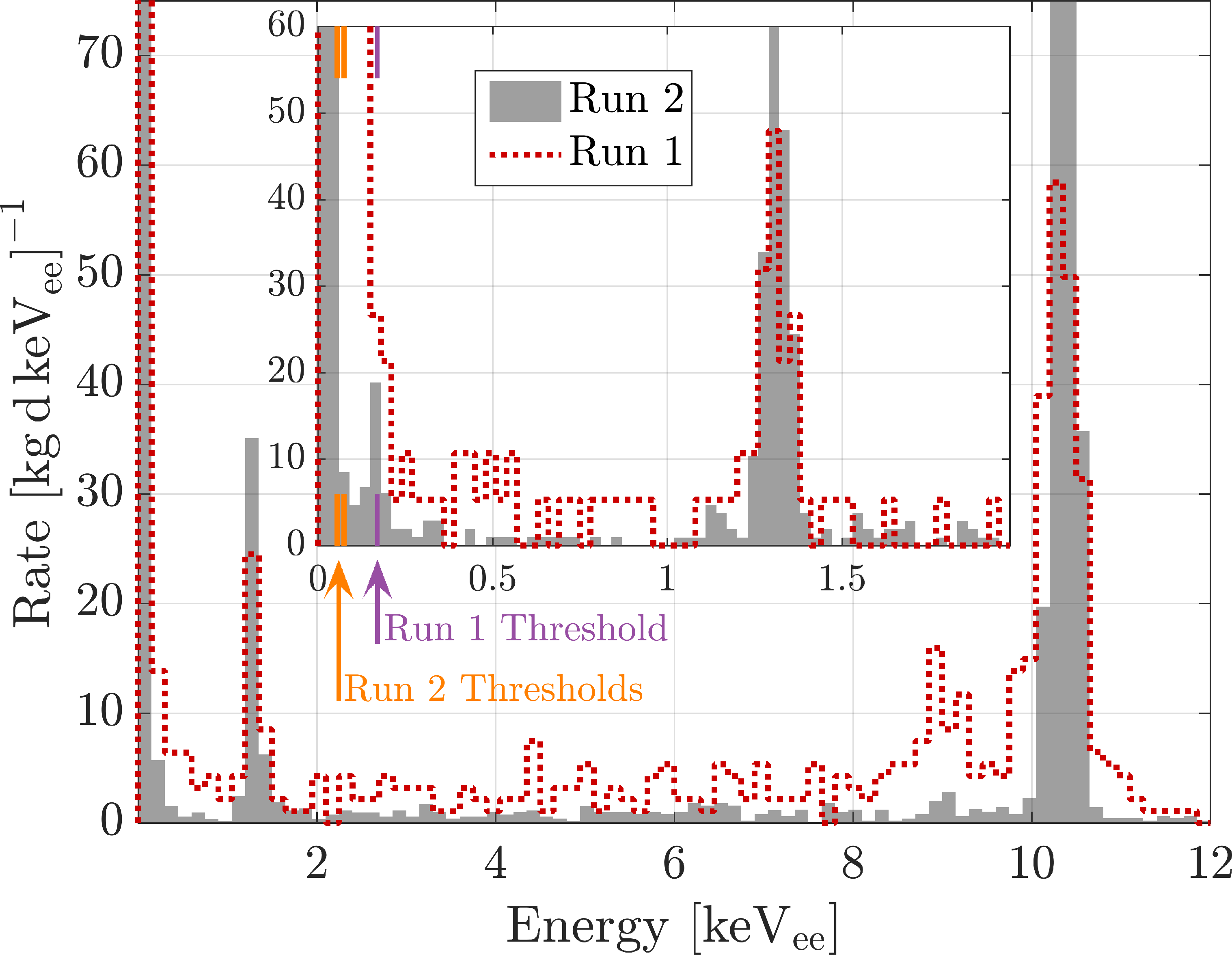}
	\caption{Measured efficiency-corrected spectra for \runOne (red dotted curve) and \runTwo (gray shaded area).  The $^{71}$Ge-activation peaks at 10.37 and 1.30~\kevee are prominent in both spectra, and the peak at 0.16~\kevee is additionally visible in the \runTwo spectrum.  The $^{65}$Zn $K$-shell electron-capture peak is also visible at 8.89~\kevee in the \runOne spectrum.  \textit{Inset:} an enlargement of the spectra below 2~\kevee with bins five times smaller and the runs' analysis thresholds given by the extended and labeled tick marks.}
	\label{fig:spectra}
\end{figure}

The total efficiency and spectrum from \runOne are shown in \FIGS\ref{fig:effs} and \ref{fig:spectra} respectively.  In addition to the $^{71}$Ge-activation peaks, the $K$-shell activation peak from $^{65}$Zn is visible in the \runOne spectrum at 8.89~\kevee~\cite{Bearden1967}.  The $^{65}$Zn was created by cosmic-ray interactions, with production ceasing once the detector was brought underground in 2011, and decayed with a half-life of $\tau_{1/2}\approx244~\text{d}$~\cite{Browne2010}.  The analysis threshold was set at 170~\evee to maximize dark matter sensitivity while avoiding noise at low energies (see \SEC\ref{sec:lfnoise}).  To compute upper limits, the conversion from \kevee to \kevnr was performed using the standard Lindhard-model $k$ value (\EQ\ref{eq:lindhard}) of 0.157.  Limits were also computed using $k=0.1$ and 0.2, chosen to represent the spread of experimental measurements~\cite{Barker2012,Soma2016,Jones1975,Barbeau2007,Scholz2016}, to bound the systematic due to the energy-scale conversion.  As shown in \FIG\ref{fig:limits_si}, this uncertainty has a large effect at the lowest WIMP masses.

In \runTwo, the detector was operated with a bias of $-70$~V, the analysis threshold was further reduced because of improved noise rejection, and a novel fiducial-volume criterion was introduced to reduce backgrounds.  The total efficiency and spectrum from this run are compared to those of the first run in \FIGS\ref{fig:effs} and \ref{fig:spectra}.  Because of the lower analysis threshold, decreased background, and a larger exposure of 70.10 kg\,d (80.25 kg\,d raw), the experiment yielded even better sensitivity to the SI interaction than \runOne\cite{Agnese2016}, as shown in \FIG\ref{fig:limits_si} (labeled ``\runTwo'').  The second run was split into two distinct data periods (see \SEC\ref{sec:lfnoise}), labeled ``\perOne'' and ``\perTwo,'' that had analysis thresholds of 75 and 56 \evee, respectively.

For the \runTwo result, the uncertainties of the analysis were propagated into the final limit by simulating 1000 pseudoexperiments and setting a limit with each.  The median and the central 95\,\% interval from the resulting distribution of limits, at each WIMP mass, are taken as the final result given in \FIG\ref{fig:limits_si}.  For each pseudoexperiment, the \kevee energy of the events and thresholds were constant. The analysis efficiencies, as indicated by the band in \FIG\ref{fig:effs}, were sampled, as was the Lindhard-model $k$ within a range of $0.1\leq k \leq 0.2$.  The uncertainty in the energy conversion dominates the band in \FIG\ref{fig:limits_si}, with the next-largest uncertainty being that of the fiducial-volume acceptance efficiency (\SEC\ref{sec:run2fid}).

\subsection{Pulse fitting and energy measurement}
\label{sec:PulseFit}

Several improvements were made in the analysis of \runTwo data, compared to that of the \runOne data, by the introduction of a new data-reduction algorithm used to extract energy and position information about scatters in the detector.  To motivate and understand this new algorithm, the dynamics of phonon detection and the older algorithms, which are still used for many parts of the analyses, are first discussed.

The phonon sensors cover only ${\sim}5\,\%$ of the surfaces of iZIP detectors. Phonons have a ${\sim}$40\,\% probability of absorption when they strike an aluminum sensor fin\footnote{This value of 40\,\% is determined by tuning a phonon simulation in a detector to match recorded pulses.  Specifically, how quickly pulses return to their baseline values is sensitive to this absorption probability.} but are reflected when striking an uninstrumented surface. The phonons continue to rebound between surfaces of the crystal until they are absorbed by, or become lost to, the sensors~\cite{Hertel2012}.  Phonons become undetectable by the sensors either by falling below the aluminum superconducting gap energy or by being absorbed through nonsensor materials (\eg, stabilizing clamps). The small fraction of phonons striking a fin at the first surface interaction produces an early absorption signal that is concentrated close to the location of the interaction, while the majority of the phonons contribute to a later absorption signal that is mostly homogeneous throughout the detector. The phonon pulse shape thus contains both position and energy information about the initial scatter in the earlier and later portions of the signal trace, respectively.

The CDMSlite analyses employ three algorithms based on optimal filter theory (see Appendix~B of \REF\cite{Golwala2000}) to extract the position and energy information of the underlying event based on the measured pulse shapes and amplitudes.  For these algorithms, the signal trace  $S{\left(t\right)}$ is generally modeled as a template, or linear combination of templates, $A{\left(t-t_0\right)}$, which can be shifted by some time delay $t_0$, and Gaussian noise $n{\left(t\right)}$ as
\begin{equation}
	S{\left(t\right)} = aA{\left(t-t_0\right)}+n{\left(t\right)},
	\label{eq:OF}
\end{equation}
where the template is scaled by some amplitude $a$.  The optimal values of $a$ and $t_0$ are then found by minimizing, in frequency space, the $\chi^2$ between the left- and right-hand sides of \EQ\ref{eq:OF}.  The amplitude, time delay, and goodness-of-fit $\chi^2$ value are returned by the algorithms.

\begin{figure}
	\centering
	\includegraphics[width=\columnwidth]{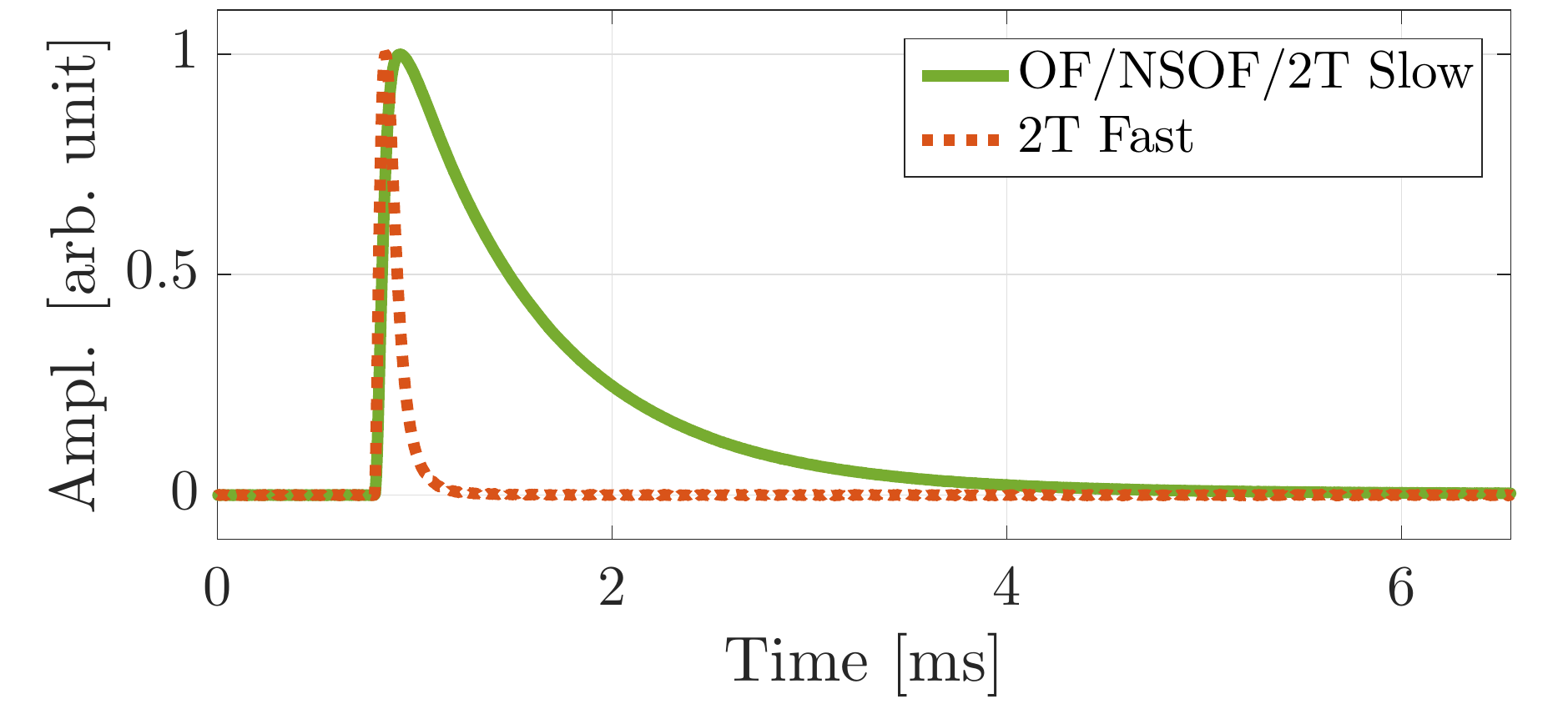}
	\caption{Templates used for the standard OF, NSOF, and 2T-fit algorithms for CDMSlite analysis.  The green solid curve is the single trace used for the OF, NSOF, and 2T-fit slow templates, which is derived from averaging high-energy traces.  In the 2T fit, the slow template's amplitude carries the main energy information.  The 2T-fit fast template (orange dotted), is derived by considering the differences between the slow template and the traces used in the slow template's derivation.   In the 2T fit, the fast template's amplitude captures the position information from the signal trace.  The maxima of the amplitudes (Ampl.) are scaled to unity in the figure.}
    \label{fig:2Ttemplates}
\end{figure}

The first algorithm is called the ``standard'' optimal filter (OF).  The OF algorithm fits a single template to a trace, as in \EQ\ref{eq:OF}, without attempting to account for the position dependence in the early portion of the trace.  The template was created by averaging a large number of high-energy traces taken from the $^{71}$Ge $K$-shell capture peak and can be seen in \FIG\ref{fig:2Ttemplates}.  The energy estimate from this fit, the amplitude $a$ in \EQ\ref{eq:OF}, has poor resolution because of the position dependence.  The position of an event's initial scatter in the detector can be estimated by fitting the traces from each individual channel of a given event and comparing the fit amplitudes among the channels: channels of which the sensors are nearer to the interaction will have a larger amplitude than those of which the sensors are farther away.

The second algorithm is called the ``nonstationary'' optimal filter (NSOF) (see Appendix~E of \REF\cite{Basuthakur2015}), and it produces an energy estimator that is less affected by the early-trace position dependence.  The NSOF uses the same single template as in the OF fit but treats the residual deviations between the trace and the template as nonstationary noise.  This procedure deweights the parts of the trace that show larger variance and results in a more accurate energy estimator.  Additionally, the NSOF fit is calculated only for the summed trace of each individual detector, which also serves to reduce, but does not completely eliminate, the effect of position dependence on the energy estimate.  The NSOF is not useful for computing position information about the initial scatter.

\begin{figure}
	\centering
	\includegraphics[width=\columnwidth]{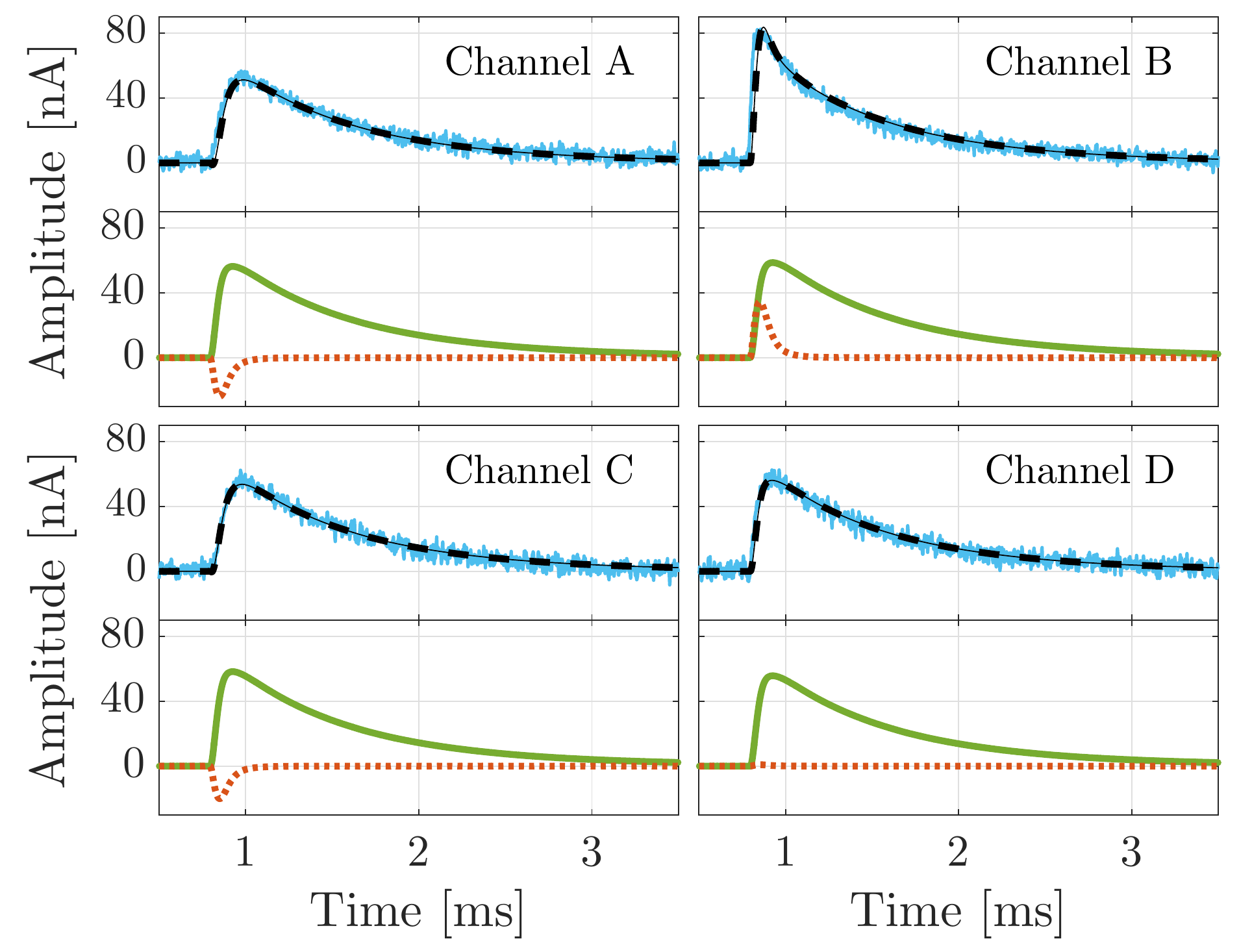}
	\caption{Results of the 2T-fit algorithm for an example event chosen from the $^{71}$Ge $L$-shell capture peak in \runTwo.  The traces and fits from all four phonon channels, labeled A--D (where channel A is the outer ring) are given. For each channel, the raw trace (blue solid) is compared to the final total fit (black dashed) which is a linear combination of the slow (green solid) and fast (orange dotted) templates.  The channel with the largest fast-template amplitude, channel B for this event, is the channel of which the sensors are closest to the initial recoil.}
    \label{fig:2Texample}
\end{figure}

The third algorithm, utilized for the first time with CDMSlite \runTwo data, is called the ``two-template'' optimal filter (2T fit) (see Appendix~E of \REF\cite{Basuthakur2015} and Chap.~10 of \REF\cite{Pepin2016}).  The 2T fit uses a linear combination of two different templates, replacing $aA{\left(t-t_0\right)}$ with $\sum_{i=s,f}a_iA_i{\left(t-t_0\right)}$.  The two templates are shown in \FIG\ref{fig:2Ttemplates} and are labeled the ``slow'' and ``fast'' templates.  The slow template is the same template used in the OF and NSOF fits.  The fast template is derived by considering the differences between the slow template and the traces used to define it, termed the residual traces.  To calculate this template, the residuals with negative amplitude are inverted before all residuals are averaged.  The inversion conserves the shape and is needed because the average of the residuals without the inversion is zero by definition.  The 2T fit returns an energy estimator---the amplitude of the slow template---which, like the NSOF, is less affected by the position of the initial scatter than the OF fit, but it also returns the amplitude of the fast template which encodes position information.  The 2T fit is applied to each individual channel's trace as well as the  summed trace.  An example of this fit is shown in \FIG\ref{fig:2Texample}. Negative fast-template amplitudes are expected in fit results and indicate greater distance from the initial scatter.

\begin{figure}
	\centering
	\includegraphics[width=\columnwidth]{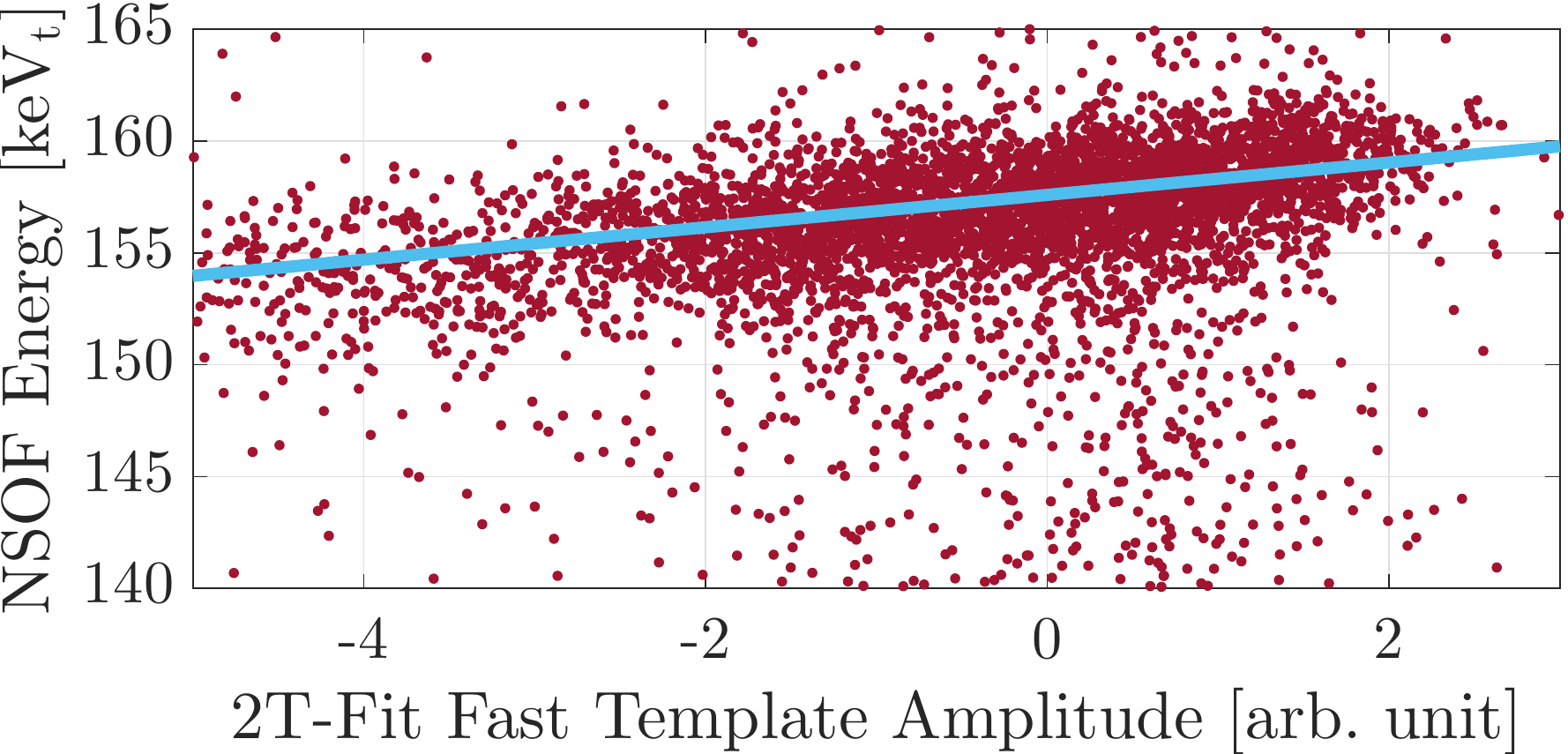}
	\caption{NSOF-fit energy estimator as a function of the 2T-fit fast-template amplitude from the summed trace.  The high-density band of events is the $^{71}$Ge $K$-shell activation line. Residual position dependence is reflected in the slope of the band. This dependence is corrected according to the straight-line fit shown by the solid line.  The location of the peak at ${\sim}155\kevteq$ is discussed in \SEC\ref{sec:totEscale}.}
	\label{fig:resAmpVsE}
\end{figure}

In the \runOne analysis, the energy estimator from the NSOF algorithm was used without any further corrections for position dependence.  For the \runTwo analysis, the NSOF energy estimator was again used, but an additional position correction was applied based on the 2T fit information.  As shown in \FIG\ref{fig:resAmpVsE}, a correlation between the fitted NSOF energy estimate and 2T-fit fast-template amplitude is observed.  The linear fit to this correlation is used for the correction.\footnote{The energy estimator extracted from the slow-template amplitude of the 2T fit has more position dependence than that of the NSOF, manifesting itself in a stronger correlation with the 2T-fit fast-template amplitude. After correcting for this correlation, the performance is very similar with a marginally better resolution of the NSOF-based algorithm in the $^{71}$Ge $K$-shell peak.}  In the \runTwo analysis, a cut was placed to remove events for which the NSOF fit returned large $\chi^2$ values to ensure that the energy estimator was reliable.  Such a cut removes events that have more than one pulse in the trace, or that exhibit a distorted pulse shape due to TES saturation.  The signal efficiency for the cut is near 100\,\% as computed via a pulse simulation that is described in \SEC\ref{sec:deltaX2}.  No poorly fit events were observed above threshold in the smaller \runOne WIMP-search data set, and thus such a cut was unnecessary.

\section{Study and removal of noise}
\label{sec:noise}

Understanding the noise in the readout wave forms is crucial for optimizing the low-energy analysis and achieving the desired low-energy thresholds using the CDMSlite technique.  Studies from both runs showed that the noise depended on both bias voltage and time.  Most crucially, cryocooler-induced \lf noise was present and limited the \runOne threshold.  A combination of timing correlations with the cryocooler and pulse-shape fitting was used in \runTwo to reject this background.

\subsection{Dependence of noise on bias potential}
\label{sec:biasscan}

\begin{figure}
	\centering
	\includegraphics[width=\columnwidth]{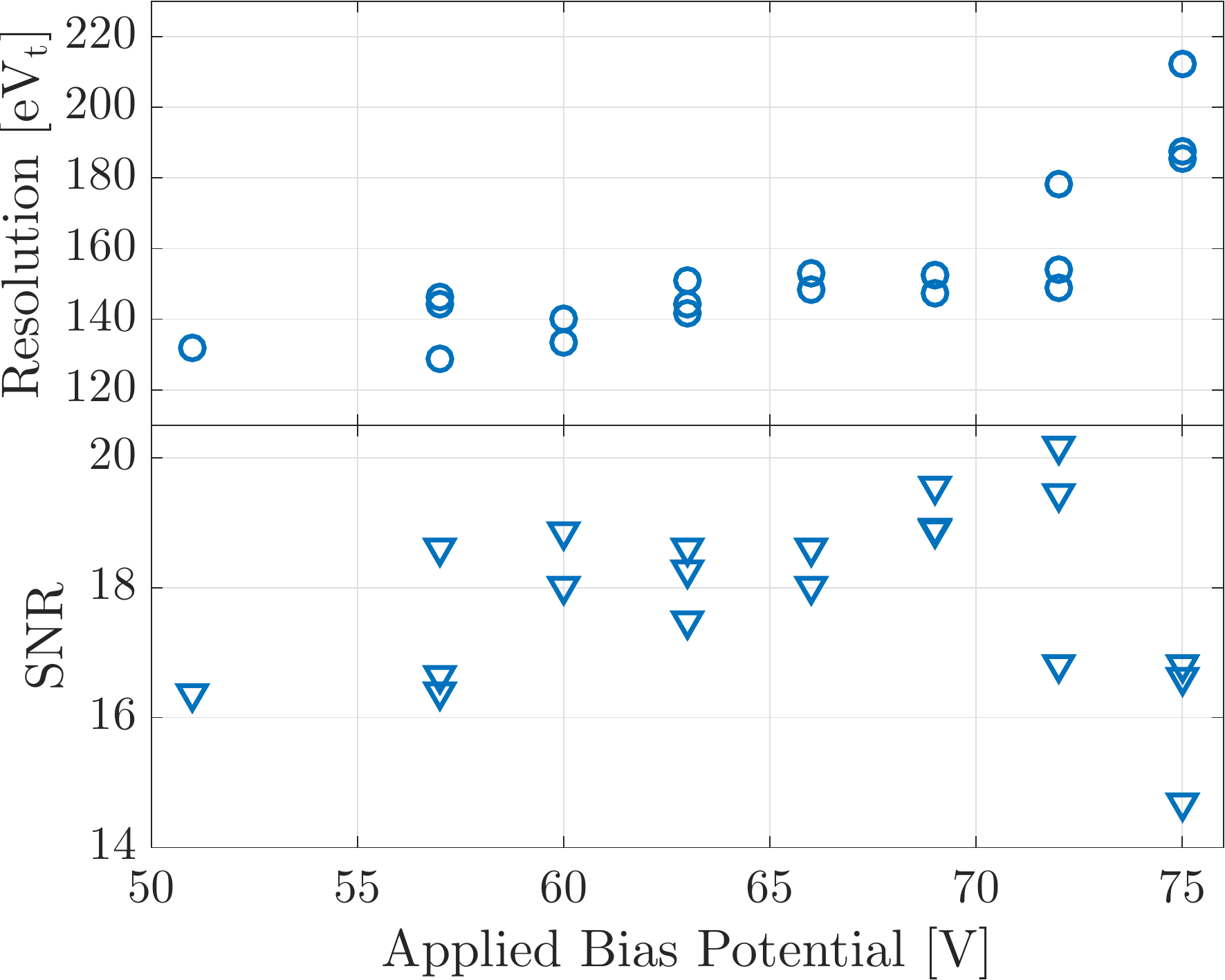}
	\caption{Baseline resolution (top) and the corresponding SNR (bottom) as a function of the applied bias potential.  Each point represents a single 3~h long data set taken prior to \runTwo.  The resolution and SNR increase and decrease, respectively, past ${\sim}$70~V in applied bias.  The average uncertainty for each point is 3.6~\evt for the resolution and 0.39 for the SNR.  The additional variation seen at a given bias is likely a result of time dependence of the noise.  For reference, $1~\kevteq\approx66~\eveeeq$.}
	\label{fig:biasScan_run2}
\end{figure}

The operating potential difference for each run was determined by studying the noise as a function of the applied potential difference.  The baseline resolution as a function of this potential difference is shown in \FIG\ref{fig:biasScan_run2} for data taken prior to \runTwo.  The resolution slowly increased until the potential difference passed ${\sim}$70~V, where a larger increase was observed.  Taking the potential difference up to 85~V resulted in greatly increased noise signaling the start of detector breakdown.  A recoil-energy-independent signal-to-noise ratio (SNR) was also considered by comparing the measured signal and noise to the $\Vbeq=0~\text{V}$ case.  The signal, according to \EQ\ref{eq:totAmp} (assuming a yield of unity), was then $1+e\Vbeq/\epgeq$.  The noise was the measured resolution in \FIG\ref{fig:biasScan_run2} divided by an assumed zero-volt resolution of 120~\evt.  The SNR is also shown in \FIG\ref{fig:biasScan_run2}, with a peak SNR at ${\sim}$70~V.  These studies were used to determine the operating potential differences of 69 and 70~V for the two runs respectively.

o
\subsection{Time dependence of noise}
\label{sec:prebias}

For iZIP detectors, the charge collection efficiency deteriorated after being biased and operated for longer than ${\sim}$3~h.  This decrease in collection efficiency was caused by charges becoming trapped on impurity sites in the crystal instead of drifting fully to the electrodes~\cite{Abrams2002}.  To avoid the collection efficiency loss, data were taken in 3~h long periods called ``series.''  At the end of each series, the detectors were grounded and exposed to photons from light emitting diodes.  These photons created excess electron-hole pairs that neutralized the impurity sites.  This light exposure increased the temperature of the detectors, and a 10~min cool-down period was required before beginning the next series.  In detectors operated in CDMSlite mode, trapped charges resulted in excess noise, and steps were developed to minimize this effect.

\begin{figure}
	\centering
	\includegraphics[width=\columnwidth]{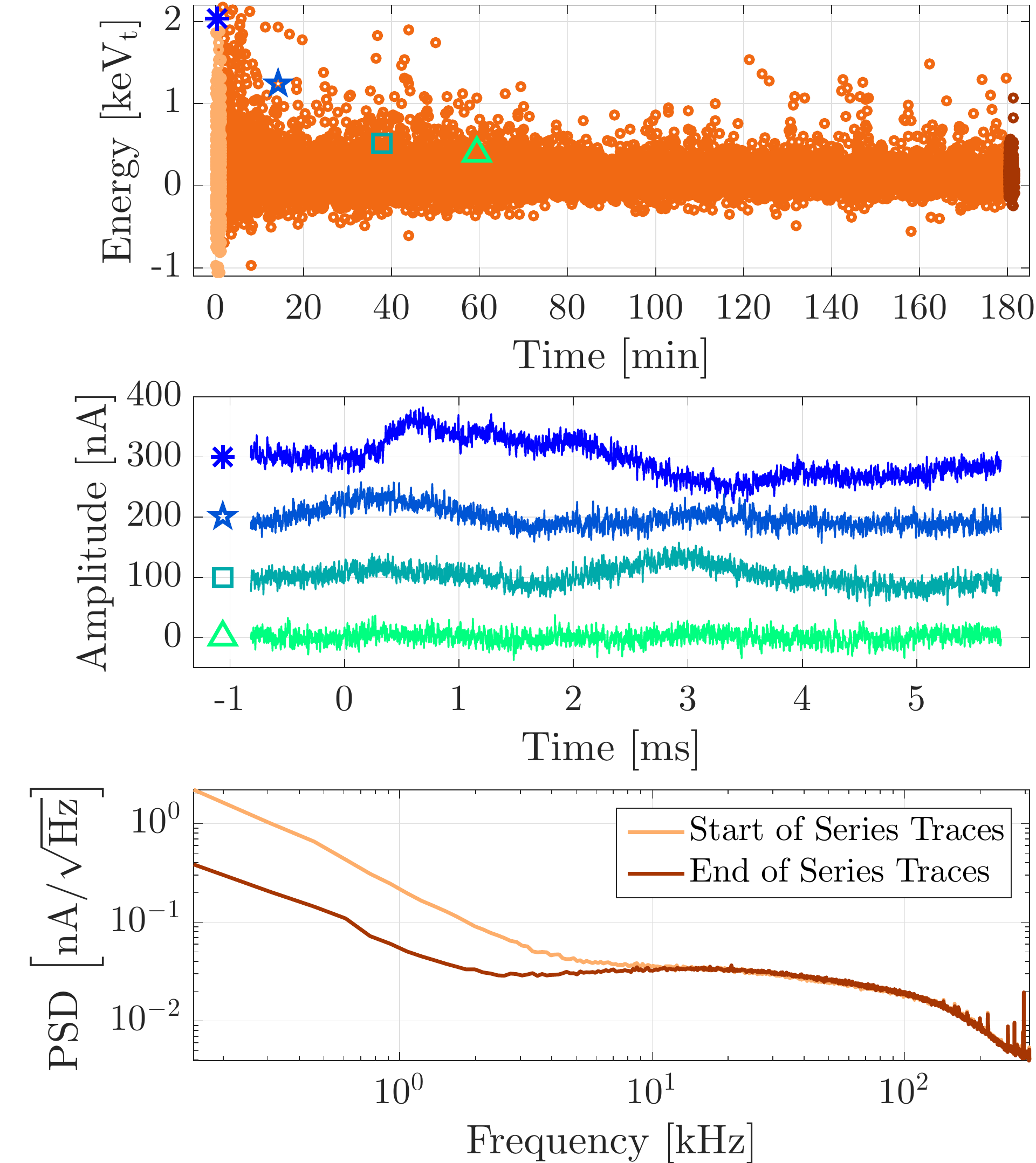}
	\caption{\textit{Top:} total phonon energy, or noise, as a function of time since biasing in \runOne. The noise decays quasiexponentially with time; four example events are given by noncircular markers.  The first and last 500 traces are highlighted in light and dark orange, respectively.  The noise distribution is offset from 0~\kevt as the energy-estimating algorithm tends to fit to upward noise fluctuations.  For reference, $1~\kevteq\approx66~\eveeeq$.  \textit{Middle:} raw traces of the events marked in the top panel.  Traces are shifted by 100~nA with respect to each other for clarity.  \textit{Bottom:} power spectral densities (PSDs) for the noise at the start (light orange) and end (dark orange) of the series.  The earlier traces have more power below ${\sim}$10~kHz.}
    \label{fig:noise_exp}
\end{figure}

During \runOne operation, the noise in the CDMSlite detector was seen to be excessively high immediately after the detector was biased to its fixed operating point at the start of a series. The noise decayed quasiexponentially with time, presumably due to the tunneling of trapped charges, until an asymptotic level was achieved (see Appendix~B of \REF\cite{Basuthakur2015}).  Noise-trace data from a typical series are shown in \FIG\ref{fig:noise_exp}, where the reconstructed energy has higher rms earlier in the series. The excess noise amplitude decayed with an exponential time constant $\tau\sim10~\text{min}$. In \runOne, the data taken during the first 4$\tau$ following the application of the bias voltage were discarded, as a balance between live time and optimal baseline resolution. Thus, in \runOne, only ${\sim}70\,\%$ of the data collected could be used for the analysis.

In \runTwo, the high initial noise was avoided by holding the detector at a larger potential difference than the operating voltage prior to the start of each series, after which the bias was dropped to the operating voltage.  Under the assumption that the initial noise is due to the release of trapped charges, this initial bias at higher potential difference allows for all traps accessible at the lower potential difference to be cleared.  This operational procedure is termed ``prebiasing'' and the SuperCDMS data acquisition system(DAQ) was configured to prebias before each data series in \runTwo.  The prebiasing procedure was as follows:
\begin{itemize}
	\item At the end of each series, ground the detector while it is exposed to the photons from the light emitting diodes.
	\item During the necessary 10~min cool-down period, hold the detector at a potential difference of $-80$~V.
	\item After the cooldown, lower the potential difference to the $-70$~V operating voltage, and begin data taking for the next series.
\end{itemize}

\begin{figure}
    \centering
    \includegraphics[width=\columnwidth]{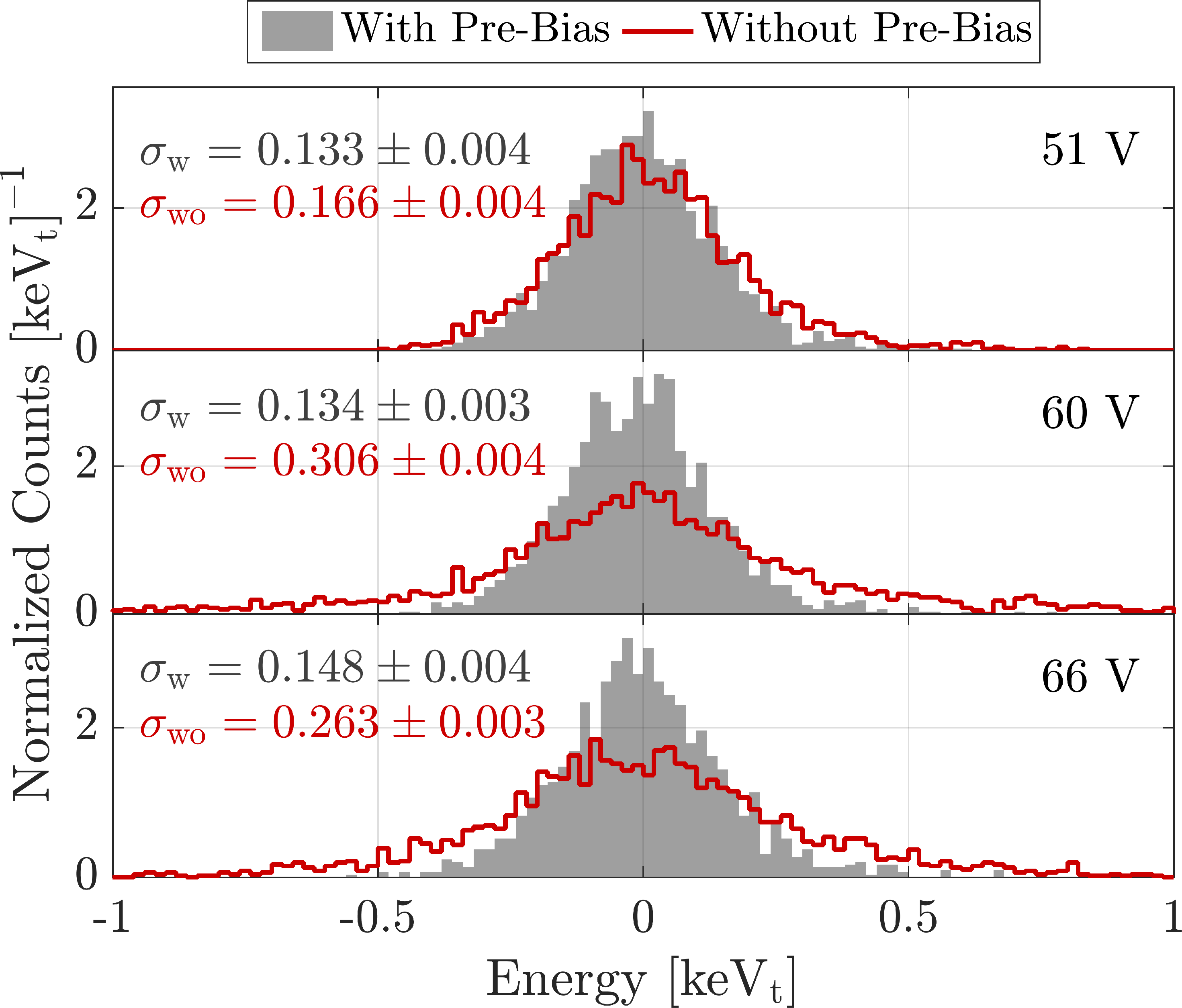}
    \caption{Baseline noise distribution for series that were prebiased (gray area) and series that were not prebiased (red curve) taken at potential differences of 51/60/66~V (top/middle/bottom).  The Gaussian-equivalent widths (see \SEC\ref{sec:run2res}) of the distributions with $\sigma_ {\text{w}}$ and without $\sigma_{\text{wo}}$ prebiasing are also given, in \kevt.  The thinner distribution widths for prebiased series compared to nonprebiased series demonstrates the effect of prebiasing.  For reference, $1~\kevteq\approx66~\eveeeq$.}
    \label{fig:biasScan_noiseWidth}
\end{figure}

The effectiveness of prebiasing can be seen in \FIG\ref{fig:biasScan_noiseWidth}, which compares the baseline noise distributions for series which were, or were not, prebiased.  The series were taken during the bias scan prior to \runTwo, described in \SEC\ref{sec:biasscan}, and were thus taken at various biases (the data in \FIG\ref{fig:biasScan_run2} were prebiased).  The widths of the distributions which were prebiased are smaller than those which were not, as shown by the values in the figure.

\subsection{Low-frequency noise}
\label{sec:lfnoise}

\begin{figure}
	\centering
	\includegraphics[width=\columnwidth]{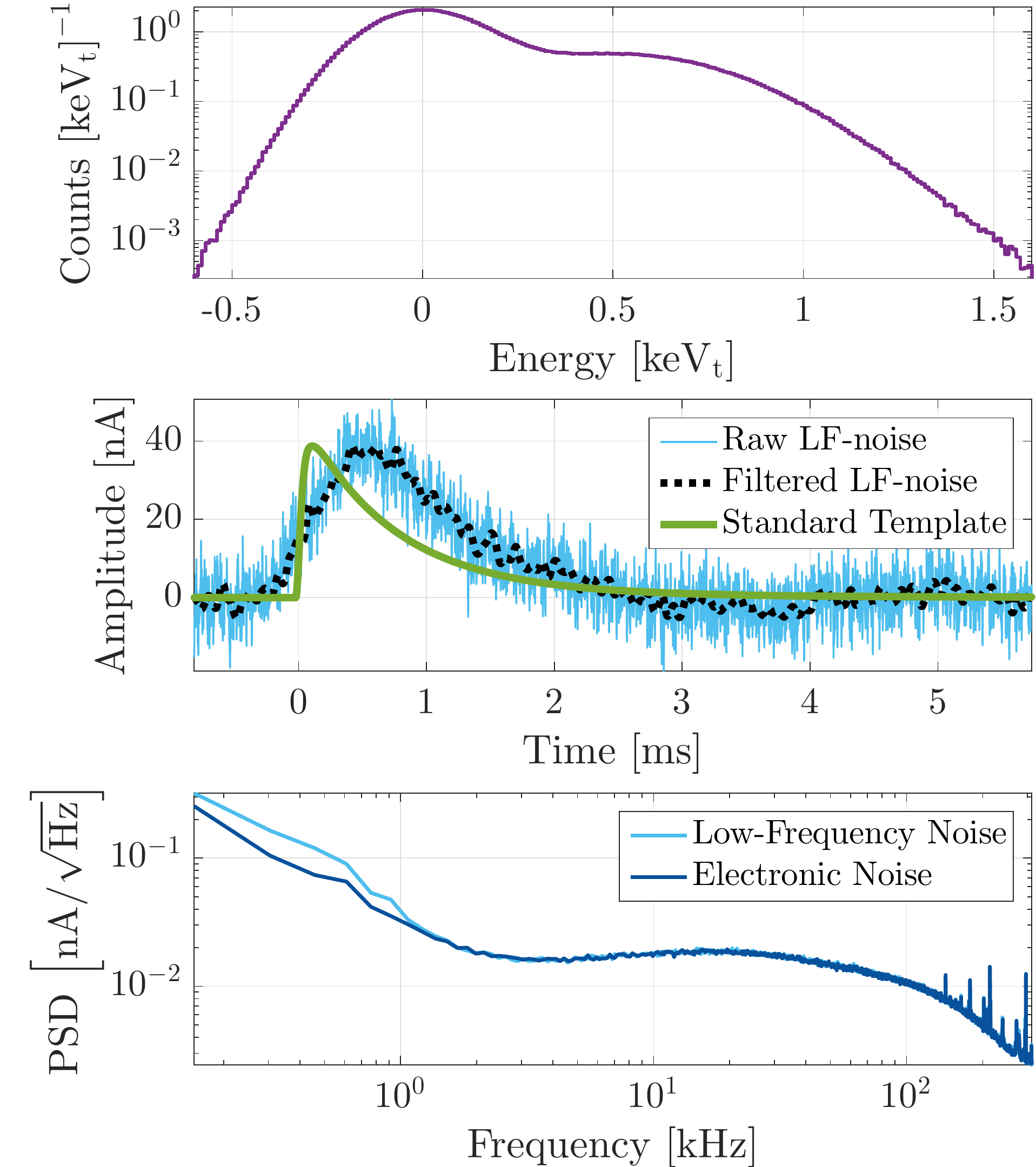}
	\caption{\textit{Top:} \runTwo noise distribution.  The electronic-noise distribution is centered at ${\sim}$0~\kevt while the \lf noise distribution dominates from 0.5--1.5~\kevt.  For reference, $1~\kevteq\approx66~\eveeeq$.  \textit{Middle:} raw (thin light blue solid) and filtered (thick black dotted) trace from a typical \lf noise event compared to the standard-event template (thick green solid), derived from high-energy $^{71}$Ge $K$-shell events.  The difference in pulse shape is most evident between 0 and 2~ms.  \textit{Bottom:} power spectral densities (PSDs) for 500 \lf (light blue) and electronic (dark blue) noise traces.  The \lf noise population has more power below ${\sim}$1~kHz.}
	\label{fig:LFintro}
\end{figure}

In \runOne, the baseline noise resolution was 14~\evee and the detector had 50\,\% trigger efficiency at 108~\evee.  The analysis threshold was set at 170~\evee to avoid being overwhelmed by a source of ${\sim}$kHz noise (labeled ``\lf'') that dominated the triggered-event rate below ${\sim}$200~\evee.  The primary source of this \lf noise was identified as vibrations from the Gifford-McMahon cryocooler used to intercept heat traveling down the electronics stem via the readout cables. The cryocooler cycled at ${\sim}1.2~\text{Hz}$, but stimulated higher-frequency vibrations that produced phonons in the detectors, including the CDMSlite detector, that were observable as \lf signals in the read-out traces.  The \lf noise was also present in \runTwo, as shown in the top panel of \FIG\ref{fig:LFintro}.  The electronic noise distribution is centered at 0~\kevt, and the \lf noise distribution is dominant from 0.5--1.5~\kevt.  These events were identified as noise by studying their pulse shape compared to the OF algorithm template as shown in the middle panel of \FIG\ref{fig:LFintro}.  In comparing the noise power spectral densities from 500 events (each) of \lf and electronic noise (bottom panel of \FIG\ref{fig:LFintro}), the \lf noise events have more power below ${\sim}$1~kHz.

The push to reject \lf noise, and subsequently reach a lower analysis threshold, for \runTwo occurred in two steps.  The first step was to characterize the \lf noise with regard to the timing of the cryocooler and identify blocks of calendar time that had similar \lf noise behavior (\SEC\ref{sec:noiseScore}).  The second step was to define a rejection criterion based on the pulse shape of individual events and to tune the position of the rejection threshold individually between the different calendar blocks (\SEC\ref{sec:deltaX2}).

\subsubsection{Cryocooler timing characterization}
\label{sec:noiseScore}

\begin{figure}
    \centering
    \includegraphics[width=\columnwidth]{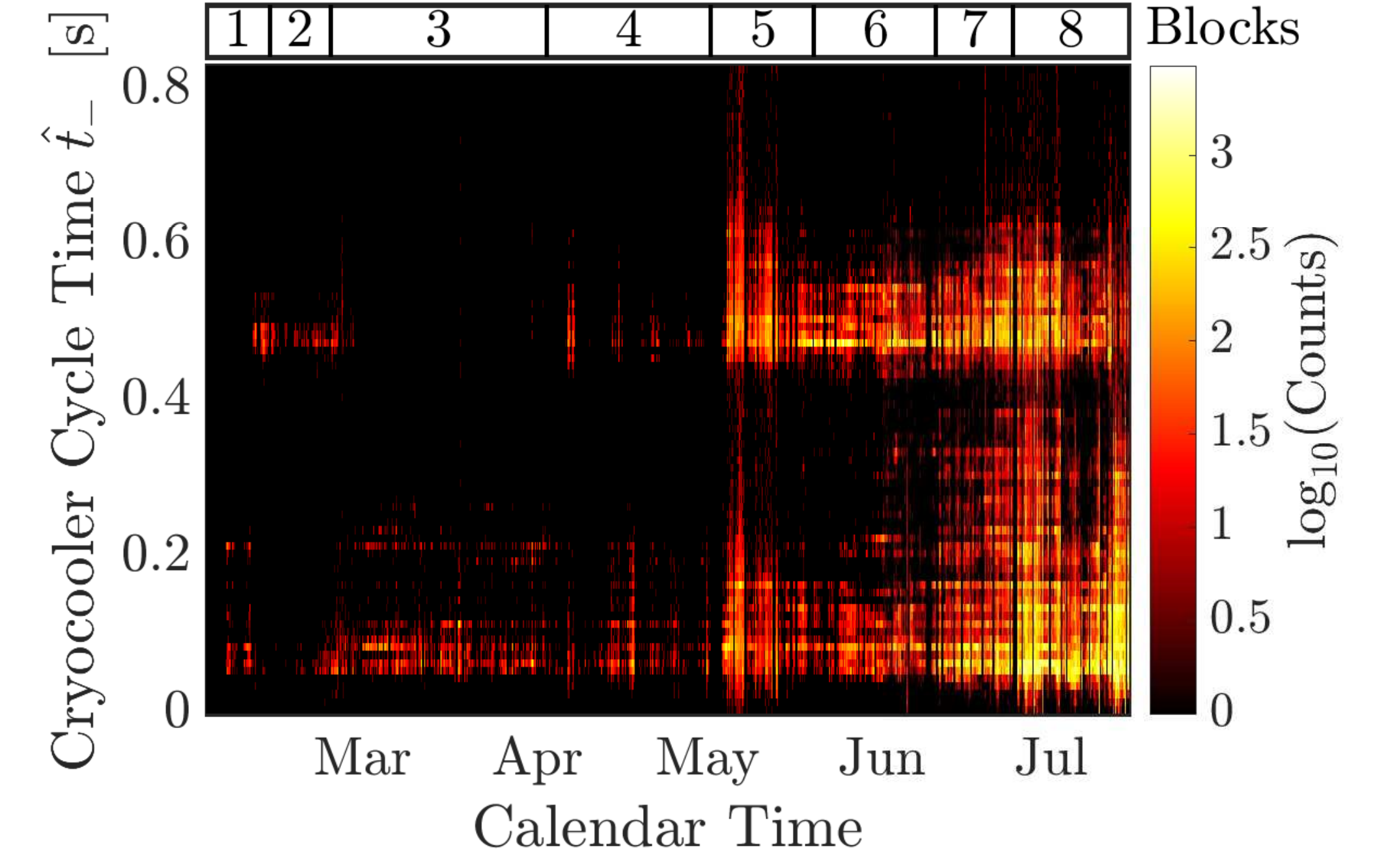}
    \caption{Number of low-energy triggered events for \runTwo \perOne in the two-dimensional plane of cryocooler time, $\hat{t}_-$, and calendar time in 2014.  The color scale is logarithmic with empty bins mapped to black.  The rate of \lf noise injection evolved  throughout the run because of the deterioration of the cryocooler, ranging from 0 to ${>}1000$ counts per bin.  The boundaries of the eight time blocks defined after applying a smoothing filter to the histogram are given in the bar labeled ``Blocks'' above the plot.}
    \label{fig:noiseScoreCnts}
\end{figure}

For \runTwo, two accelerometers were placed on and near the cryocooler to monitor vibrations.  Custom processing electronics were also installed to record the cryocooler cycle in the DAQ~\cite{Basuthakur2015,Pepin2016}.  Comparing the time stamps of recorded events to those of the cryocooler gives, for each event, the time since the start of the previous cryocooler cycle $\hat{t}_-$.  The precision of $\hat{t}_-$ is 3~ms and is dictated by the precision of the accelerometer read-out.  The cryocooler cycle (${\sim}$830 ms) starts with a compression event, which causes the largest amount of vibrational noise, and includes an expansion phase, ${\sim}$400 ms after the compression, which also causes noise.  These two parts of the cryocooler cycle are distinctly observed in \FIG\ref{fig:noiseScoreCnts}, which histograms the number of low-energy triggered events (dominated by \lf noise) in both $\hat{t}_-$ and calendar time.

During the course of \runTwo, the cryocooler degraded further, and the rate of events triggered by \lf noise greatly increased.  The rate increase was accompanied by a change in the \lf noise induction pattern as seen on the right side of \FIG\ref{fig:noiseScoreCnts}.  During this part of the run, \lf noise appeared throughout the entirety of the cryocooler cycle.  This obvious deterioration demanded a room-temperature warm-up of the experiment for servicing of the cryocooler cold head, and divided the run into the aforementioned Periods 1 and 2.

The \lf noise induction was characterized by developing and applying a smoothing filter to the histogram in \FIG\ref{fig:noiseScoreCnts}~\cite{Pepin2016}.  As the average number of particle interactions expected in each bin is \Om{10^{-3}}, bins with $10^2$--$10^3$ counts are clear outliers due to \lf noise. Correlations between neighboring bins are also indicators of \lf noise, as the noise typically occurs in bursts in calendar time and cryocooler time.  Applying a smoothing filter then deemphasizes true noise fluctuations, high-count bins surrounded by low-count bins, and allows better identification of times with a high \lf noise rate.  Using the filtered data, eight blocks in calendar time were defined such that the \lf noise behavior within each block was roughly consistent. These time blocks are indicated at the top of \FIG\ref{fig:noiseScoreCnts}.

In \perTwo of \runTwo, the accelerometers were not configured in the DAQ.  This oversight was not discovered until after the end of the run and thus the cryocooler timing information was not available in \perTwo.  Instead, four time blocks were  defined in \perTwo based on shifts in the energy scale and general noise environment.  The first two blocks occurred during the end of September and the beginning of October.  The energy scale noticeably shifted between these periods (see \SEC\ref{sec:run2gainCorr} and \FIG\ref{fig:eScaleCorr}).  The last two blocks, taken at the end of October and beginning of November, each contained a small amount of live time and coincided with a number of unrelated calibration and noise studies.  Small shifts in the noise environment were observed between these blocks.  In total, \runTwo was divided into 12 nonoverlapping time blocks.

\subsubsection{Pulse-shape discrimination}
\label{sec:deltaX2}

\begin{figure}
    \centering
    \includegraphics[width=\columnwidth]{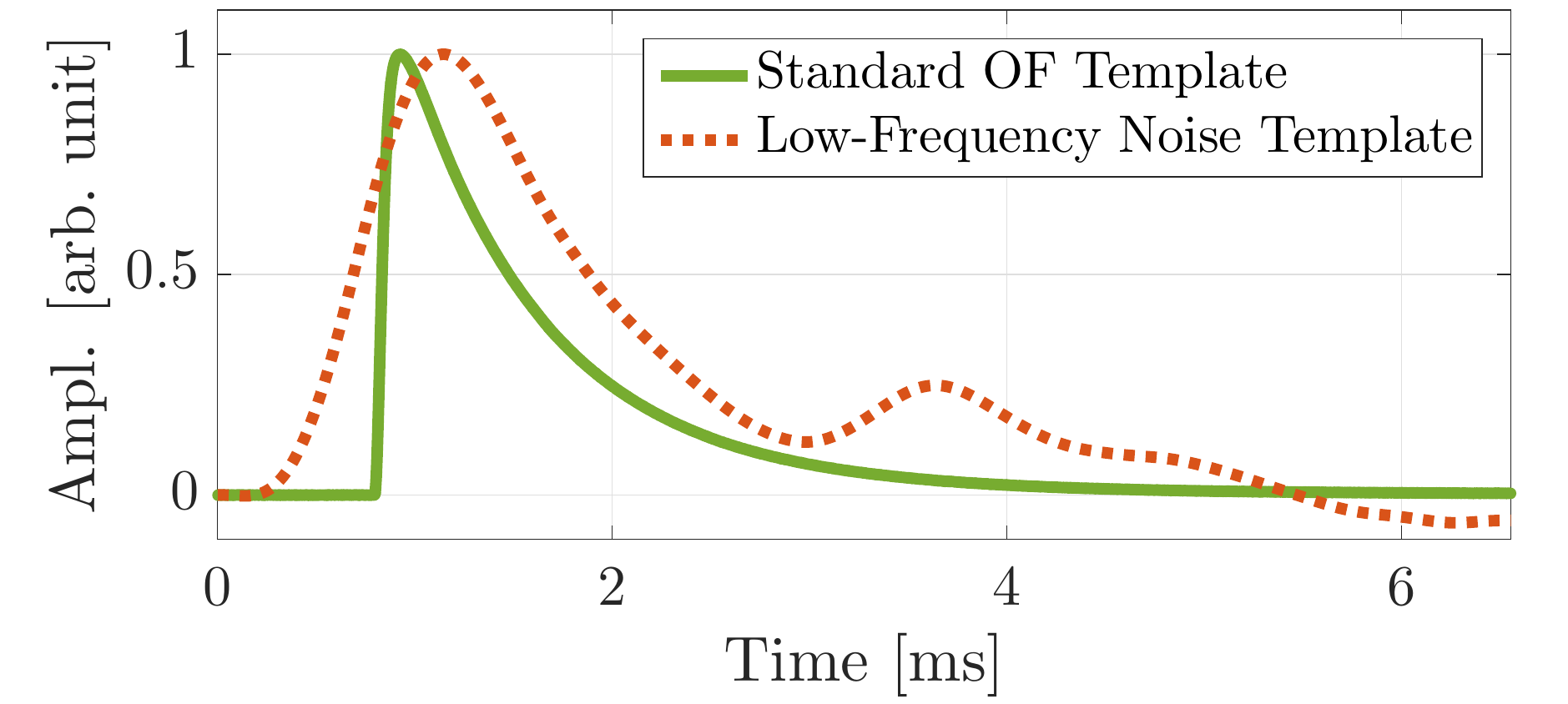}
    \caption{Template traces for the standard OF (green solid) and low-frequency noise (orange dotted) fits.  The templates were generated by averaging many events' pulse shapes, which removed uncorrelated noise.  Details of the \lf noise template generation are discussed in the text, and the standard OF template definition is discussed in \SEC\ref{sec:PulseFit}.  The maxima of the amplitudes (Ampl.) are scaled to unity in the figure.}
    \label{fig:LFnoiseTemplate}
\end{figure}

The criterion that was ultimately used to remove \lf noise from the data set was based on pulse shape, tailored to the different time blocks.  A new trace template was created by averaging a large number of \lf noise events; these traces were identified as those which triggered the detector, were in the energy range characteristic of \lf noise, and took longer than 1~ms to reach their maximum value.   This template is compared to the standard OF template in \FIG\ref{fig:LFnoiseTemplate}.  This new template was then fit to every trace using the single-template OF algorithm described in \SEC\ref{sec:PulseFit} (\ie, using the new template for $A{\left(t\right)}$ in \EQ\ref{eq:OF}), returning a goodness-of-fit parameter $\chi^2_{\text{LF}}$.  A discrimination parameter \dXLF was then defined as
\begin{equation}
	\dXLFeq\equiv\chi^2_{\text{OF}}-\chi^2_{\text{LF}},
\end{equation}
where $\chi^2_{\text{OF}}$ is the goodness-of-fit parameter from the single-template OF algorithm using the standard template.

\begin{figure}
    \centering
    \subfloat{
        \includegraphics[width=\columnwidth]{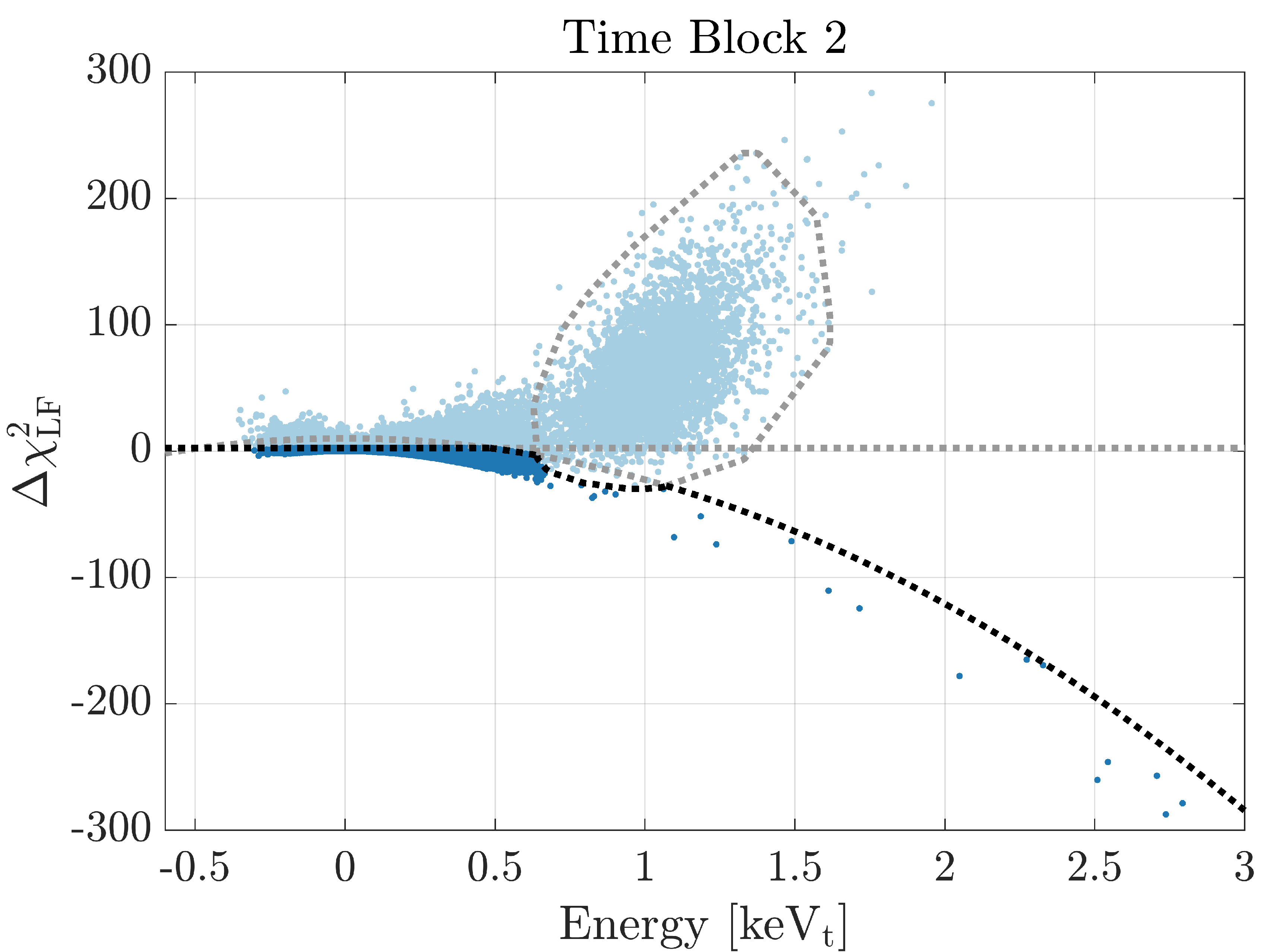}
        \label{subfig:LFnoiseFull2}
    }
    \quad
    \subfloat{
        \includegraphics[width=\columnwidth]{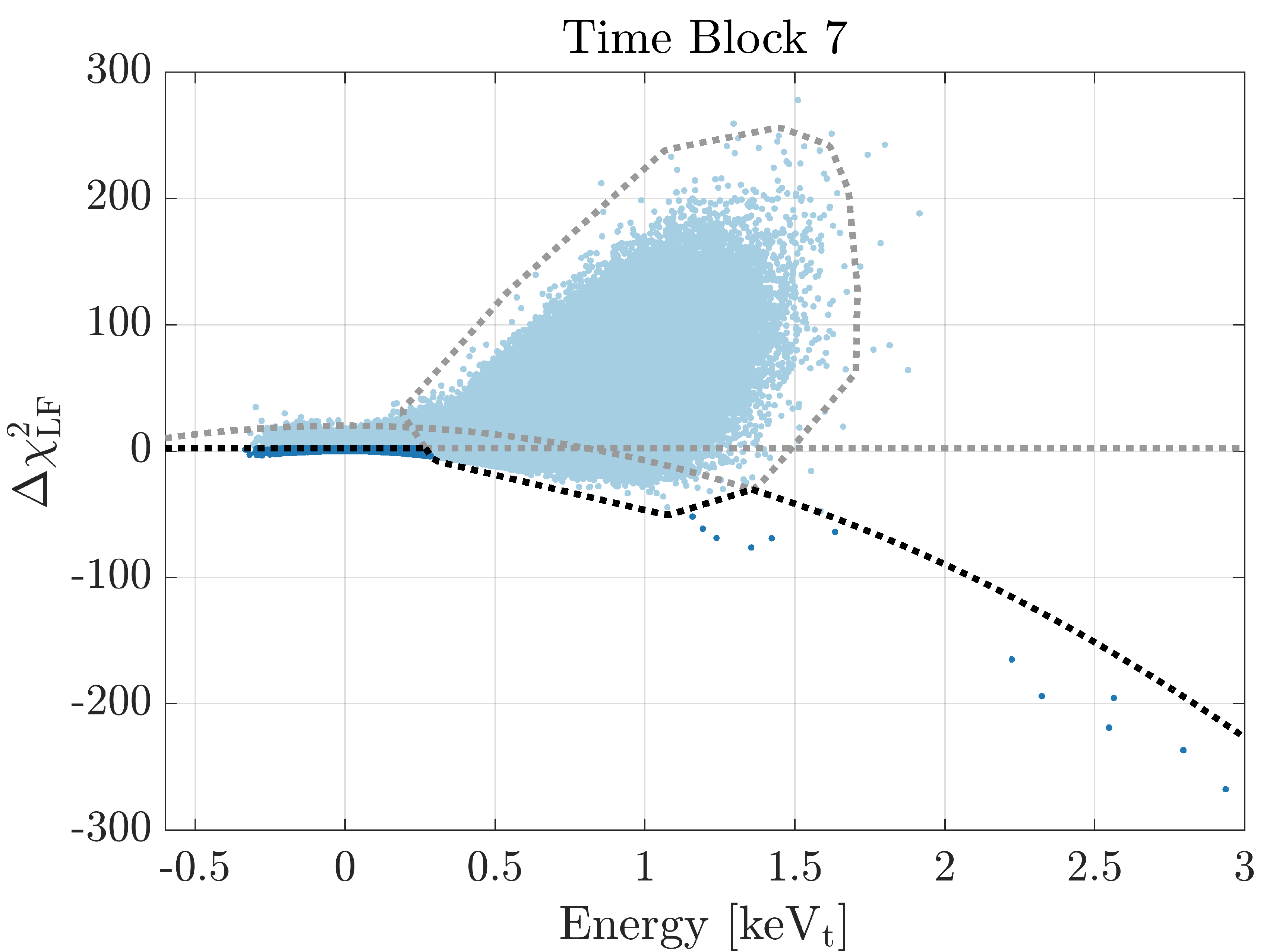}
        \label{subfig:LFnoiseFull7}
    }
	\caption{\dXLF as a function of total phonon energy for time blocks 2 (top) and 7 (bottom) showing the three portions of the \lf noise rejection cut (dotted) with the defining portion at any given energy darkened.  Low-frequency noise events cluster near ${\sim}$1~\kevt, while good events fall on a downward opening parabola.  The major difference between the two subplots is the difference in \lf noise: time block 2 shows low noise, while time block 7 is more noisy.  Events above any portion of the cut are rejected (light blue), while those below are retained (dark blue).  Time block 2 is relatively less noisy, while time block 7 is relatively more noisy.  The contour portion in block 2 cuts more loosely ($2.5\sigma$) than in block 7 ($5\sigma$) because of the changing low-frequency noise environment throughout the run.  A preselection cut removing events with unusually high NSOF $\chi^2$ values has been applied in these figures and, for reference, $1~\kevteq\approx66~\eveeeq$.}
    \label{fig:LFnoiseFull}
\end{figure}

Example planes of \dXLF versus energy are given in \FIG\ref{fig:LFnoiseFull} for time blocks 2 and 7, both from \perOne.  Pulse shapes that better fit the standard OF template have negative \dXLF and lie on a downward opening parabola, while those which better fit the \lf noise shape have positive \dXLF.  The cut was tuned piecewise with three components.  The first is a flat portion tuned to reject the worst (based on \dXLF) ${\sim}$10\,\% of the electronic noise distribution.  The second component was tuned on the good-event parabola, where the mean $\mu$ and width $\sigma$ of the \dXLF distribution in a number of energy bins extending to 400~\kevt were computed and the threshold fit to the $\mu+5\sigma$ points from each bin.  The $\mu+\sigma$ values were used to ensure a loose cut at high energies where no \lf noise is expected.  However, in order for the threshold to be tight enough to exclude the \lf noise distribution at low energies, an additional constraint of an upper bound on the $y$-intercept was also required.  The third component was based on a two-dimensional kernel-density estimate~\cite{Botev2010} of the \dXLF and energy of low-energy triggers (dominated by the \lf noise).  The threshold was taken as a convex hull around the largest $n\sigma$ contour from the estimate, where $n$ varied from 2.5--5 in steps of 0.5.  The tuning of this position was set individually for each time block based on a manual scan of borderline traces; \ie, if any trace that appeared to be contaminated by \lf noise was found, $n$ was increased.  Thus, the cut was tighter in time blocks of greater \lf noise rate and looser in time blocks with a lower \lf noise rate.  The time blocks shown in \FIG\ref{fig:LFnoiseFull} represent examples of low and high cryocooler-induced triggered noise rates, with looser and tighter cut thresholds, respectively.

\begin{figure}
	\centering
	\includegraphics[width=\columnwidth]{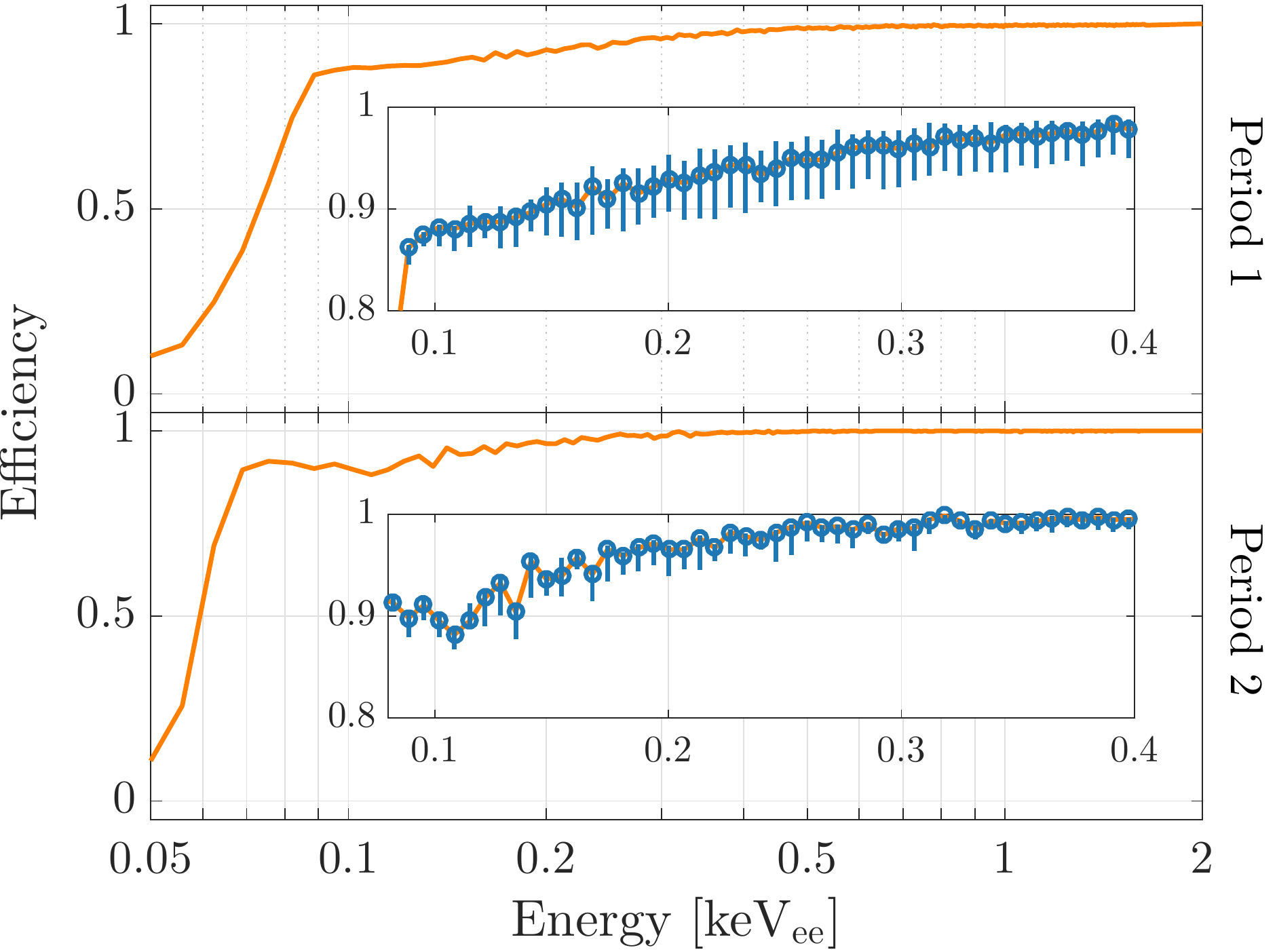}
	\caption{Efficiency of the pulse-shape based cuts for \runTwo \perOne (top) and \perTwo (bottom) as a function of electron-equivalent energy.  Almost all loss in efficiency is due to the \lf noise cut, with the sharp drop in efficiency below 100~\evee due to the kernel-density-estimate portion of that cut.  The insets give an enlargement in the $\mathcal{O}{\left(100~\text{\evee}\right)}$ range, where the systematic uncertainty from varying the pulse shape, shown by the error bars, is largest.  The average statistical uncertainty for each bin, due to the number of traces simulated, is 1.2\,\%.}
	\label
{fig:LFnoiseEff}
\end{figure}

The joint efficiency of three pulse-shape-based cuts, including the \lf noise cut, was determined by generating simulated traces, applying the same pulse-fitting techniques as the experimental data, and computing the fraction of simulated events that pass the cuts as a function of energy. Efficiency was also assessed for cuts that remove events with high NSOF-returned $\chi^{2}$ values and electronic-glitch events, which are events with pulses that have uncharacteristically fast fall times.  The simulated traces were constructed by combining a measured noise trace, selected from those recorded routinely throughout the WIMP search, and a noiseless template scaled to a desired amplitude.  The procedure was repeated using three templates of different shapes to assess the systematic uncertainty of the efficiency due to pulse shape.  The templates were the standard OF-fit template and two new templates defined as $T_{\pm}=T_s\pm\alpha T_f$, where $T_{s/f}$ are the slow and fast templates from the 2T fit (\FIG\ref{fig:2Ttemplates}).  $\alpha$ was chosen to be 0.125 to encompass the observed fast-to-slow template ratio of events in the $^{71}$Ge $K$-shell peak.  The efficiency of these cuts is shown in \FIG\ref{fig:LFnoiseEff}, including the uncertainty from varying the template shape.  The loss in efficiency due to the non-\lf noise cuts is ${<}$5\,\% at any given energy bin.  The large decrease below 100~\evee is where the kernel-density-estimate portions of the \lf noise cut are active.  The sharp onset of this decrease differs by time block, while the more gradual decrease seen in the figure (particularly for \perOne) is due to averaging over all time blocks.  Also note that, while the cut thresholds, such as those shown in \FIG\ref{fig:LFnoiseFull}, are defined in the \kevt energy scale, the efficiency must be evaluated in the energy scale used in the final analysis, \kevee.

\section{\runTwo Energy resolution and threshold}
\label{sec:threshAndRes}

The \lf noise cut described in the previous section allowed the event selection in \runTwo to avoid events resulting from known noise sources.  The remaining noise distribution was studied to measure the baseline resolution of the detector, which in turn was used to model the detector's energy resolution.  The analysis threshold, however, was constrained by the detector's efficiency for triggering on low-energy events, \ie, the trigger threshold.

\subsection{\runTwo energy resolution model}
\label{sec:run2res}

The total energy resolution $\sigma_{\text{T}}{\left(\Ereeeq\right)}$ for the detector was modeled as 
\begin{align}
    \sigma_{\text{T}}{\left(\Ereeeq\right)}&= \sqrt{\sigma^2_{\text{E}} + \sigma^2_{\text{F}}{\left(\Ereeeq\right)} + \sigma^2_{\text{PD}}{\left(\Ereeeq\right)}} \\
    &= \sqrt{\sigma^2_{\text{E}} + B\Ereeeq + \left(A\Ereeeq\right)^2},
    \label{eq:resModel}
\end{align}
where $\sigma_{\text{E}}$ is the baseline resolution caused by electronic noise, $\sigma_{\text{F}}{\left(\Ereeeq\right)}$ describes the additional width due to electron-hole pair statistics including the Fano factor~\cite{Fano1947}, and $\sigma_\text{PD}{\left(\Ereeeq\right)}$ is the broadening due to position dependence.  The electronic noise is energy independent.  The variance due to electron-hole pair statistics can be written as $F\epgeq\Ereeeq\equiv B\Ereeeq$, where $F$ is the Fano factor.  Previous measurements at higher temperatures give $F=0.13$~\cite{Alig1980}, and using $\epgeq\simeq3$~eV~\cite{Emery1965,*Pehl1968} per \eh pair gives an expectation of $B=0.39~\text{\evee}$.  Finally, variations due to position dependence are expected to be proportional to energy; this final term may also include other effects that scale with energy.

\begin{figure}
	\centering
    \includegraphics[width=\columnwidth]{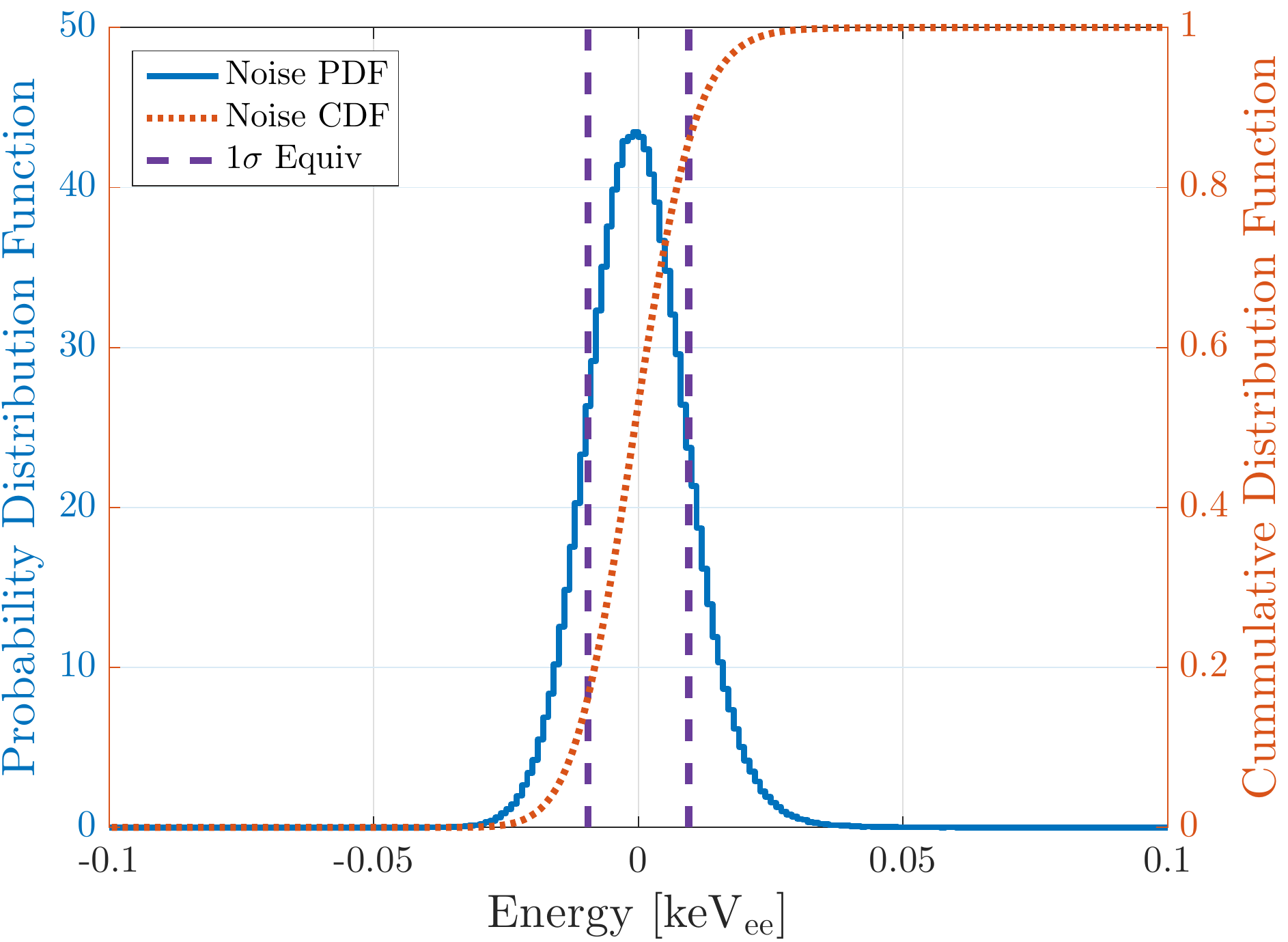}
    \caption{Reconstructed energy probability distribution function (PDF) of noise-only events in \runTwo (blue solid, left vertical axis) with the corresponding cumulative distribution function (CDF) (orange dotted, right vertical axis).  The 1$\sigma$-equivalent is taken as half the distance between the 15.87th and 84.13th percentiles (dark purple dashed) and is $9.26\pm0.11~\text{\evee}$.}
    \label{fig:baseRes}
\end{figure}

The baseline resolution can be measured using the reconstructed energy of noise-only events taken throughout the run.  When applied to noise traces, the algorithms described in \SEC\ref{sec:PulseFit} tend to fit to the largest noise fluctuation, which biases the fit toward nonzero amplitudes.  This is undesirable for characterizing the baseline noise distribution; for this study, the time delay is forced to be zero, and the corresponding energy distribution for \runTwo is shown in \FIG\ref{fig:baseRes}.  To avoid efficiency effects, no cut against \lf noise was applied, and thus the distribution is slightly skewed to positive energy.  A simple Gaussian fit would not be representative of the distribution; the resolution is determined via a Gaussian-equivalent computation: the 1$\sigma$-equivalent is taken as one-half the energy between the 15.87th and 84.13th percentiles (the $\mu\pm\sigma$ values for a normal distribution). Repeating the procedure for a variety of histogram bin sizes gives an estimate of the uncertainty.  The baseline resolution determined in this way is $9.25\pm0.11~\text{\evee}$.

\begin{figure}
 	\centering
    \includegraphics[width=\columnwidth]{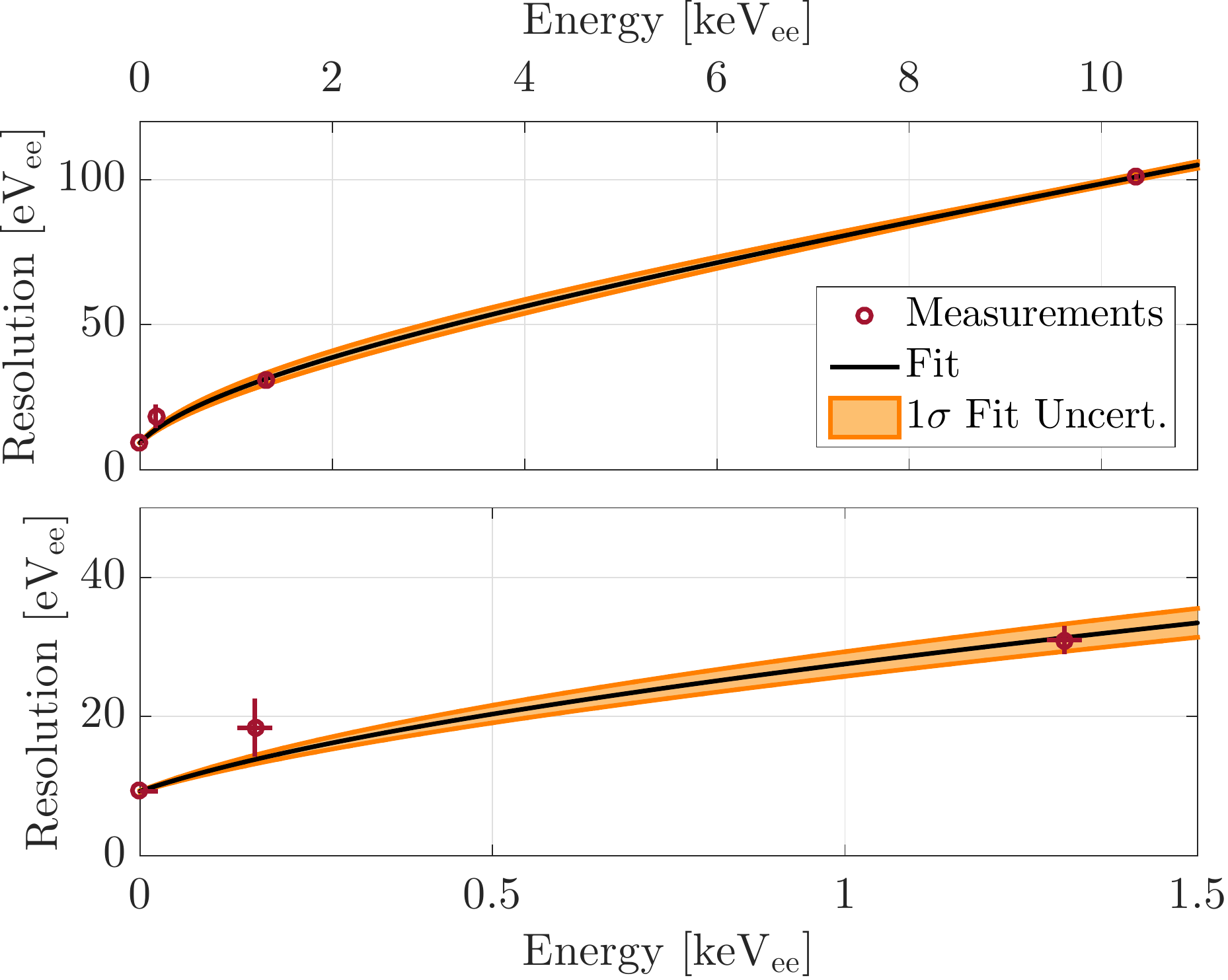}
    \caption{Width of four points in the \runTwo energy spectrum (red points), the best-fit curve (black), and 68\,\% uncertainty band (orange).  The bottom panel is an enlargement of the top panel below 1.5~\kevee.}
    \label{fig:resFit}
\end{figure}

The resolution model of \EQ\ref{eq:resModel} with parameters $\sigma_E$, $B$, and $A$ was fit to the peaks, weighted by their uncertainties, at four different energies: the zero-energy baseline distribution and the three $^{71}$Ge-activation peaks at 10.37~\kevee ($K$ shell), 1.30~\kevee ($L$ shell), and 0.16~\kevee ($M$ shell).  The resolution of each of these peaks is given in \TAB\ref{tab:run2Res}.  The final fit is given in \FIG\ref{fig:resFit} with a goodness-of-fit per degree of freedom $\chi^2/\text{dof}=1.22$.  Because of the small uncertainty on the baseline resolution, and the weighting of the fit, $\sigma_{\text{E}}=9.26\pm0.11~\text{\evee}$ is very similar to the measured value.  The best-fit Fano coefficient is $B=0.64\pm0.11~\text{\evee}$, while the position-dependence coefficient is $A=\left(5.68\pm0.94\right)\times10^{-3}$.  The last two parameters are strongly anticorrelated with a Pearsons product-moment correlation coefficient of $\rho_{\text{AB}}=-0.984$.  Repeating the fit with $B$ fixed to the expected value gives $A=\left(7.53\pm0.13\right)\times10^{-3}$, with a goodness-of-fit per degree of freedom of $\chi^2/\text{dof}=3.77$.  The larger deviation of the M-shell measurement from the fit function is still compatible with statistical fluctuations.  The free fit is chosen as the final result to allow for the possibility of temperature dependence in the Fano factor and any other unaccounted effects.

\begin{table}
	\centering
	\begin{tabular}{@{}cd{2.2}p{3.2}@{}}
		\hline \hline
		Peak		& \multicolumn{1}{c}{Energy $\left[\keveeeq\right]$}	& \multicolumn{1}{c}{Resolution $\left[\eveeeq\right]$}	\bigstrut	\\
		\hline
		Baseline	& 0,0		& 9.25,0.11	\bigstrut[t]	\\
		$M$ Shell	& 0,16	& 18.6,4.2			\\
		$L$ Shell	& 1,30	& 31,2			\\
		$K$ Shell	& 10,37	& 101,1			\\
		\hline \hline
	\end{tabular}
	\caption{Peak resolutions from \runTwo for the baseline noise and three $^{71}$Ge-activation peaks.}
	\label{tab:run2Res}
\end{table}
 
\subsection{\runTwo trigger efficiency and threshold}
\label{sec:run2Trigger}

During WIMP-search data taking, the traces from all detectors were recorded when the experiment triggered. For calibration data, only the detectors in the same tower as the triggering detector were recorded.  Recall that the experiment triggered if the analog sum of any detector's phonon traces exceeded a user-set hardware threshold.  In anticipation of better \lf noise rejection, the hardware trigger threshold was lowered for \runTwo compared to \runOne, and again within \runTwo, between \perOne and \perTwo.

For \runTwo, the analysis thresholds were defined as the energy at which the detector's trigger efficiency reached 50\,\%. The trigger efficiency for a given detector $D$ was determined using events that triggered one of the other detectors and may or may not have deposited energy in detector $D$. The efficiency at a given energy $E$ was then given by the fraction out of all events with energy $E$ in detector $D$ that also generated a trigger in detector $D$.  The $^{252}$Cf calibration data set, which has more recorded events than the WIMP-search data set, was  used to measure trigger efficiency, with strict cuts applied to remove nonparticle interactions that also caused triggers, \ie, due to noise or detector cross-talk.

Two cuts were used to remove \lf noise, which triggered the detector at a high rate and could bias the trigger efficiency calculation, from the calibration data.  The first was a pulse-shape cut based on the \dXLF parameter defined in \SEC\ref{sec:deltaX2}, and the second was based on the cryocooler timing discussed in \SEC\ref{sec:noiseScore}.  The \dXLF-based cut was independent of energy and tighter than the energy-independent portions of the WIMP-search-data specific cut of \SEC\ref{sec:deltaX2}.  A tighter cut was used to be particularly cautious against using \lf noise in the calculation.

\begin{figure}
	\centering
    \includegraphics[width=\columnwidth]{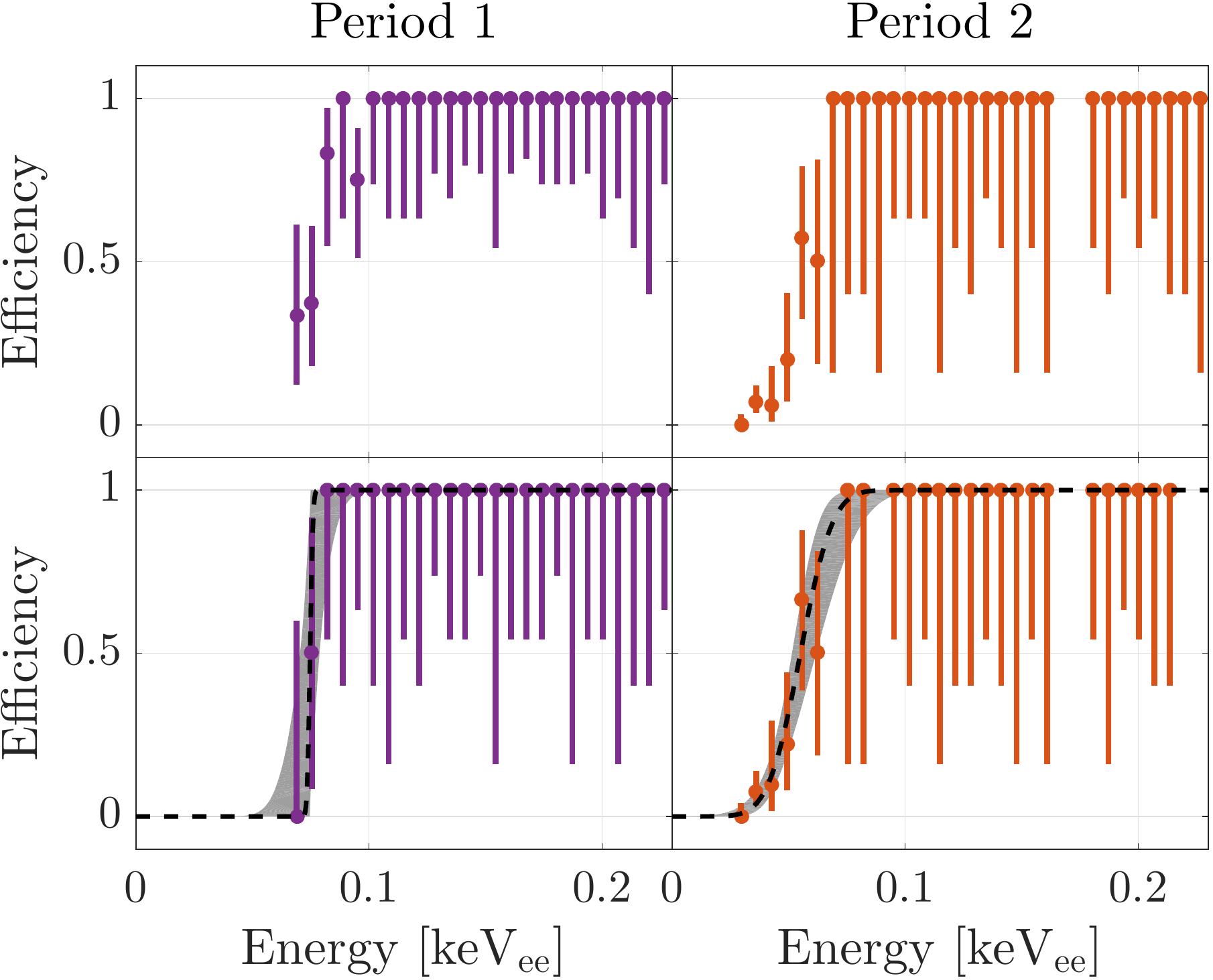}
    \caption{Binned trigger efficiency without (top) and with (bottom) a cut on cryocooler timing for \runTwo Periods 1 (left) and 2 (right).  Using the cryocooler information noticeably improved the \perOne measurement while marginally improving that for \perTwo.  The best-fit error function (black dashed curve) and its 68\,\% uncertainty (gray shaded) are given in the bottom row for each period.}
    \label{fig:trigEff}
\end{figure}

The binned trigger efficiency shown in the top row of \FIG\ref{fig:trigEff} is the result of using the pulse-shape-based cut alone.  The highest-energy nonunity bin in \perOne is at 95~\evee.  The highest-energy events that failed to trigger the detector in \perOne were found to coincide with the high-rate periods of the cryocooler cycle; \ie, they were contaminated with \lf noise and therefore are not representative of true physical events.  The second row in \FIG\ref{fig:trigEff} shows the binned efficiency after applying the second cut against \lf noise, removing the high-rate periods of the cryocooler cycle.  After this second cut, the highest-energy nonunity bin in \perOne shifts to 82~\evee.

The absence of accelerometer data in \perTwo was discovered very soon after the end of the run.  Given the utility of the cryocooler timing information in determining the \perOne trigger efficiency, a dedicated \perTwo $^{252}$Cf calibration was performed with the accelerometers properly configured.  The binned \perTwo trigger efficiency is shown in the right panels of \FIG\ref{fig:trigEff}.  The difference between applying the cryocooler timing or not is marginal, retrospectively unsurprising considering the better state of the cryocooler following the repair.  The highest-energy nonunity bin for the final \perTwo calculation is at 62~\evee.  As a verification, the computation was repeated, for both \perOne and \perTwo, using the lower-rate WIMP-search data, and consistent results were found.

The final 50\,\% trigger efficiency points come from fitting the resulting events' energy to an error function by maximizing an unbinned log-likelihood function which contains a rising error function for events that do trigger the CDMSlite detector and a falling error function for those that do not.  Both functions are needed as the event energies themselves are used in the fit as opposed to a binned passage fraction.  The log-likelihood function is
\begin{equation}
	\ln{\mathcal{L}{\left(\mu,\sigma\right)}}=\sum_i^{N_+}\ln{f_+{\left(E_i;\mu,\sigma\right)}}+\sum_j^{N_-}\ln{f_-{\left(E_j;\mu,\sigma\right)}},
\end{equation}
where $N_{\pm}$ is the number of events passing/failing the trigger condition on the CDMSlite detector and
\begin{equation}
	f_{\pm}{\left(E_i;\mu,\sigma\right)}=0.5\left[1\pm\text{erf}{\left(\frac{E_i-\mu}{\sqrt{2}\sigma}\right)}\right],
\end{equation}
where $E_i$ is the total phonon energy of the given event and $\mu$ and $\sigma$ are the 50\,\% point and width of the error function, respectively.  A Markov chain Monte Carlo simulation was used to scan the parameter space, with a log-normal prior on $\sigma$ and flat prior on $\mu$.  The prior on $\sigma$ was required as the turn on is very sharp in \perOne; the log-normal prior inputs knowledge of the detector's resolution to prevent fits with an unphysical turn on.  The best-fit values give thresholds of $\mu=75^{+4}_{-5}$ and $56^{+6}_{-4}$~\evee for the two periods with the corresponding curves and 68\,\% uncertainty bands shown in the bottom panel of \FIG\ref{fig:trigEff}.

\section{Effects of bias voltage variation}
\label{sec:bias}

The bias applied at the detector, and therefore the NTL amplification, varied with time because of the presence of parasitic resistances in the biasing-electronics chain.  This variation affected the calibration of the ER and NR energy scales, which thus required empirical correction.  Additionally, the observed energy scale of \runTwo calls the assumed bias potential of \runOne into question, though the effect on the \runOne result is found to be small compared to other uncertainties.

\subsection{Total phonon energy scale}
\label{sec:totEscale}

The measured scale for total phonon energy \Et is determined by calibrating the TES-readout units of amperes to \kevt using calibration data taken at the standard iZIP operating bias of 4~V.  In \runOne, the location of the strong $^{71}$Ge $K$-shell activation peak at ${\sim}$120~\kevt, close to the expected 124~\kevt, was taken as confirmation of this procedure, and \Et was then converted to \Eree using \EQ\ref{eq:totAmp} with an assumed $-69$~V bias.

However, this procedure did not match the expectation in \runTwo, both for the final $-70$~V, data as well as initial $-60$~V data taken during \runTwo commissioning.  The peak appears at 135 and 154~\kevt for $-60$ and $-70$~V respectively, both of which are ${\sim}23\,\%$ higher than expected.  This is now understood as the effect of a bias-dependent ionization extraction and collection efficiency.  For these detectors, the collection efficiency is ${<}100\,\%$ at 4~V, while being at or above $100\,\%$ at CDMSlite biases (${>}100\,\%$ is possible because of impact ionization~\cite{Phipps2016a}).  These effects were not well understood at the time of \runOne.  For \runTwo, the calibration from \Et to \Eree was thus performed empirically by scaling the energy such that the $K$-shell peak appeared at the expected 10.37~\kevee (see \SEC\ref{sec:run2gainCorr}).

The \runTwo study thus implies a problem with the interpretation of the data from the first run, as the observed NTL amplification in the second run was noticeably higher than in the first run though the nominal bias voltages were similar at $-69$ and $-70$~V.   In \runTwo, the high-voltage power-supply current was measured, verifying that the bias at the detector was close to the nominal 70~V. However, such a measurement was not done during \runOne, and postrun inspections of the high-voltage biasing board indicated deterioration of a sealant epoxy, originally applied to the biasing electronics to prevent humidity-related effects.  Thus, it is possible that a significant leakage current across the bias resistor, which would have reduced the effective bias voltage at the detector, went undetected. Assuming that the ionization collection efficiency was the same for both runs, and using the energy calibration from \runTwo, the \runOne peak location indicated that the effective bias potential was approximately $-55~\text{V}$.  This ${\sim}20\,\%$ difference in NTL gain affected the final \runOne results, and is considered in the next section.

\subsection{Effect of gain variation on nuclear-recoil energy scale in \runOne}
\label{sec:gainvsbias}

\begin{figure}
	\centering
	\includegraphics[width=\columnwidth]{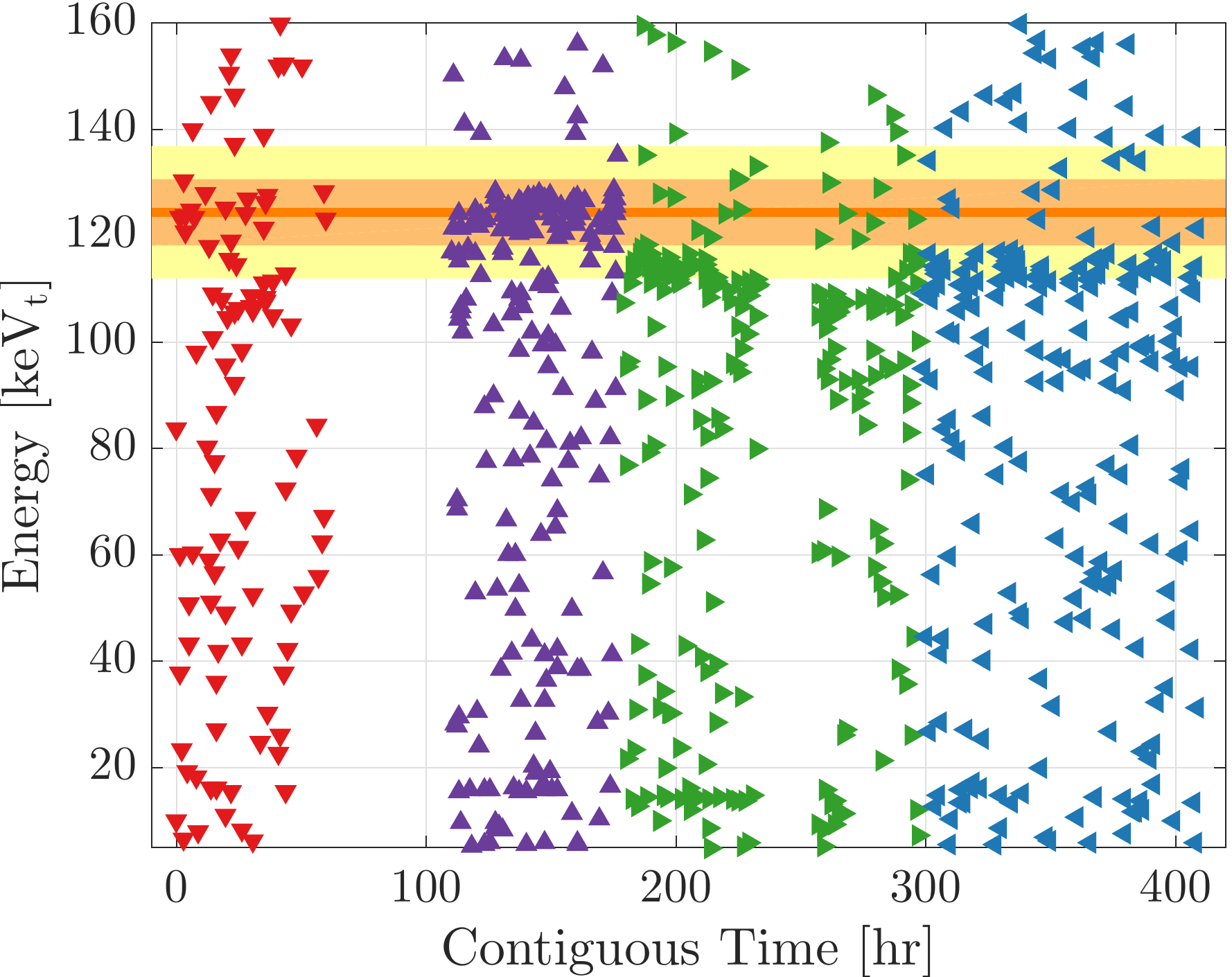}
	\caption{Phonon energy as a function of run time for \runOne. The overdensity around 120~\kevt is from the 10.37~keV $K$-shell electron-capture products.  Gaps exist because of unstable conditions.  The different colors/orientations of the triangles indicate the four time periods which were fit to independent polynomials in the gain-correcting piecewise fit. The horizontal line indicates the peak's expected location (under the assumptions made for the \runOne analysis; see text) with departures of 5 and 10\,\% indicated by the bands. The measured energy of the line shows up to 15\,\% variation over the course of the run.}
	\label{fig:pt_time_1}
\end{figure}

The NTL-amplification gain was measured by tracking variations of the total phonon energy of the 10.37~keV activation line with time. The line's intensity decreased exponentially with an 11.43~d half-life~\cite{Hampel1985} and increased whenever a \cf calibration was performed. This activation line is shown as a function of time during \runOne in \FIG\ref{fig:pt_time_1}. The measured energy of this line shows variations up to 15\,\%.  In the \runOne analysis, this variation was corrected for by an empirical piecewise polynomial fit to the $K$-shell peak.  The different colors in \FIG\ref{fig:pt_time_1} indicate the parts of the run that were fit with independent polynomials.

These variations of the total phonon energy scale, from the inferred 20\,\% correction due to calibration and the observed time dependence, necessarily affect the nuclear-recoil energy scale, and hence the threshold and final limit. As described \SEC\ref{sec:intro}, the effect of varying the threshold can be non-negligible. Thus, it is imperative to understand what a 10\,\%--20\,\% variation in total phonon energy implies for the nuclear-recoil energy scale.

The effect of reducing the potential difference, compared to the assumed 69~V, is estimated by considering the relation between the reconstructed energies \Ernr and \Eree as given by \EQ\ref{eq:ee2nr}.  At any given \Eree, \Ernr is calculated, assuming the standard Lindhard yield model, for both the original 69~V and at the reduced potential difference.  A 10\,\%--20\,\% reduction in potential difference has minimal effect on the nuclear-recoil energy scale.  The maximum fractional change at the \runOne threshold for gain drops of 10\,\%, 15\,\%, and 20\,\% are $\left|\delta\Ernreq\right|/\Ernreq{\left(170~\eveeeq,69~\text{V}\right)}=1.7\,\%,\ 2.7\,\%\text{\ and\ }3.8\,\%$ respectively. In terms of absolute energy scale, these correspond to a variation  of ${<}$5~\evnr at threshold.  Reevaluating the \runOne result assuming a $-55$~V bias, as indicated in the previous section, leads to a 2.7\,\% drop in threshold, which in turn leads to an improvement of the sensitivity for lower-mass \ws of up to 12\,\%, while the sensitivity to higher-mass \ws decreases by about 2\,\%.  This is less than the uncertainty due to the ionization yield model as shown in \FIG\ref{fig:limits_si}. In conclusion, a 10\,\%--20\,\% drop in gain, even if unaccounted for, does not significantly impact the interpretation of the \runOne result in terms of the sensitivity to low-mass \ws.

\subsection{Gain correction in \runTwo}
\label{sec:run2gainCorr}

Laboratory testing after \runOne revealed that the bias variations were likely due to humidity on the high-voltage biasing board, leading to varying parasitic resistances $R_p\sim\text{\Om{10~\text{M}\Omega}}$, parallel to a biasing resistance of $R_b\sim400~\text{M}\Omega$. A new circuit was designed with a biasing resistance of $R_b\sim200~\text{M}\Omega$. The board was specially treated in an ultrasonic bath, baked, and layered with HumiSeal$^{\circledR}$ (HumiSeal, Westwood, MA), reducing the effects of parasitic resistances under humid conditions to $R_p\gtrsim\text{\Om{1~\text{G}\Omega}}$.  See Appendix~A of \REF\cite{Basuthakur2015} for details of the biasing board.

For \runTwo, the DAQ was configured to record the bias \Vb and current $I_{\text{b}}$ of the high-voltage power supply for each event.  Changes in the current are indicative of changes in total resistance encountered by the power supply, \ie some combination of $R_b$ and $R_p$. The recorded current was then used to correct the energy scale on an event-by-event basis as 
\begin{equation}
	\Eteq^{\text{Corr}} = \Eteq\cdot\frac{1+eV_b/\epgeq}{1+e\left(V_b-I_bR\right)/\epgeq},
	\label{eq:biasCorrect}
\end{equation}
where $R$ is the encountered resistance.  A fit of \Et \vs $I_b$ demonstrated that $R\approx R_b$; \ie, $R_p$ is much greater than $R_b$, is parallel to the detector, and is downstream of $R_b$.  Based on this fit and a measured bias current $I_b\lesssim10$~nA, a ${\lesssim}$2\,\% correction was applied.

\begin{figure}
	\centering
	\includegraphics[width=\columnwidth]{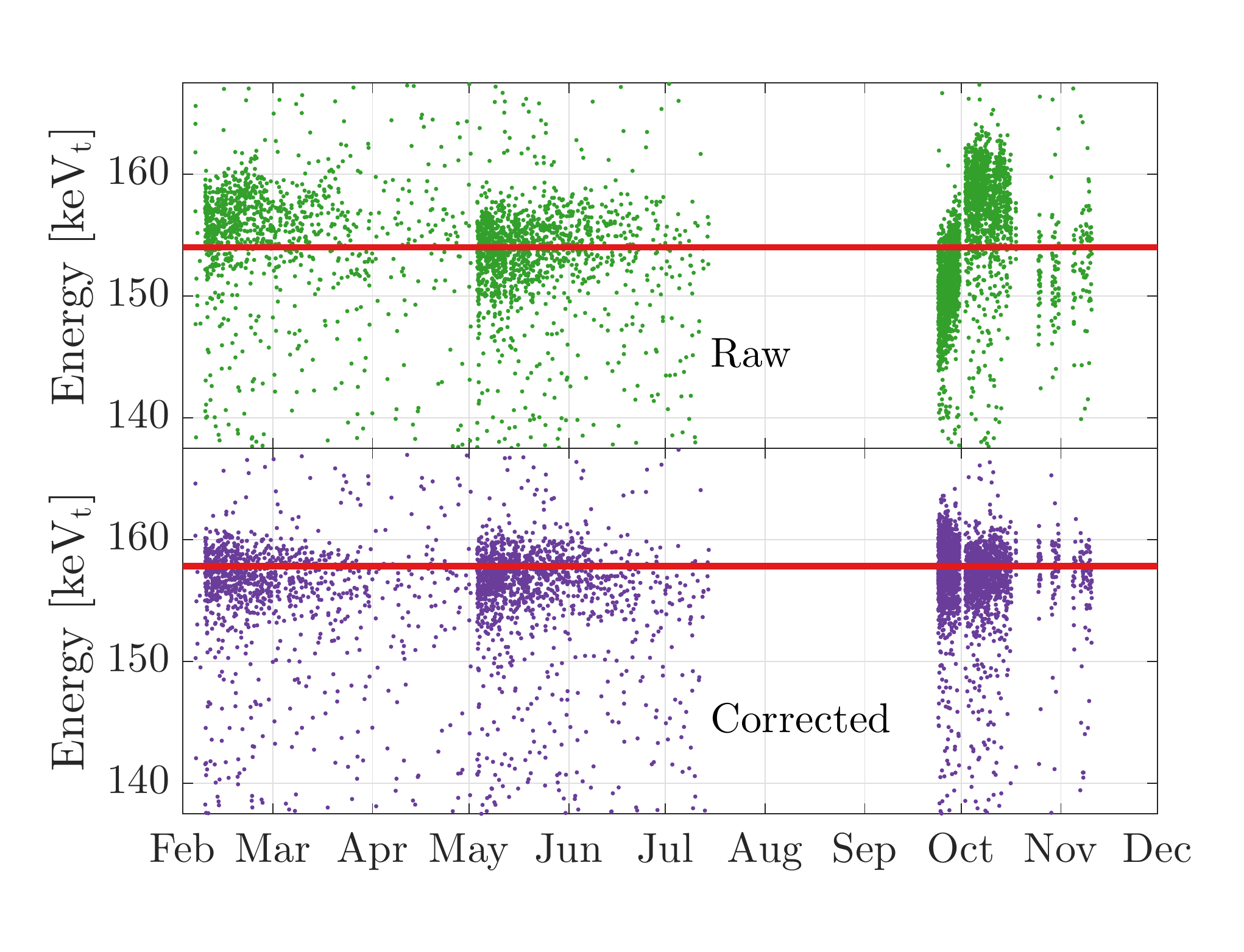}
	\caption{$K$-shell activation peak (cluster at 150--160~\kevt) in Run 2 as a function of time without (top) and with (bottom) corrections for gain variations.  \cf calibrations occurred in February, May, and September/October. The horizontal lines indicate the means of the two peak distributions.}
	\label{fig:eScaleCorr}
\end{figure}

In addition to the position dependence mentioned in \SEC\ref{sec:PulseFit}, which gave a correction of 0\,\%--3\,\%, two other sources of gain variation were identified in \runTwo: the cryostat base temperature and discrete shifts that were possibly caused by changes in the noise environment. The base temperature of the experiment ranged from 47--52~mK and was recorded by the DAQ for each event.  These temperature differences caused a ${\lesssim}3\,\%$ variation in the energy scale that was corrected using the recorded temperature. After correcting for leakage current and base temperature, the mean value of the $^{71}$Ge $K$-shell peak was consistent in time throughout \perOne.  However, there were two distinct populations in \perTwo, one lower than \perOne by 2.87\,\%, and the other higher than \perOne by 0.81\,\%.  The origin of these shifts was not identified. They were corrected for by scaling the means of the activation peak distributions to match that of \perOne.  A comparison of the initial to final \kevt energy scale over the duration of \runTwo is given in \FIG\ref{fig:eScaleCorr}.  The mean of the final distribution was then used to scale to the \Eree energy scale.

\section{CDMSlite backgrounds}
\label{sec:backfid}

CDMSlite is an ER background-limited search because it cannot discriminate between ER and NR events. However, efforts have been made to understand and reduce the overall background rate in order to extend sensitivity to smaller WIMP scattering cross sections.  Operating a SuperCDMS iZIP detector in CDMSlite mode required grounding one side of the detector, which created an asymmetric electric-field geometry.  This geometry was studied in simulation to understand how it affects ER background modeling.  Motivated by this understanding of the electric field, a fiducial volume was defined in \runTwo to remove areas of the detector where the electric-field configuration led to reduced signal amplification and therefore a higher background rate at low energies.  Defining a fiducial volume thus significantly reduced the background rate in \runTwo.

\subsection{\runTwo radial fiducial-volume cut motivation}
\label{sec:fidMot}

The two primary reasons to apply a radial fiducial-volume cut are to remove events of which the energy reconstruction is inaccurate and to remove low-energy background events (\eg, $^{222}$Rn daughters on the detector surfaces and surrounding material).  Such a cut was not applied in the \runOne analysis as the small data set did not allow the impact of the cut to be properly assessed.  With the larger \runTwo exposure, however, a radial fiducial-volume study became possible.  The \runTwo cut was particularly motivated by further study of the CDMSlite electric-field configuration and an unexpected instrumental background population.

\subsubsection{Improved understanding of electric-field effects}
\label{sec:electricfields}

A copper detector housing enclosed the crystal radially with a small gap between the detector edge and the grounded housing.  Such an arrangement, coupled with the asymmetric biasing configuration, led to an inhomogeneous electric field. The field geometry was modeled by finite-element simulation using \textsc{COMSOL Multiphysics}$^{\circledR}$ software (COMSOL, Inc., Burlington, MA).  The simulation only included a single detector, and thus any effects from the biased detectors above and below the CDMSlite detector were not included.  The resulting electric field showed in which parts of the detector freed charges were attracted to the sidewall, and the grounded housing outside, rather than the grounded flat face. These regions experienced reduced NTL phonon emission and therefore a reduced reconstructed energy compared to events of the same initial-energy deposition in the bulk of the detector.

\begin{figure}
	\centering
	\includegraphics[width=\columnwidth]{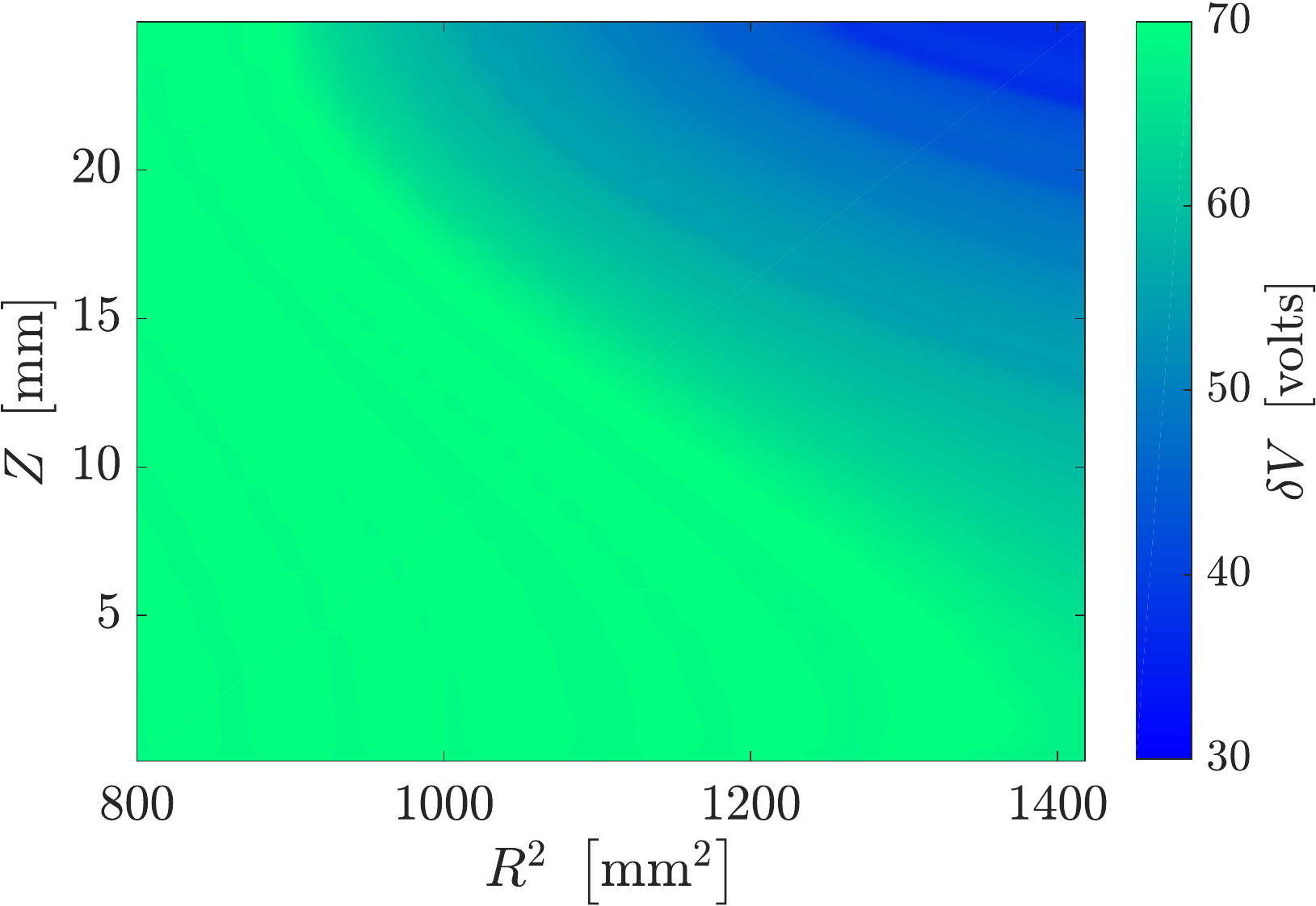}
	\caption{Difference in electric potential between the final locations of electrons and holes (color map), after propagating through the crystal, as a function of their initial position in the detector.  A single vertical slice of the detector, perpendicular to the circular top and bottom faces (see \FIG\ref{fig:CDMSlite_Det}) and along an arbitrary radius ($R$ coordinate, with 0 at the center of the detector) is shown.  To uniformly cover the crystal, the squared radius is sampled, and thus $R^2$ is plotted.  The top of the crystal (along the $Z$ coordinate) is at 70~V, and the bottom is at 0~V.  The copper housing (not shown at high $R^2$) surrounding the detector is also at 0~V, and a small gap exists between it and the sidewall.  This causes the total potential difference experienced by drifting charges to be ${<}$70~V in regions where field lines terminate on the sidewall.  Radii with $R^2<800$~mm$^2$ experience the full 70~V potential difference and are not shown.}
	\label{fig:onesideE}
\end{figure}

To further quantify the position-dependent effective bias voltage due to field inhomogeneities, a Monte Carlo simulation was performed of the detector crystal considering the calculated field map.  In this simulation, electron-hole pairs were placed at various points throughout the detector volume and allowed to propagate according to the electric-field map.\footnote{The electrons travel along the direction of the field at high bias voltages. Thus, oblique propagation and internally scattering mechanisms were disabled in order to increase the efficiency of the simulation.}  The difference in electric potential at the final positions of the charge carriers was recorded for each pair, allowing for the construction of a potential difference map $\delta V = f{\left(x,y,z\right)}$.  A slice of this map is given in \FIG\ref{fig:onesideE} and shows the region of reduced potential near the sidewall and the biased face.

\begin{figure} 
	\centering
	\includegraphics[width=\columnwidth]{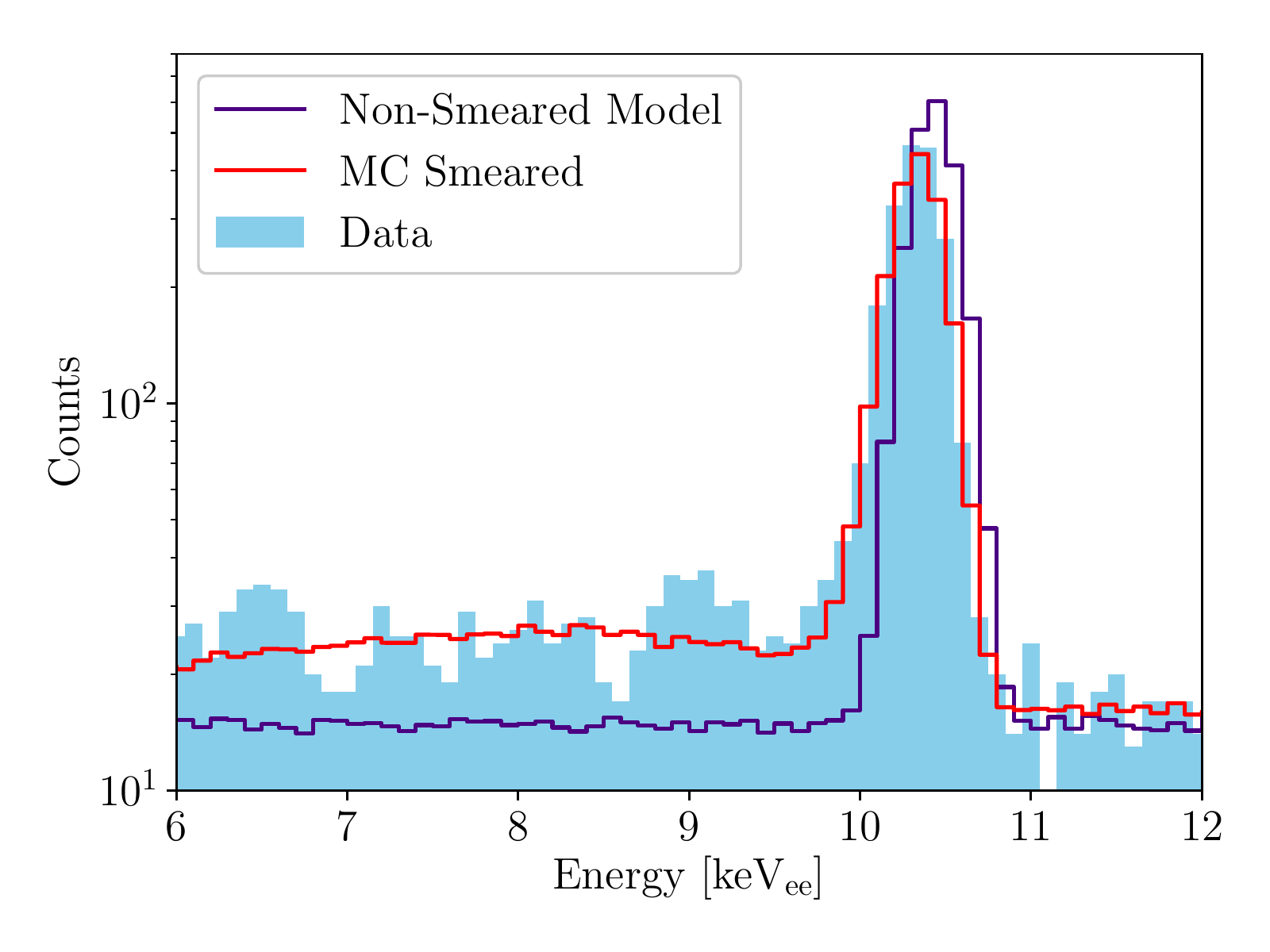}
	\caption{$^{71}$Ge $K$-shell peak in the \runTwo data, with no fiducial-volume cut, compared to the results of the electric-field study.  The study simulates peak events on top of a flat Compton background before applying a smearing function. The smeared low-energy tail observed in the data is replicated in the simulation.}
	\label{fig:smearedSpectrum}
\end{figure}

The reduced NTL phonon emission at the edge of the detector has the effect of smearing the energy response to lower energies. Of particular interest is the effect on the $^{71}$Ge $K$-shell peak, which has visible smearing in the nonfiducialized \runTwo data as shown in \FIG\ref{fig:smearedSpectrum}.  To estimate this smearing, sample events were drawn from a flat spectrum to model the Compton background, plus a Gaussian peak distribution, with the rate, mean, and width of the distributions chosen to match the observed spectrum.  Next, a position was uniformly selected in the crysta,l and the corresponding potential drop from $\delta V = f{\left(x,y,z\right)}$ was used. For every sample from the initial spectrum, $E_i^{\text{init}}$, the energy $E_i^{\text{final}}$ expected to be measured for an interaction at the respective position in the detector was calculated as
\begin{equation}
	E_{i}^{\text{final}} = E_{i}^{\text{init}} \times \frac{1+e\delta V_i/\epgeq}{1+e\Vbeq/\epgeq},
\end{equation}
where \Vb is the applied 70~V bias.  The result of this smearing is also shown in \FIG\ref{fig:smearedSpectrum}.  The asymmetric peak observed in the data, as expected from the reduced NTL gain, is matched by the smeared simulation.  The smearing also partially explains the rise in counts below the peak.

The \runOne analysis did not apply a cut to remove events from this region of the detector; nor did it account for this smearing in the assumed WIMP-recoil spectrum used for deriving the published upper limit.  The effect on the \runOne result was studied postpublication by considering the fractional change of the cumulative above-threshold WIMP spectrum due to smearing the spectrum.  The smear decreased the expected above-threshold WIMP spectrum by ${\lesssim}$5\,\% for WIMP masses above 3~\gev.  The change to the published results would thus be well within the uncertainty associated with the ionization yield model shown in \FIG\ref{fig:limits_si}.

The simulation and study performed here are sufficient to identify the electric field as the source of the observed spectral smearing.  They are insufficient, however, for use in the analysis of the measured data, as they cannot inform how to remove the low-gain events.  Regions at high radius are clearly seen to be most affected. However, a map of the true physical location as derived from accessible position-dependent analysis parameters is not known \textit{a priori}, requiring an in-depth simulation of the phonon propagation and signal formation in the detector. Such a simulation is under development by SuperCDMS~\cite{Leman2012}. The underlying physics is understood and implemented in these simulations, but work is still needed to match simulated pulses to data.  Thus, these simulations could not be used for the studies presented here.

\subsubsection{Localized instrumental background}
\label{sec:run2spot}

\begin{figure}
	\centering
	\includegraphics[width=\columnwidth]{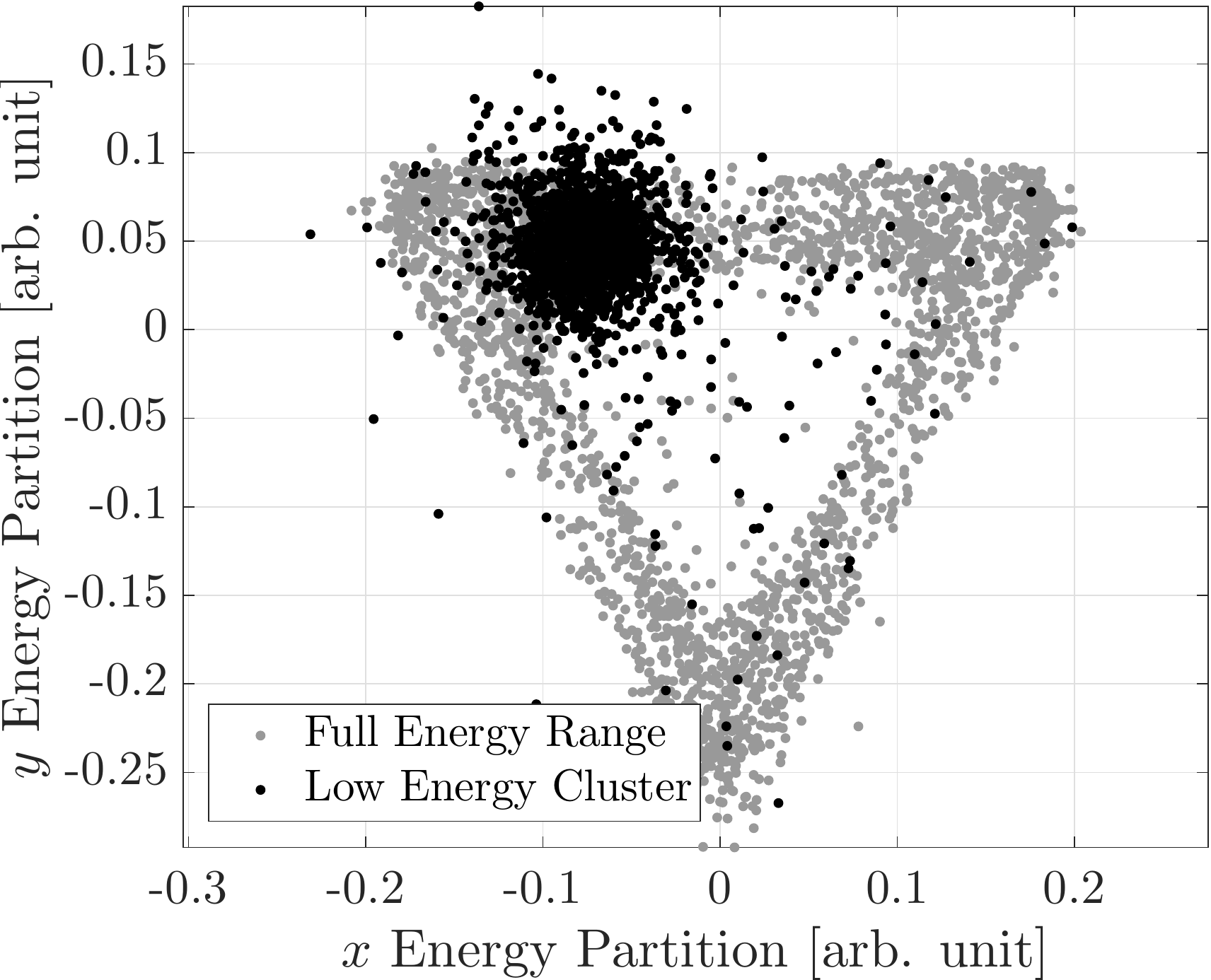}
	\caption{Position of \runTwo events using the energy partition coordinates.  Events in the full energy range are gray, while those between 100 and 200 \evee are highlighted in black.  The population at low energy is clearly clustered in position.}
	\label{fig:spot}
\end{figure}

In \perTwo of \runTwo, an instrumental background appeared at 100--200~\evee.  These events are identifiable as background as they are localized in time, only occurring during \perTwo, and position. This position localization can be seen in an $x$-$y$-plane representation shown in \FIG\ref{fig:spot}, where the positions $X_{\text{OF}}$ and $Y_{\text{OF}}$ are computed by the partition of energy between the three inner channels as
\begin{align}
    X_{\text{OF}}&= \frac{\cos\left(30\text{\textdegree}\right)D_{\text{OF}} + \cos\left(150\text{\textdegree}\right)B_{\text{OF}}+\cos\left(270\text{\textdegree}\right)C_{\text{OF}}}{B_{\text{OF}}+C_{\text{OF}}+D_{\text{OF}}} \label{eq:xPart} \\
    Y_{\text{OF}}&= \frac{\sin\left(30\text{\textdegree}\right)D_{\text{OF}} + \sin\left(150\text{\textdegree}\right)B_{\text{OF}}+\sin\left(270\text{\textdegree}\right)C_{\text{OF}}}{B_{\text{OF}}+C_{\text{OF}}+D_{\text{OF}}} \label{eq:yPart},
\end{align}
where $B_{\text{OF}}$, $C_{\text{OF}}$, and $D_{\text{OF}}$ are the OF fit amplitudes for the three inner channels and the angles correspond to their relative locations (\textit{cf.} \FIG\ref{fig:CDMSlite_Det}); events at the corners of the triangle correspond to events that are predominately underneath a single channel's sensors.  The events in the energy range of the low-energy cluster are highlighted and localized near the top left corner, implying that they are localized in a single channel.  The exact source of these events is unknown, but their localization in time and position identifies them as an instrumental background that can be removed, as shown in the next section.\footnote{Similar instrumental backgrounds have been observed during early CDMSlite testing of other detectors.}

\subsection{\runTwo radial fiducial volume cut implementation}
\label{sec:run2fid}

A fiducial-volume algorithm was developed based on the position information from the 2T fit (defined in \SEC\ref{sec:PulseFit}).  The channel nearest the event has the highest fast-amplitude contribution (see \FIG\ref{fig:2Texample}) and the earliest pulse onset. These features are used to define a new radial parameter with improved position resolution, which is used to exclude events at high radius~\cite{Pepin2016}. The parameter was derived in several steps: 
\begin{enumerate}
	\item Correct for time variations: correct the energy-carrying slow-template amplitude for each channel in the same manner as described in \SEC\ref{sec:run2gainCorr}. Derive the corrected fast amplitude $N^{\text{Corr}}_f$ (where $N$ stands for the channel labels $A$--$D$) by applying these same correction factors to the fitted fast-template amplitude.
    \item  Correct for spatial variations: for channel $N$ calculate a relative calibration coefficient $\xi_{N,\text{2T}}$ by normalizing the average of the slow-template amplitude over all good pulses in the energy region of interest to the respective average of channel $A$. This ensures that the energy scale is the same in all sensors.
    \item  Determine a weight factor for each channel. This is done in three steps:
    \begin{enumerate}
	\item Determine peakiness: For channel $N$, the peakiness $P_N$ is given by the corrected fast amplitude $N^{\text{Corr}}_f$ scaled by the relative calibration factor $\xi_{N,\text{2T}}$ of that channel normalized by the total energy of the event \Eree as defined in \SEC\ref{sec:run2gainCorr}:
	\begin{equation}
		P_N = \xi_{N,\text{2T}}\cdot N^{\text{Corr}}_{f}/\Ereeeq
	\end{equation}
	$P_N$ will be high for channels close to the interaction point.
	\item Determine the delay: For channel $N$, the delay $\Delta_N$ is given by the difference of the 2T-fit delay parameters for that channel, $\delta_{N,\text{2T}}$ and for the total phonon pulse, $\delta_{tot,\text{2T}}$:
	\begin{equation}
		\Delta_N = \delta_{N,\text{2T}} - \delta_{tot,\text{2T}}
	\end{equation}
	$\Delta_N$ will be low for channels close to the interaction point.
	\item The weight factor $W_N$ for channel $N$ is now defined as the difference between the delay and the peakiness:
	\begin{equation}
		W_N = \Delta_N - P_N
	\end{equation}
	$W_N$ will be low for channels close to the interaction point.
    \end{enumerate}
	\item Construct a preliminary radial parameter $R_{0,\text{2T}}$ as the difference between the weight of the outer channel and that of the inner channel that is closest to the interaction point:
	\begin{equation}
		R_{0,\text{2T}} = \text{min}{\left(W_B,W_C,W_D\right)} - W_A
	\end{equation}
$R_{0,\text{2T}}$ is low for events in the center of the detector and high for events near the edge.
	\item  Construct $x$- and $y$-positions $X_{\text{2T}}$ and $Y_{\text{2T}}$ in the same manner as the numerators of \EQS\ref{eq:xPart} and \ref{eq:yPart} using the weights derived here instead of the OF-fitted amplitudes.
    \item  Derive the final radial parameter $R_{\text{2T}}$ by correcting for a systematic dependence on angular position, reflecting the threefold symmetry of the sensor layout, that is observed in the $X_{\text{2T}}$ \vs $Y_{\text{2T}}$ plane. 
\end{enumerate}

\begin{figure}
	\centering
    \includegraphics[width=\columnwidth]{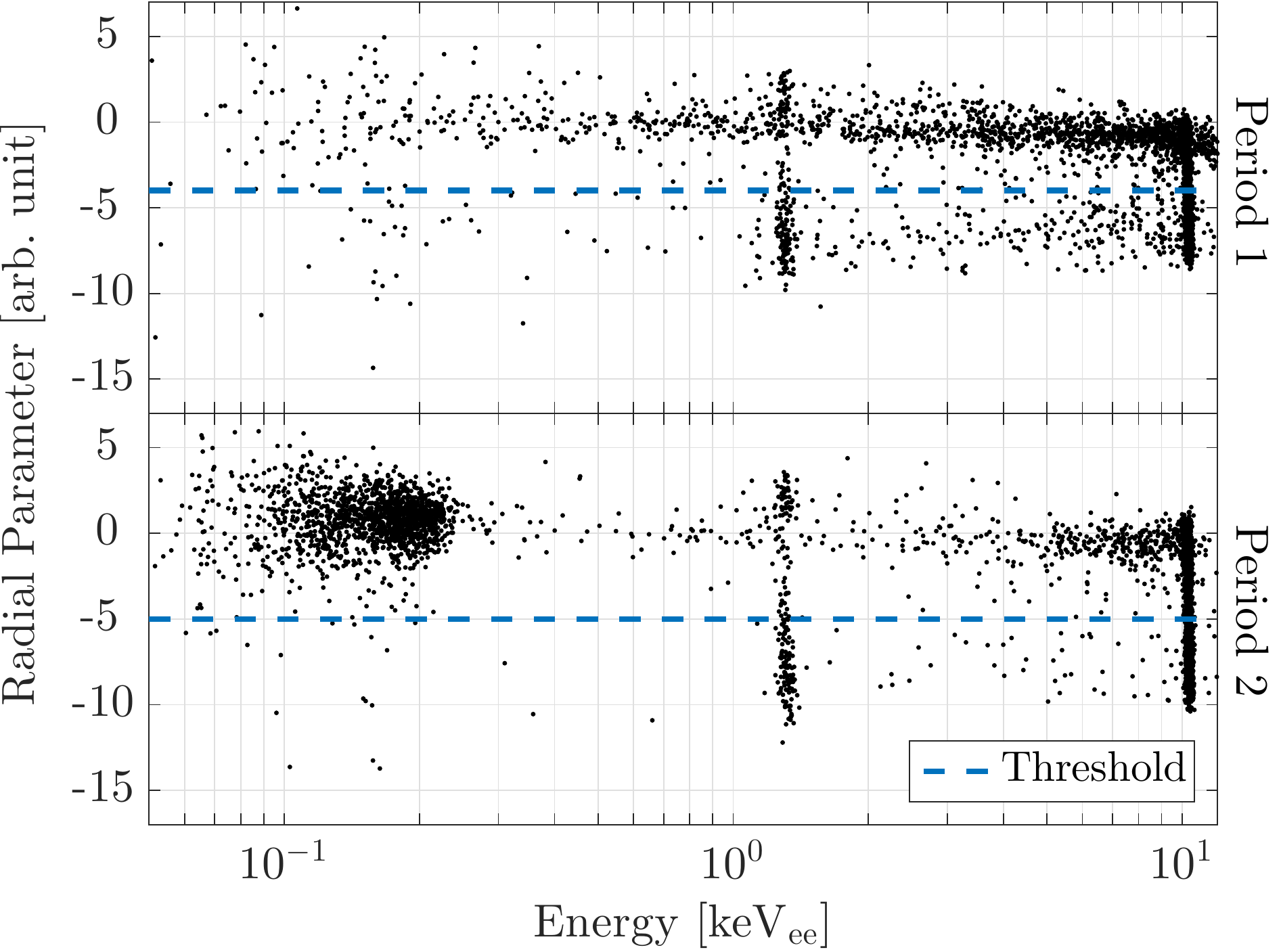}
    \caption{2T-fit-based radial parameter as a function of energy for \runTwo \perOne (top) and \perTwo (bottom).  The vertical clusters are the $^{71}$Ge-activation lines, and the horizontal band at high radius contains reduced-amplification events.   The radial cut thresholds are indicated by the blue dashed line, effectively removing events at high radius, including the low-energy cluster seen in \perTwo.}
    \label{fig:2Trad}
\end{figure}

Figure~\ref{fig:2Trad} shows the final $R_{\text{2T}}$ as a function of reconstructed energy.  A higher density of events is seen at higher radius, and the $^{71}$Ge-activation peaks are visible as vertically oriented populations at 1.30 and 10.37~\kevee.  The low-energy instrumental background in \perTwo is also visible, localized at high radial parameter.  Note that events from within the cluster were not used in defining the radial parameter.   It is obvious that $R_{\text{2T}}$ is a nonlinear function of the true radius; the event density in the activation lines (particularly the $L$-shell peak) shows a clear decrease with increasing radius and then rises when the edge events begin to contribute. The cut threshold in the radial parameter, given by the dashed horizontal lines in \FIG\ref{fig:2Trad}, was chosen empirically on the falling edge of the radial distribution of the inner events of the $L$-shell peak, maximizing the efficiency while removing the low-energy cluster along with essentially the entire edge-event distribution.  The radial distributions of the two periods differ somewhat, leading to slightly different choices of cut threshold values between the periods.

\begin{figure}
	\centering
	\includegraphics[width=0.7\columnwidth]{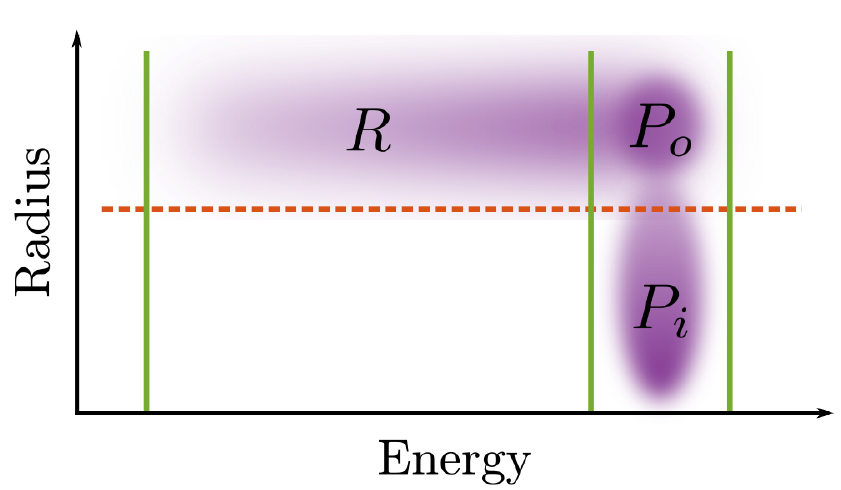}
	\caption{Diagram showing the morphology of the expected event distribution in the radial-parameter \vs reconstructed-energy plane from a monoenergetic homogeneously distributed source. The distribution is split (vertical solid lines) into nonpeak events $R$, with reduced NTL amplification, and peak events $P$. The latter group is further separated into inner peak events $P_i$, that pass the cut threshold (horizontal dotted line), and outer peak events $P_o$, that do not.  In practice, the $^{71}$Ge-activation peaks were considered and can be separated from background because of the known half-life of the isotope.}
	\label{fig:radEffCartoon}
\end{figure}

The signal efficiency of the radial cut was determined using the known 11.43~day half-life~\cite{Hampel1985} of the $^{71}$Ge produced \textit{in situ} during neutron calibrations, together with a pulse-simulation technique. The expected distribution of events from a monoenergetic and uniformly distributed source in the plane of radial parameter \vs reconstructed energy is sketched out in \FIG\ref{fig:radEffCartoon}.  The population is divided into two groups: events with reduced NTL amplification due to field variation ($R$) and those with full amplification that appear in the peak ($P$). The peak population is further split into two sub-groups: inner events that pass the radial cut ($P_i$) and outer events that do not ($P_o$).  The signal efficiency $\mathcal{E}$ of the radial cut is defined by the probability that an individual event of the population passes the cut and appears at the expected energy: 
\begin{equation}
	\mathcal{E} = \frac{P_i}{R+P} = \frac{P}{R+P}\cdot\frac{P_i}{P}.
\end{equation}
The second step separately calculates the fraction of events that have full NTL amplification, $\mathcal{E}_1=P/\left(R+P\right)$, and the fraction of events with full amplification that pass the radial cut, $\mathcal{E}_2 = P_i/P$.  These two factors are determined separately, taking into account the presence of background events that are not associated with the $^{71}$Ge decay.

To compute $\mathcal{E}_1$, the plane spanned by the radial and energy parameters was separated into several two-dimensional bins with notably different concentrations of $K$-shell capture events. The event distribution as a function of time was then fit, within each of these bins, with the sum of a constant and an exponential with an 11.43~day half-life, to separate the background from the $^{71}$Ge contribution.  The known ratio of $K$- to $L$-capture events, together with the assumption that the energy reduction is based on the electric-field geometry and thus proportional to the recoil energy, was used to identify the distribution of $K$-capture events at energies below the $L$-capture line.  Following the steps outlined in this paragraph gives $\mathcal{E}_1=86\pm0.9\,\%$, where the uncertainty is statistical, and due to the finite number of events in each radius \vs energy bin.  For the chosen cut position, more than 90\,\% of the events with reduced energy are removed. This calculation also provides $\mathcal{E}_2$ for the $K$-shell activation line as $\mathcal{E}_2=54.5\pm1.9$\,\% and $49.8\pm1.7$\,\% for Periods 1 and 2, respectively.  The total signal efficiency at the $K$-shell peak is then $\mathcal{E}=47.3\pm1.7$\,\% for Period 1 and $43.2\pm1.6$\,\% for Period 2.

To determine $\mathcal{E}_2$ at lower energies, a pulse-simulation method was implemented. All events from the $L$-peak were converted to quasi-noise-free pulses by combining the fast and slow templates from the 2T fit according to their respective fit amplitudes for each of the phonon channels. The $K$-peak would have provided considerably more events; however, because of saturation of the 2T-fit--fast-template amplitude in the outer channel above $\sim$2~\kevee,\footnote{The onset of this saturation was used to determine the upper energy threshold for events used in the final WIMP results.} these were not a good representation of the low-energy events, and thus could not be used for this study. The noise-free pulses were then scaled to each of 13 different energies between 0.04 and 1.30~\kevee before measured noise traces were added.  The full $L$-shell population was scaled to each energy, as opposed to using subpopulations for each, because of the limited number of peak events.  In each case, the measured noise was taken from the same time period as the original pulse.  At each scaled energy, the same combination of the $L$-peak event and noise event was used.  By using the measured 2T-fit fast/slow amplitude ratio for the simulated pulses, the radial distribution of the $L$-shell peak events was simulated at each energy.

\begin{figure}
	\centering
	\includegraphics[width=\columnwidth]{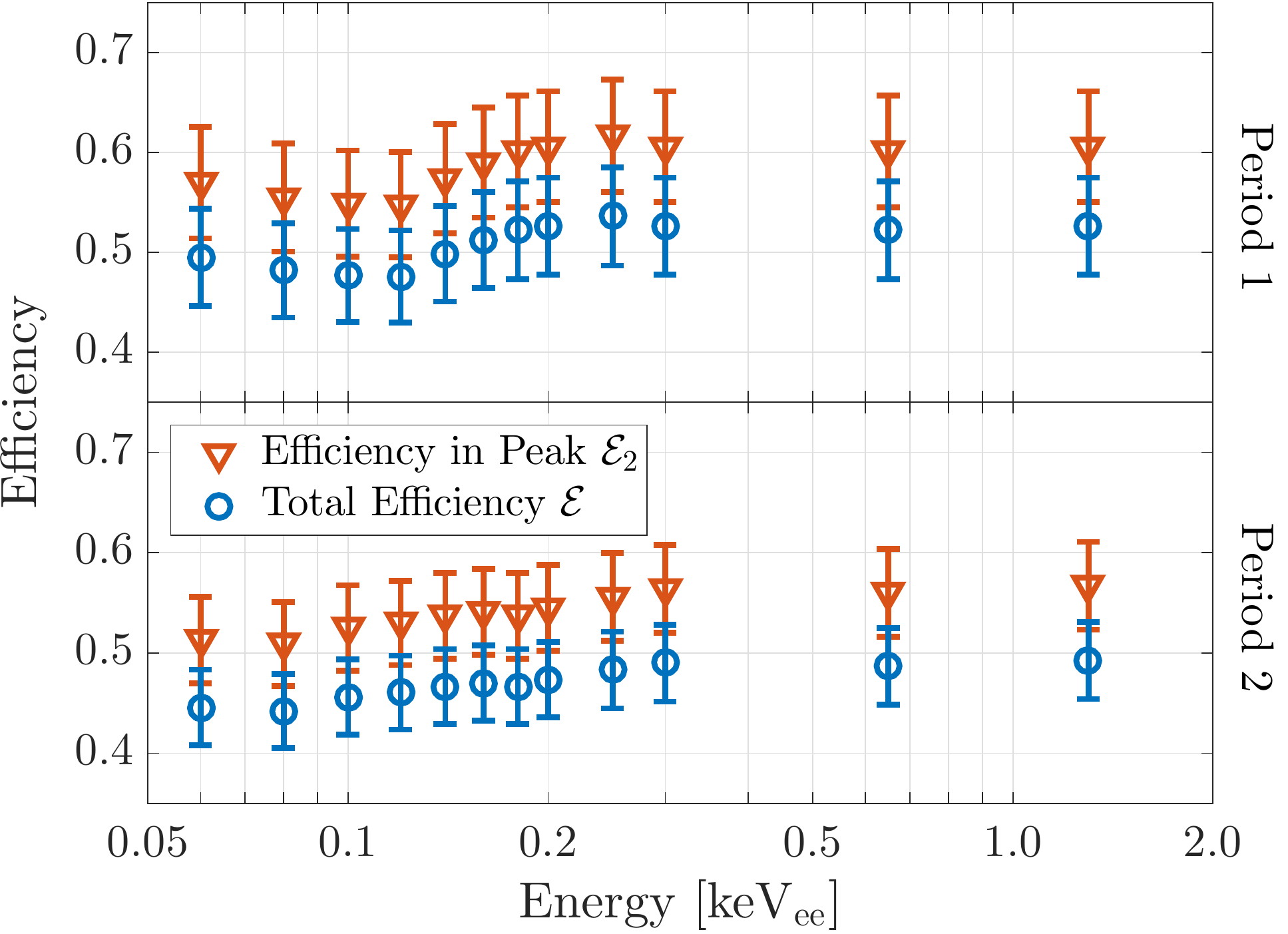}
	\caption{Radial fiducial-volume cut efficiency below 2~\kevee for \perOne (top) and \perTwo (bottom).  The efficiency at full NTL amplification $\mathcal{E}_2$ (orange triangles) as well as the total efficiency $\mathcal{E}$ (blue circles) are shown along with their respective uncertainties.  The error bars on $\mathcal{E}_2$ encompass statistical uncertainty due to the available number of $L$-shell peak events used as simulation inputs (same for each energy simulated), statistical uncertainty due to the number of simulated events passing the cut (different for each energy simulated), and a systematic uncertainty due to the estimate of nonpeak background events simulated (same for each energy simulated).  The error bars on $\mathcal{E}$ additionally contain a small statistical uncertainty from the computation of the efficiency to have full NTL amplification (same for each energy simulated).}
	\label{fig:radEff}
\end{figure}

The cut efficiency was then measured by applying the chosen radial cut to the distribution of artificial events at each energy, accounting for the radial distribution of signal and background as measured in and around the $L$-peak.  At lower scaled energies, some events which were close to, and on one side of, the cut threshold in the original $L$-shell sample moved to the other side because of the added noise.   However, threshold crossing occurred in both directions; therefore, the overall cut efficiency stayed almost constant down to the lowest energies tested, as shown in \FIG\ref{fig:radEff}.  The uncertainty on $\mathcal{E}_2$ contains statistical uncertainty due to the limited number of $L$-shell peak events (same for each energy simulated), statistical uncertainty due to the number of simulated events that passed the cut (different for each energy simulated), and a systematic uncertainty on the estimate of nonpeak background events simulated (same for each energy simulated).

\subsection{Effect of the delay parameter in the radial efficiency calculation}
\label{sec:radEffRedo}

As discussed in the previous section, the radial parameter was constructed from a combination of 2T-fit amplitude differences and relative delay of the outer and primary inner phonon channels. The pulse simulation used to compute the radial cut efficiency, described in the previous section and implemented for the original publication of the \runTwo data \cite{Agnese2016}, only considered the relative amplitude of the input $L$-shell events without including the relative delay. In order to confirm that this omission did not introduce any significant systematic uncertainty, a new version of the pulse simulation that included this relative delay of the input pulses was tested. The largest change  between the original implementation and the improved version of the pulse simulation is seen at 60~\evee, just above threshold in \perTwo, where the central value of the efficiency drops by about 6\%. However, all changes are well within the statistical uncertainties (typically ${\pm}$10\,\%--15\,\%). Given the lack of statistical significance, this modification was not propagated into any final results.

\subsection{Background rates and energy dependence}
\label{sec:background}

The effectiveness of the \runTwo radial fiducial-volume cut in reducing the background rate can be seen by comparing the resulting spectrum to that of \runOne (\FIG\ref{fig:spectra}). These spectra show the energy of events that scatter only in the CDMSlite detector, called ``single scatters.''  Single-scatter events are of interest as \ws are expected to scatter extremely rarely, whereas photons and electrons often scatter multiple times in the detector array giving ``multiple scatters.''  Multiple-scatter events were removed from the analysis of both data sets to reduce the background rate, with a loss of ${<}$2\,\% in signal efficiency for both analyses.

In both spectra, the germanium activation lines are seen to be on top of a continuous background, primarily from Compton scattering $\gamma$'s.  The average rate between the various activation peaks and analysis thresholds are given in \TAB\ref{tab:specRates} for both analyses.  The \runTwo rate above the $K$-shell peak is reduced by a factor of 6 from the \runOne rate by the fiducial-volume cut.  The \runTwo rates are also significantly reduced at lower energies compared to those of \runOne, though some energy dependence is seen.

\begin{table}
	\centering
	\begin{tabular}{h{4.3}p{3.3}p{3.3}p{4.4}p{3.3}}
		\hline \hline
		\multicolumn{1}{c}{Range}	& \multicolumn{1}{c}{\runOne Rate}	& \multicolumn{3}{c}{\runTwo Rate $\left[\keveeeq\,\text{kg}\,\text{d}\right]^{-1}$}	\bigstrut[t] \\
		\multicolumn{1}{c}{$\left[\keveeeq\right]$}	& \multicolumn{1}{c}{$\left[\keveeeq\,\text{kg}\,\text{d}\right]^{-1}$}	& \multicolumn{1}{c}{Full}	& \multicolumn{1}{c}{\perOne}	& \multicolumn{1}{c}{\perTwo}	\bigstrut[b] \\
		\hline
		0.056,0.14	& \multicolumn{1}{c}{-} 	& 16,8	& 2.5,1.3	& 26,10	\bigstrut[t]	\\
		0.17,1.1	& 5.5,1.0			& 1.1,0.2	& 1.2,0.2	& 0.86,0.43			\\
		1.5,7.5	& 2.7,0.3			& 0.97,0.07	& 0.95,0.08	& 1.1,0.2			\\
		12,22		& 1.5,0.2			& 0.25,0.03	& 0.26,0.03	& 0.20,0.06	\bigstrut[b]	\\
		\hline \hline
	\end{tabular}
	\caption{Average single-scatter event rate for energy regions between the activation lines in \runOne, the full \runTwo exposure, and the two periods within \runTwo.  All errors contain ${\pm}\sqrt{N}$ counting uncertainties, and the \runTwo values additionally include uncertainty from the analysis efficiency (negligible in \runOne).  For \runTwo \perOne, the first energy bin cuts off at that period's threshold of 75~\evee.  See the text for discussion on the various rates.}
	\label{tab:specRates}
\end{table}

Previous measurements of the Compton background at higher energies indicated a flat rate of  ${\sim}$1.5~\cpdee\cite{Ahmed2010c}.  As shown in \TAB\ref{tab:specRates}, this rate was confirmed above the  $K$-shell activation line in \runOne.  Additionally, the measurements show that, below this peak, the overall background rate increased toward lower energy in both analyses.  The increase in rate going from above to below the $K$-shell peak can be explained by the decay of cosmogenic isotopes within the detector and, for the \runOne spectrum, $^{71}$Ge events with reduced NTL amplification (see \SEC\ref{sec:electricfields}).

\begin{figure}
	\centering
	\includegraphics[width=\columnwidth]{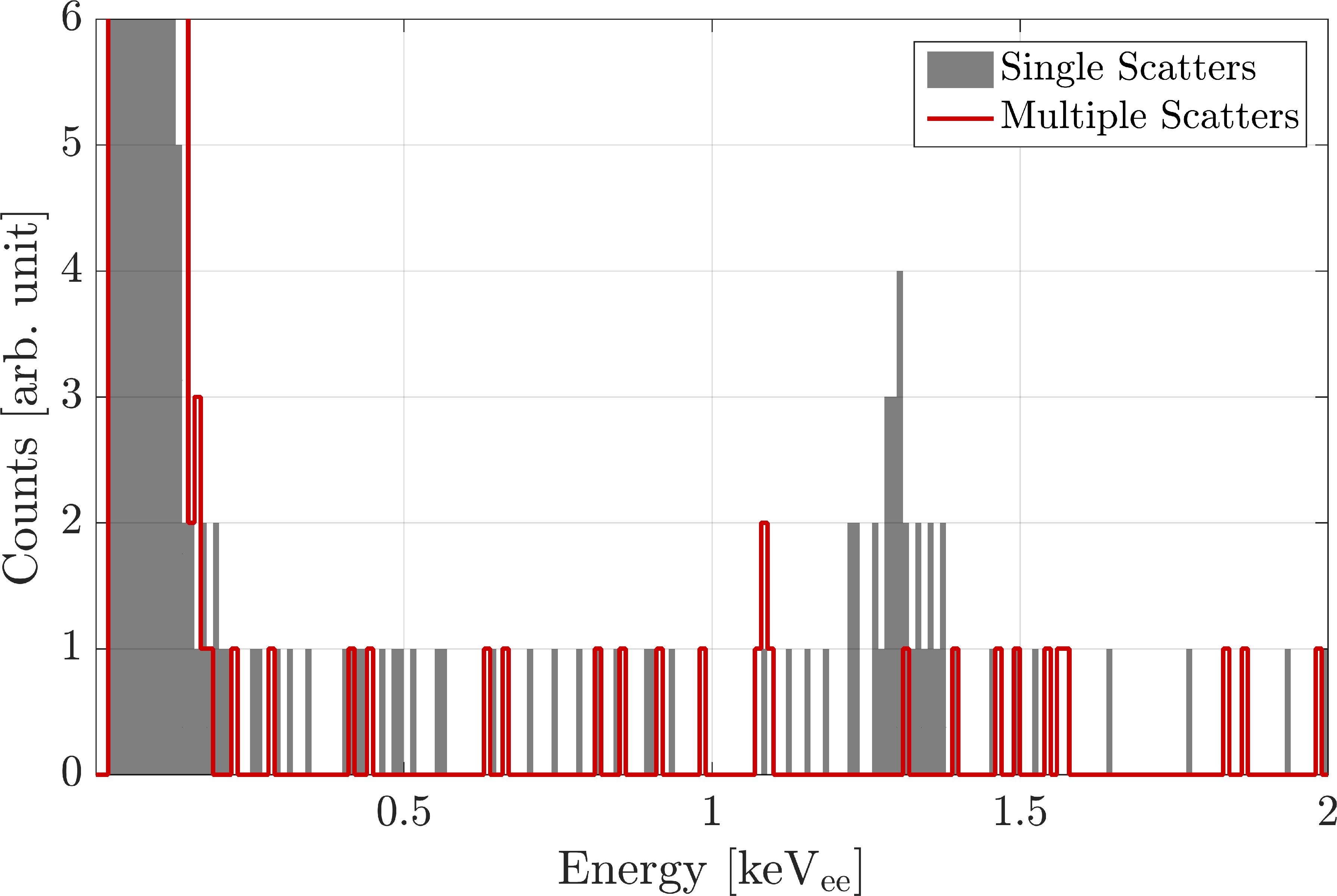}
	\caption{\runOne low-energy spectrum showing both single- (gray shaded) and multiple-scatter (red line) events.  Below the $L$-shell peak, the shape of the multiple-scatter spectrum is statistically compatible with the shape of the single-scatter spectrum.}
	\label{fig:Eree_singmult_raw}
\end{figure}

The \runOne spectrum shows a further increase in rate below the $L$-shell peak.  A statistical test to compare the single- and multiple-scatter spectra was performed to understand this energy region.  The \runOne multiple-scatter spectrum is shown together with the single-scatter spectrum below 2~\kevee in \FIG\ref{fig:Eree_singmult_raw}.  These two spectra were compared by performing a Kolmogorov-Smirnov (KS) test using the energies for events between the $L$-shell peak and threshold.  The test accepts the hypothesis that these two spectra are drawn from the same underlying probability distribution functions, giving a p-value of $79.24\,\%$ that is considerably above the standard 5\,\% hypothesis acceptance limit for a KS test.  This shows that the shape of the single-scatter spectrum is consistent with that of the WIMP-free multiple-scatter spectrum, and thus the increase at low energy cannot be taken as indication of a WIMP signal.  This is further supported by the fact that the single-scatter rates above and below the $L$-shell peak in the \runTwo spectrum are statistically compatible with each other.

The \runTwo spectrum shows an increase in rate going from above to below the $M$-shell peak.  Comparing the two periods of \runTwo in this energy range gives insight into this excess.  For all energy regions above the $M$-shell peak, the two periods' rates are statistically consistent.  Below the $M$-shell peak, however, the rate in \perTwo is dramatically higher compared to \perOne.  This indicates that the increase in rate is likely due to background events leaking past the selection cuts.  Such leakage is generally expected at lower energies, and leakage of the localized instrumental background in \perTwo (\SEC\ref{sec:run2spot}) can explain the difference between the periods.

Further studies of the rate require a detailed knowledge of the shape of all expected background distributions.  The spectral shape of Compton recoils at very low energies is actively being studied.  A recent simulation study of the effects of atomic shell structure using \geant~\cite{Agostinelli2003,*Allison2006} has shown that the Compton spectrum should not be expected to be flat~\cite{Barker2016a}.  Tritium and other low-energy background sources (\eg, $^{210}$Pb daughters) will additionally modify the expected spectral shape, and are still being studied with simulations. Future analyses will attempt to take this information into account.

\section{New \runTwo dark matter results}
\label{sec:results}

This section presents new results based on the \runTwo analysis, including the effect of varying astrophysical parameters on the spin-independent limit, as well as limits on spin-dependent interactions.

\subsection{Effects of varying astrophysical parameters}
\label{sec:halo}

The astrophysical description of the \w halo described in \SEC\ref{sec:intro} enters the differential WIMP-rate expression through the halo-model factor $\mathcal{I}_{\text{halo}}$, which depends on the velocities of the \ws $\bm{v}$, the velocity of the Earth \wrt the halo $\bm{v}_E$, and the local dark matter mass density $\rho_0$.  As defined in \EQ\ref{eq:astroIdef}, this factor is an integral over the assumed velocity distribution of the halo \wrt the Earth $f{\left(\bm{v},\bm{v}_E\right)}$.

The limits computed for both Runs 1 and 2 assume the standard halo model (SHM) for the dark matter spatial and velocity distributions.  The SHM assumes an isotropic, isothermal, and nonrotating sphere of dark matter in which the Galaxy is embedded.  The velocity distribution associated with this model is a Maxwellian distribution boosted to the lab frame of the Earth as
\begin{equation}
	f{\left(\bm{v},\bm{v}_{\text{E}}\right)}\propto\exp{{\left(-\left|\bm{v}+\bm{v}_E\right|^2/2\sigma_v^2\right)}},
\end{equation}
where the proportionality constant has already been subsumed into \EQ\ref{eq:astroIdef} and the velocity dispersion is $\sigma_v^2=v_0^2/2$, where $v_0$ is the large-radius asymptotic Galactic circular velocity.  It is typically assumed that this asymptotic value has been reached at the Sun's position~\cite{Donato1998}, giving $v_0=\Theta_0\equiv\left|\bm{\Theta}_0\right|$.  $\bm{\Theta}_0$ is the Galactic local standard of rest (LSR), corresponding to the average circular orbital velocity at the Sun's distance from the Galactic center.\footnote{The LSR is of interest to astronomers regardless of whether this assumption is true, and thus the $\Theta_0$ notation, common in the astrophysical literature, is used for the LSR and its equality to $v_0$ only taken when specifically referring to the SHM.}  The Earth's velocity is decomposed as $\bm{v}_E=\bm{\Theta}_{0}+\bm{v}_{\odot}+\bm{v}_{\oplus}$, where the other velocities are $\bm{v}_{\odot}$, the solar peculiar velocity \wrt neighboring stars, and $\bm{v}_{\oplus}$, the Earth's orbital velocity around the Sun.  The Earth's orbital velocity is assumed to average to zero over a year.  Integrating this distribution over the range of velocities described in \SEC\ref{sec:intro} gives \EQ\ref{eq:astroIsolve}.  Note that the maximum velocity used in the integration, which is related to the Galactic escape velocity $v_{\text{esc}}$, truncates the theoretical distribution which would otherwise extend to infinite velocities.

The direct-detection experimental community has been using a uniform set of measurements for each of these parameters in its analyses: $\rho_0=0.3~\text{GeV\,c$^{-2}$\,cm$^{-3}$}$~\cite{Olive2014}, $\bm{\Theta}_{0}=220\pm20~\text{km\,s}^{-1}$~ in the direction of Galactic rotation~\cite{Kerr1986}, $v_{\text{esc}}=544^{+64}_{-46}~\text{km\,s}^{-1}$~\cite{Smith2007}, and $\bm{v}_{\odot}=\left(11.0\pm1.2,12.24\pm2.1,7.25\pm1.1\right)~\text{km\,s}^{-1}$, where the first component is the radial velocity toward the Galactic center, the second component is in the direction of Galactic rotation, and the third component is the vertical velocity (out of the Galactic plane)~\cite{Schonrich2010}.  It is well known that the uncertainties in these values, in particular $\bm{\Theta}_0$ and $v_{\text{esc}}$, can have significant effects on computed WIMP exclusion limits~\cite{McCabe2010,*Green2012,*Ohare2016}, and thus astrophysical uncertainties are also expected on the CDMSlite \runTwo spin-independent result.  Although the local dark matter density is also uncertain~\cite{Salucci2010}, all experiments are equally affected by its value, so the effect of its uncertainty on the \runTwo limit is not considered further.

For this astrophysical-parameter discussion, the \runTwo analysis uncertainties are not considered.  Upper limits are computed using the central efficiency curve in \FIG\ref{fig:effs} and the standard Lindhard model with $k=0.157$: a set of parameters labeled ``best fit.''\footnote{Calling this the ``best fit'' is a slight misnomer as no actual fitting was performed to obtain the values.}  All other assumptions about the rate discussed in \SECS\ref{sec:intro} and \ref{sec:runs} are left unchanged, and the optimum interval method~\cite{Yellin2002,*Yellin2007} is again used to compute limits.

\begin{figure}
	\centering
    \includegraphics[width=\columnwidth]{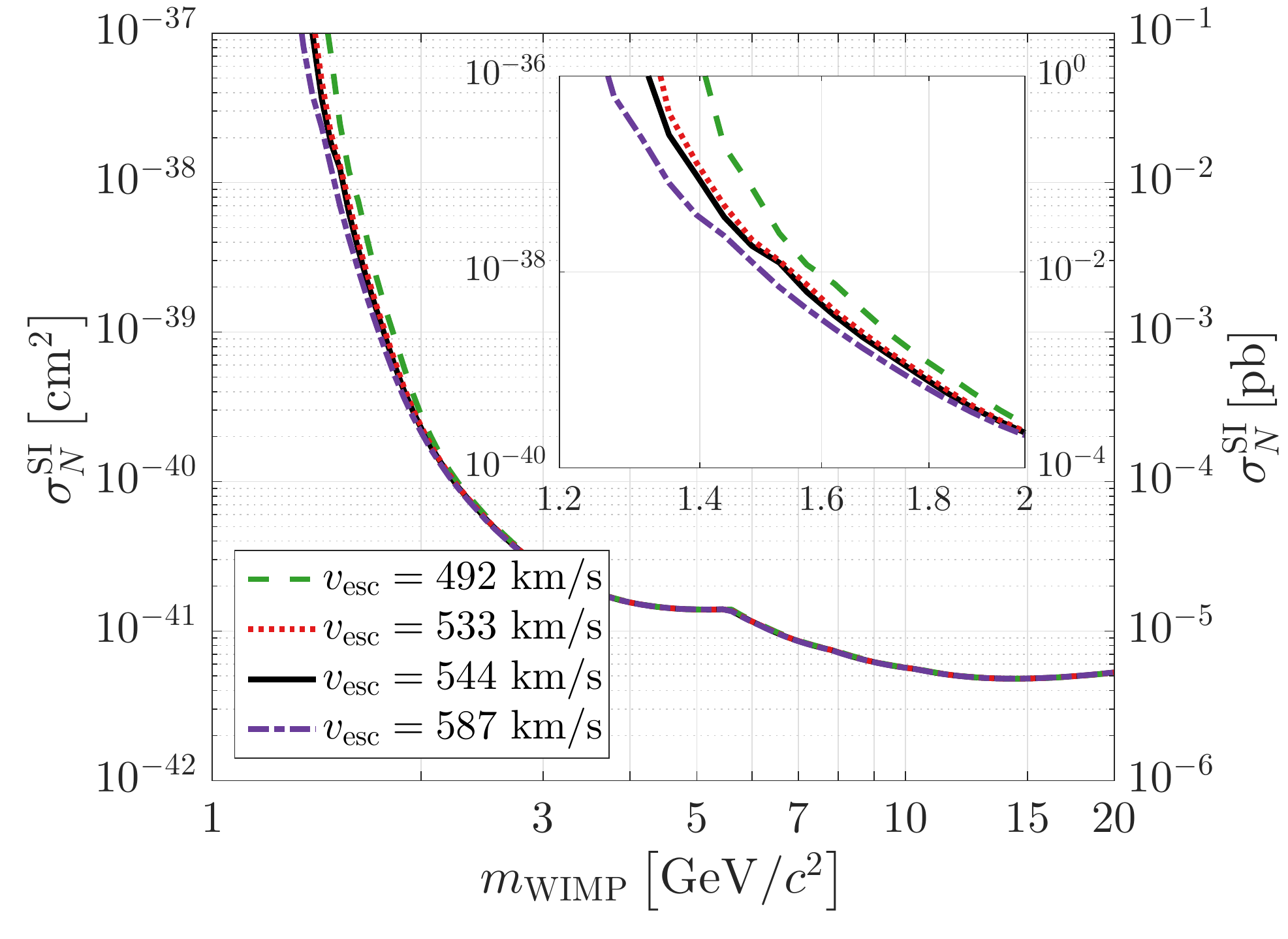}
    \caption{Effect on the \runTwo best-fit limit from varying the Galactic escape velocity $v_{\text{esc}}$ in the Maxwellian halo model while keeping all other parameters constant.  Curves shown are the median values of the 2007 and 2014 RAVE survey results at $544~\text{km\,s}^{-1}$ (black solid) and $533~\text{km\,s}^{-1}$ (red dotted), respectively, as well as the 90\,\% confidence bounds of the 2014 result at $492~\text{km\,s}^{-1}$ (green dashed) and $587~\text{km\,s}^{-1}$ (purple dot-dashed).  The inset shows an enlargement below \w masses of 2~\gev.  Varying $v_{\text{esc}}$ changes the lowest \w mass that can produce recoils above threshold, while the impact on the limit at higher masses is negligible.}
    \label{fig:varyVesc}
\end{figure}

The SHM value of $v_{\text{esc}}$ comes from the median and 90\,\% confidence region of the 2007 RAVE survey study~\cite{Smith2007}.  The RAVE survey collaboration released an updated study of the escape velocity in 2014~\cite{Piffl2014} in which they found a slightly lower median and reduced uncertainty span of $v_{\text{esc}}=533^{+54}_{-41}~\text{km\,s}^{-1}$.  Varying the escape velocity changes the lower edge of the \w-mass range, as a higher maximum halo velocity allows lower-mass \ws to deposit energy above threshold.  The effect on the \runTwo limit of varying the escape velocity while keeping all other SHM parameters constant can be seen in \FIG\ref{fig:varyVesc}.  The difference between the 2007 and 2014 RAVE medians is negligible at all but the lowest \w masses.

\begin{figure}
    \centering
    \includegraphics[width=\columnwidth]{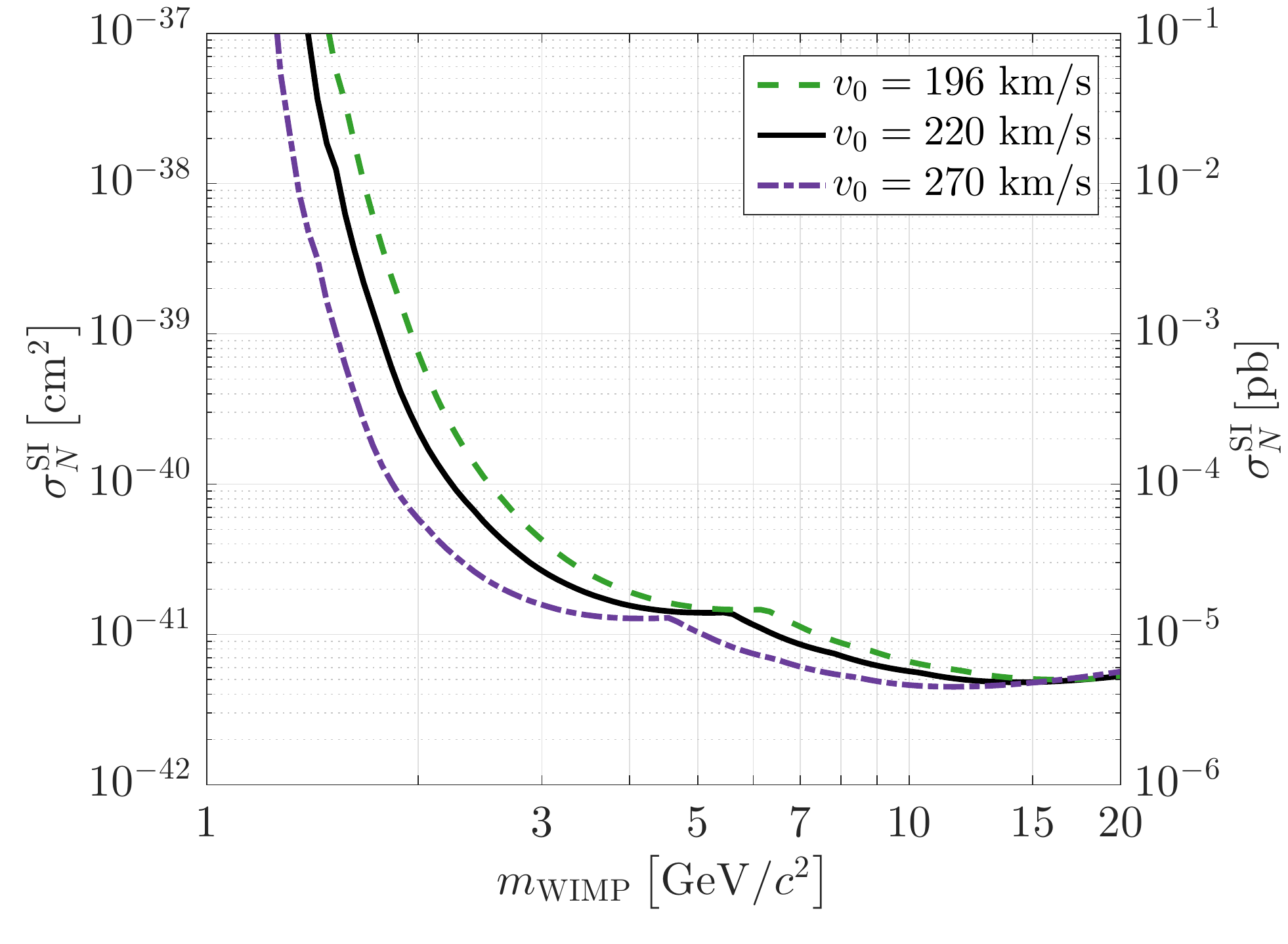}
    \caption{Effect on the \runTwo\ best-fit limit from varying the most probable WIMP velocity $\Theta_0$ in the Maxwellian halo model while keeping all other parameters constant.  Curves shown are for the SHM value of $220~\text{km\,s}^{-1}$ (black solid) and the upper and lower bounds of the measured values at $270~\text{km\,s}^{-1}$ (green dashed) and $196~\text{km\,s}^{-1}$ (purple dot-dashed).  Varying $\Theta_0$ changes where the most sensitive part of the curve lies in addition to slight changes in the lowest accessible WIMP mass.  The effect is largest for the lowest \w masses, vertically shifting the limit by up to an order of magnitude in either direction.}
    \label{fig:varyV0}
\end{figure}

Recent measurements of the magnitude of the LSR $\Theta_0$ are numerous~\cite{Reid2009,*Bovy2009,*McMillan2010,*Koposov2010,*Bobylev2010,*McMillan2011,*Bovy2012,*Carlin2012,*Honma2012,*Reid2014,*Bobylev2016} and include different approaches in measurement technique, galactic modeling, and prior assumptions.  The range that the collection of results spans, 196--270 km\,s$^{-1}$, is broader than any individual uncertainty, which indicates possible systematic uncertainties between the measurements and models.  The effect of varying $\Theta_0$ on the \runTwo limit, keeping all other halo parameters at their standard values, can be seen in \FIG\ref{fig:varyV0}.  Varying $\Theta_0$, and therefore the most probable velocity in the distribution $v_0$, changes where the most sensitive part of the curve lies in addition to changing the lowest accessible WIMP mass.  This uncertainty has a large effect at the lowest \w masses, shifting the limit on $\sigma_N^{\text{SI}}$ by up to an order of magnitude in either direction.

\begin{figure}
    \centering
    \includegraphics[width=\columnwidth]{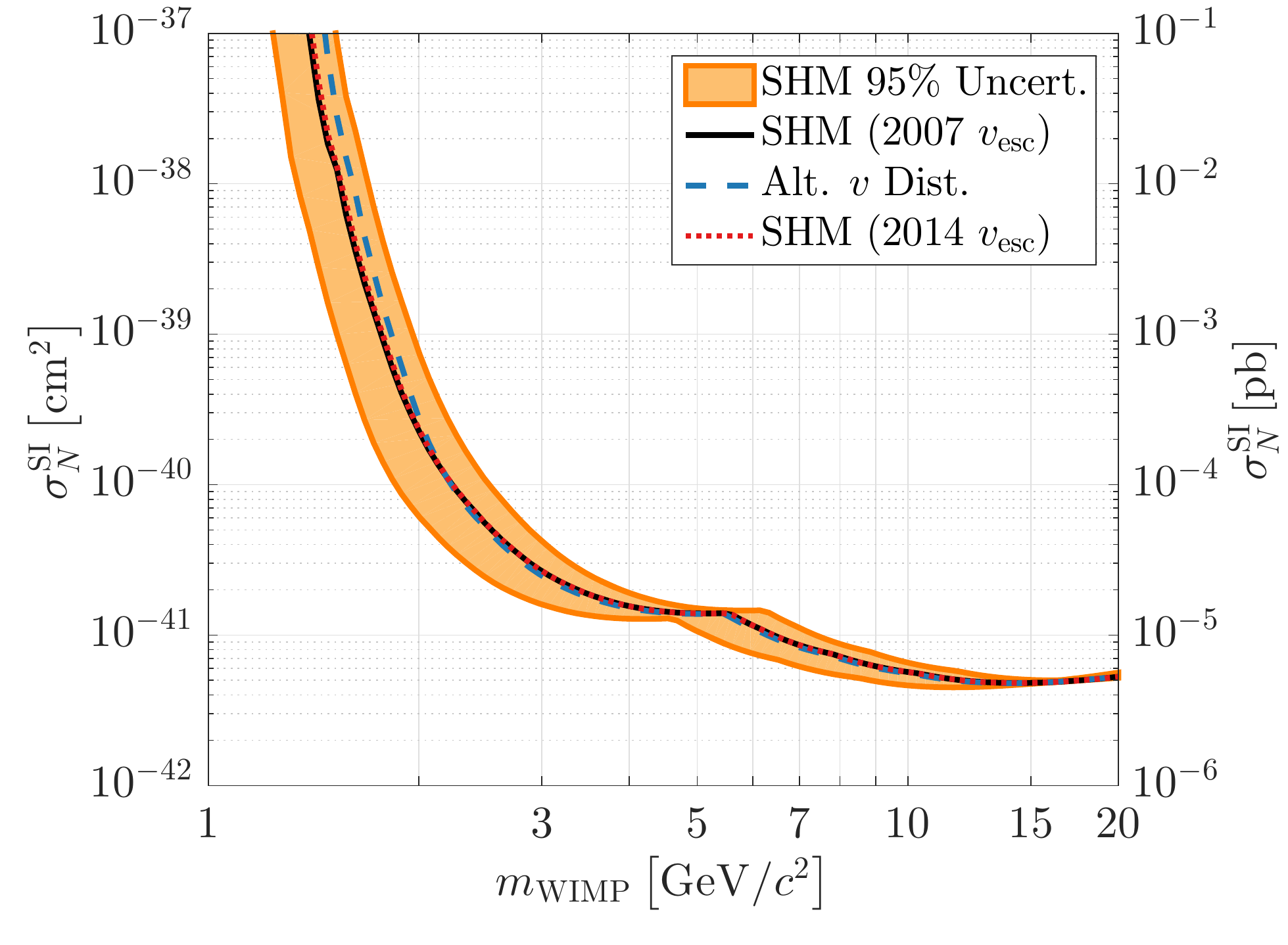}
    \caption{The 95\,\% (orange) uncertainty band on the best-fit \runTwo\ spin-independent limit (black solid) due to the uncertainties in the most probable WIMP velocity ($v_0$) and the Galactic escape velocity ($v_{\text{esc}}$) used in the SHM.  The 2014 RAVE survey $v_{\text{esc}}$ distribution is sampled, and thus the best-fit curve substituting the 2014 median value into the SHM is given for consistency (red dotted).  The black and red-dotted curves are the same as in \FIG\ref{fig:varyVesc}, where an enlargement at low \w mass is given.  The best-fit limit computed using the alternative velocity distribution of \EQ\ref{eq:Mao} is also presented (blue dashed).}
    \label{fig:limitVaryV}
\end{figure}

The effect of jointly varying $\Theta_0$ and $v_{\text{esc}}$ is considered by computing the limit 1000 times, each time selecting a different set of velocity parameters from their respective distributions.  For $\Theta_0$, a conservative flat distribution between the bounding measurements, 196--270~km\,s$^{-1}$, is sampled.  For $v_{\text{esc}}$, the probability distribution of $v_{\text{esc}}$ from the 2014 RAVE study (distribution graciously provided by the study authors) is directly sampled.  The 95\,\% central interval from the 1000 limit curves is shown in \FIG\ref{fig:limitVaryV} around the SHM-value curve.  The size of the uncertainty band is comparable to the uncertainty band on the analysis uncertainties given in \FIG\ref{fig:limits_si}.  Note also that \REF\cite{Piffl2014} demonstrates an anticorrelation between $\Theta_0$ and $v_{\text{esc}}$, meaning that the computed uncertainty band, which samples the velocity values independently, is an overestimate of the combined uncertainty.

Finally, an alternative \w velocity distribution is also considered in \FIG\ref{fig:limitVaryV}.  The model is that of Mao \etal~\cite{Mao2013,*Mao2014}, which gives, in the rest frame of the \dm,
\begin{equation}
    f{\left(v\right)} \propto e^{-v/v_a}\left(v^2_{\text{esc}}-v^2\right)^p,
    \label{eq:Mao}
\end{equation}
where $v_a$ and $p$ are parameters of the model.  Fits to a Milky-Way-like simulation with baryons give $p=2.7$ and $v_a/v_{\text{esc}}=0.6875$~\cite{Pillepich2014}.  The distribution is boosted to the lab frame via the usual $\bm{v}\rightarrow \bm{v}+\bm{\Theta}_0+\bm{v}_{\odot}+\bm{v}_{\oplus}$, where the SHM values for these astrophysical velocities are used.  This model naturally tends to $v=0$ at the escape velocity, which explains the reduced sensitivity at the lightest \w masses seen in the limit curve.

\subsection{Spin-dependent limits on WIMPs}
\label{sec:sdlimits}

While the SuperCDMS technology is most sensitive to spin-independent \w-nucleon scattering, the presence of a neutron-odd isotope, $^{73}\text{Ge}\left(N=41\right)$ with an abundance in natural Ge of 7.73\,\%, yields competitive limits for spin-dependent scattering at low \w masses~\cite{Akerib2006a}.

The differential elastic-scattering cross section for a fermionic WIMP \wrt the momentum transferred to the nucleus $q$ is given by
\begin{equation}
	\frac{\d{\sigma}^{\text{SD}}}{\d{q^2}} = \frac{8G_{\text{F}}^2}{\left(2J+1\right)v^2}S_T{\left(q\right)},
\end{equation}
where $G_{\text{F}}$ is Fermi's constant, $J$ is the total nuclear spin of the target nucleus, and $S_T{\left(q\right)}$ is the momentum-transfer-dependent spin-structure function.  $S_T{\left(q\right)}$ can be parametrized into isoscalar $S_{00}$, isovector $S_{11}$, and interference $S_{01}$ terms as
\begin{equation}
	S_T{\left(q\right)} = a_0^2S_{00}{\left(q\right)}+a_1^2S_{11}{\left(q\right)}+a_0a_1S_{01}{\left(q\right)},
	\label{eq:spinStructFunc}
\end{equation}
where the isoscalar and isovector coupling coefficients are related to the proton and neutron couplings as $a_0=a_p+a_n$ and $a_1=a_p-a_n$.  Explicit forms of $S_T{\left(q\right)}$ are obtained from detailed nuclear models for specific isotopes.

The scattering cross section is typically written in a form similar to the spin-independent case as
\begin{equation}
	\frac{d\sigma^{\text{SD}}}{dq^2} = \frac{8G_{\text{F}}^2}{\left(2J+1\right)v^2}S_T{\left(0\right)}F^2_{\text{SD}}{\left(q\right)},
\end{equation}
where $F_{\text{SD}}^2{\left(q\right)}\equiv S_T{\left(q\right)}/S_T{\left(0\right)}$ is the form factor of \EQ\ref{eq:wimpRate}, which is normalized to unity at zero momentum transfer ($q\rightarrow0$).  In that limit, the structure function is
\begin{multline}
	S_T{\left(0\right)}= \frac{\left(2J+1\right)\left(J+1\right)}{4\pi J} \\ \times\left|\left(a_0+a'_1\right)\left\langle S_p\right\rangle+\left(a_0-a'_1\right)\left\langle S_n\right\rangle\right|^2,
	\label{eq:zeroMomSpinStruct}
\end{multline}
where $a'_1=a_1\left(1+\delta a_1{\left(0\right)}\right)$ includes contributions from two-body current scattering as given by Klos \etal in \REF\cite{Klos2013}.  In two-body current scattering, the WIMP effectively interacts with two nucleons in the nucleus, via the $\delta a_1{\left(0\right)}$ term.  The expectation values of the proton and neutron groups within the nucleus $\left\langle S_{p}\right\rangle$ and $\left\langle S_{n}\right\rangle$ are computed from nuclear theory and usually $\left\langle S_{p}\right\rangle\gg\left\langle S_{n}\right\rangle$ for proton-odd nuclei and vice versa for neutron-odd nuclei.  Note that, although the spin-coupling to the even-nucleon species is weak, the inclusion of two-body currents allows for WIMP-proton-neutron effective interactions.  Thus, the odd-nucleon-species coupling dominates the scattering calculations for any coupling type.

The standard cross section  $\sigma_0^{\text{SD}}$ from \EQ\ref{eq:wimpRate} is defined as the total cross section in the $q\rightarrow0$ limit
\begin{equation}
	\sigma_0^{\text{SD}}=\frac{32}{2J+1}G_{\text{F}}^2\mu_T^2S_T{\left(0\right)}.
	\label{eq:sigma0SD}
\end{equation}
The differential cross section can then be written as
\begin{equation}
	\frac{\d{\sigma^{\text{SD}}}}{\d{q^2}}=\frac{1}{4\mu_T^2v^2}\sigma_0^{\text{SD}}F^2_{\text{SD}}{\left(q\right)},
\end{equation}
where $\mu_T=m_{\chi}m_T/\left(m_{\chi}+m_{T}\right)$ is the reduced mass of the WIMP-nucleus system.  Results are presented in the ``proton-only'' model where $a_p=1$ and $a_n=0$, implying $a_0=a_1=1$,  and the ``neutron-only'' model where $a_p=0$ and $a_n=1$, implying $a_0=-a_1=1$.  Results are also normalized to the scattering of a WIMP and a free proton/neutron as
\begin{equation}
	\sigma_0^{\text{SD}}=\frac{4\pi}{3}\frac{1}{\left(2J+1\right)}\left(\frac{\mu_T}{\mu_{p{/}n}}\right)^2S_T^{p{/}n}{\left(0\right)}\sigma_{p{/}n}^{\text{SD}},
	\label{eq:sigmaSDnorm}
\end{equation}
where $\sigma_{p{/}n}^{\text{SD}}$ is the free proton/neutron standard cross section, $\mu_{p{/}n}$ is the proton-/neutron-WIMP reduced mass, and $S_T^{p{/}n}{\left(0\right)}$ is $S_T{\left(0\right)}$ evaluated in the proton-/neutron-only models.

\begin{figure*}
    \centering
    \subfloat{
	    \includegraphics[width=0.48\textwidth]{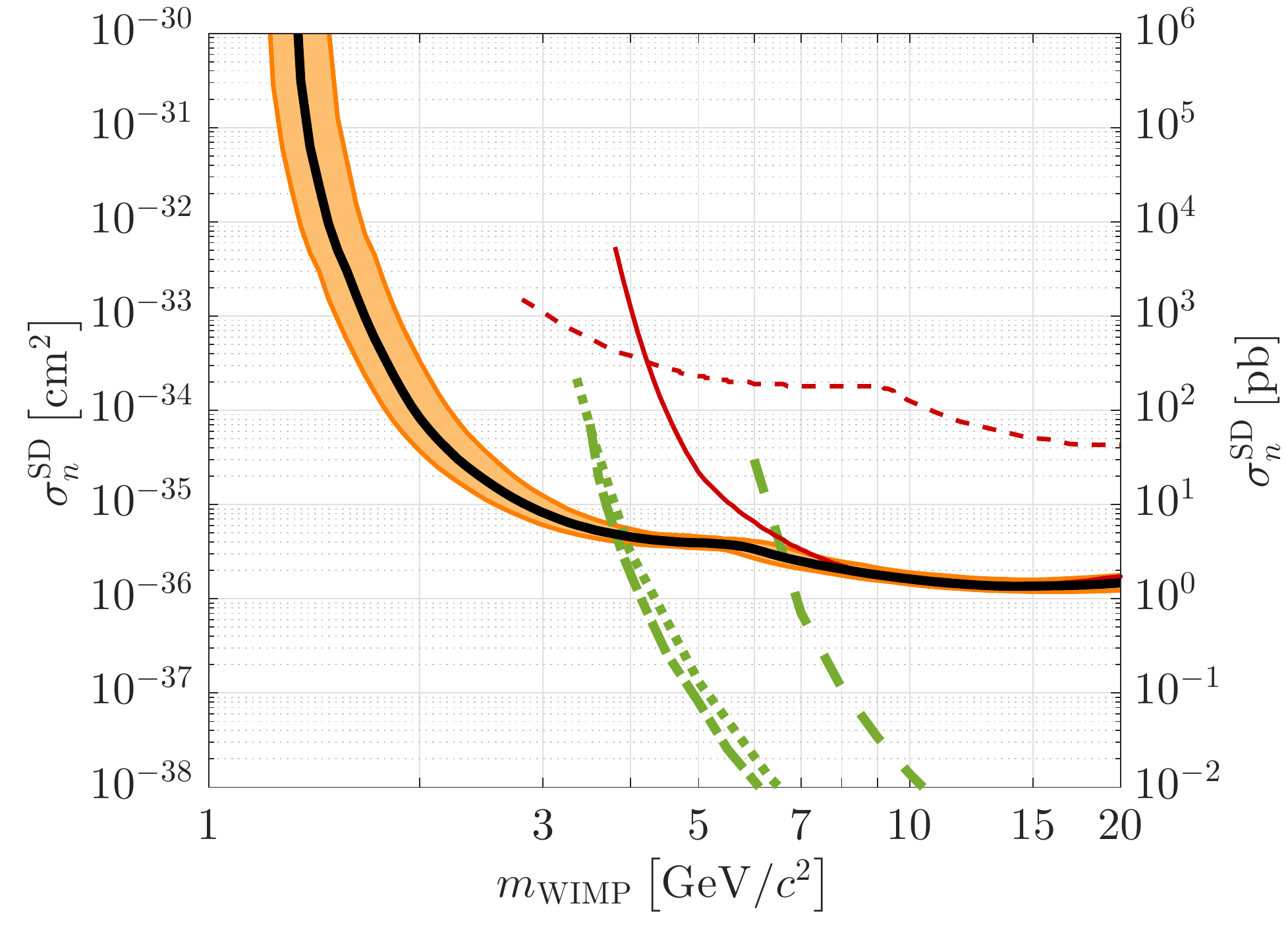}
	    \label{subfig:limitWorld_n}
    }
    \subfloat{
	    \includegraphics[width=0.48\textwidth]{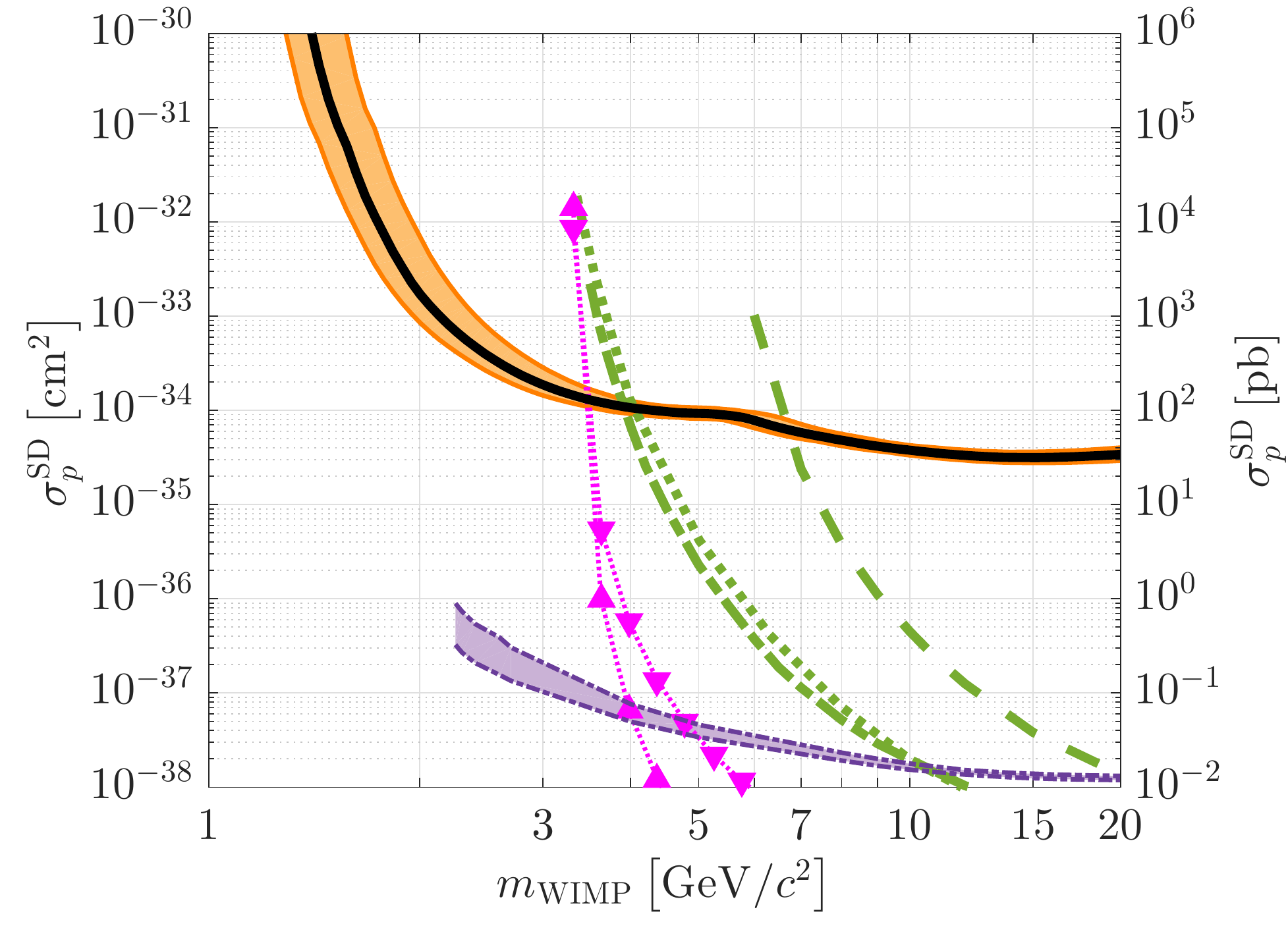}
        \label{subfig:limitWorld_p}
    }
    \caption{Upper limits on the spin-dependent free neutron $\sigma_n^{\text{SD}}$ (left) and free proton $\sigma_p^{\text{SD}}$ (right) WIMP scattering cross sections in the proton- and neutron-only models, respectively.  For both, the median (90\,\% C.L) (thick black solid curve) upper limit from CDMSlite \runTwo is compared to other selected direct-detection limits from PANDAX-II (thick-green dotted curve)~\cite{Fu2017}, LUX (thick-green dot-dashed curve)~\cite{Akerib2016}, XENON100 (thick-green dashed curve)~\cite{Aprile2016}, PICO-60 (magenta upward triangles)~\cite{Amole2017}, PICO-2L (magenta downward triangles)~\cite{Amole2016a}, PICASSO (purple dot-dashed band)~\cite{Behnke2017}, CDEX-0 (thin-red dashed curve)~\cite{Liu2014,Zhao2016a}, and CDEX-1 (thin-red solid curve)~\cite{Zhao2016a}.  The orange band surrounding the \runTwo result is the 95\,\% uncertainty interval on the upper limit.  The Run 2 limits are the most sensitive for $m_{\text{WIMP}}\lesssim4$ and $\lesssim 2$~\gev for the neutron- and proton-only models, respectively.}
    \label{fig:SDlims}
\end{figure*}

Limits set on $\sigma_{p{/}n}^{\text{SD}}$ using the \runTwo data and analysis are presented in \FIG\ref{fig:SDlims}.  The limits were computed using the same framework as the spin-independent limits that is described in \SEC\ref{sec:runs}, including using the optimum interval method~\cite{Yellin2002,*Yellin2007} and sampling the analysis uncertainties.  The median and 95\,\% uncertainty band from the resulting set of limits are shown in the figure for each model.  The low threshold of CDMSlite gives world-leading limits for WIMP masses ${\lesssim}$4 and ${\lesssim}$2~\gev for the neutron-only and proton-only models, respectively.  Limits were also computed using the older spin-structure model of \REF\cite{Dimitrov1995}, which does not include two-body currents.  In the neutron-only case, only a mild improvement of 8\,\% is seen using the newer Klos \etal model.  However, using the newer model improves the proton-only limit by a factor of ${\sim}7$, a direct consequence of the WIMP-proton-neutron two-body current increasing the proton-only structure function.

\begin{figure*}
	\centering
    \subfloat{
    	\includegraphics[width=0.35\textwidth]{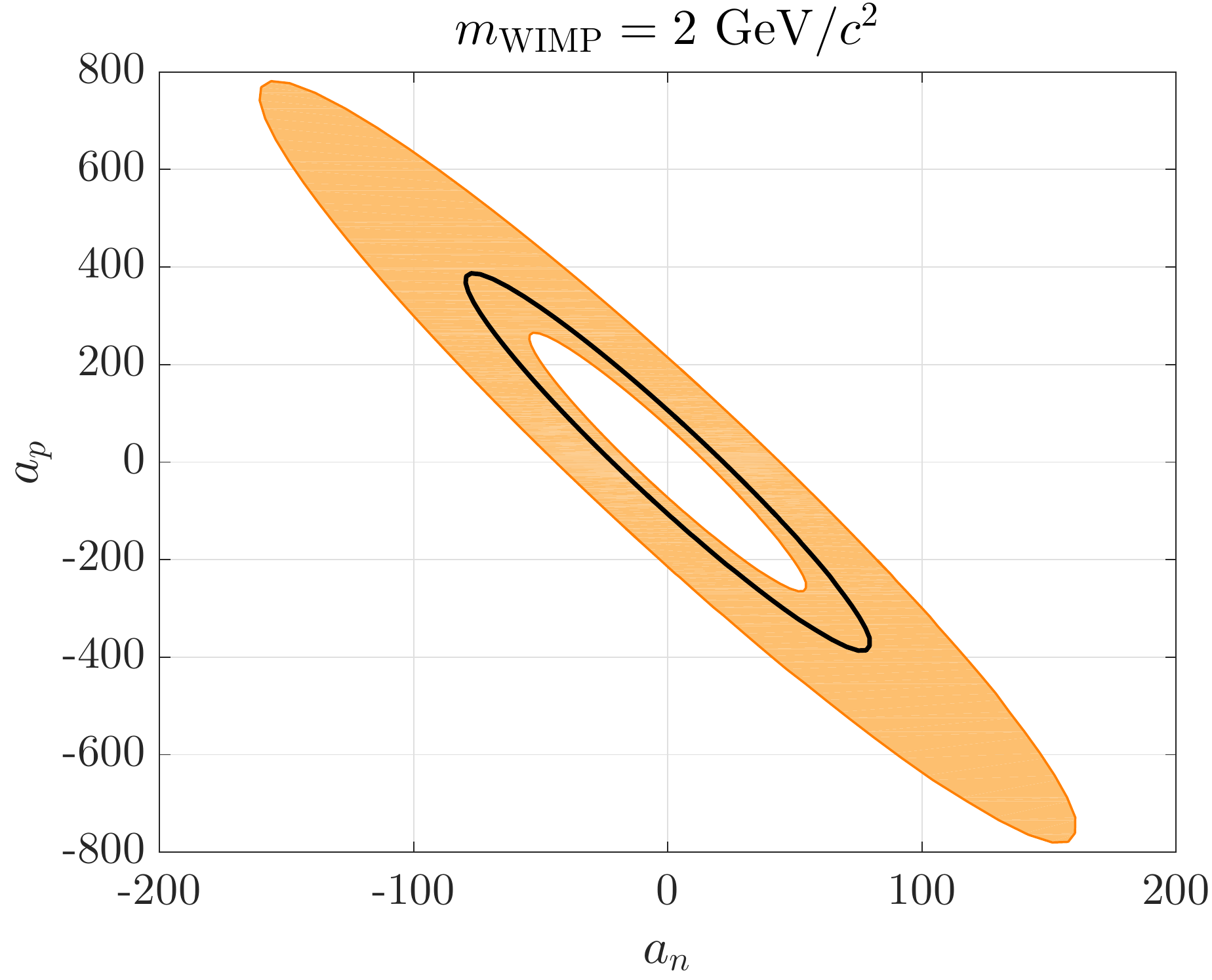}}
    \subfloat{
    	\includegraphics[width=0.35\textwidth]{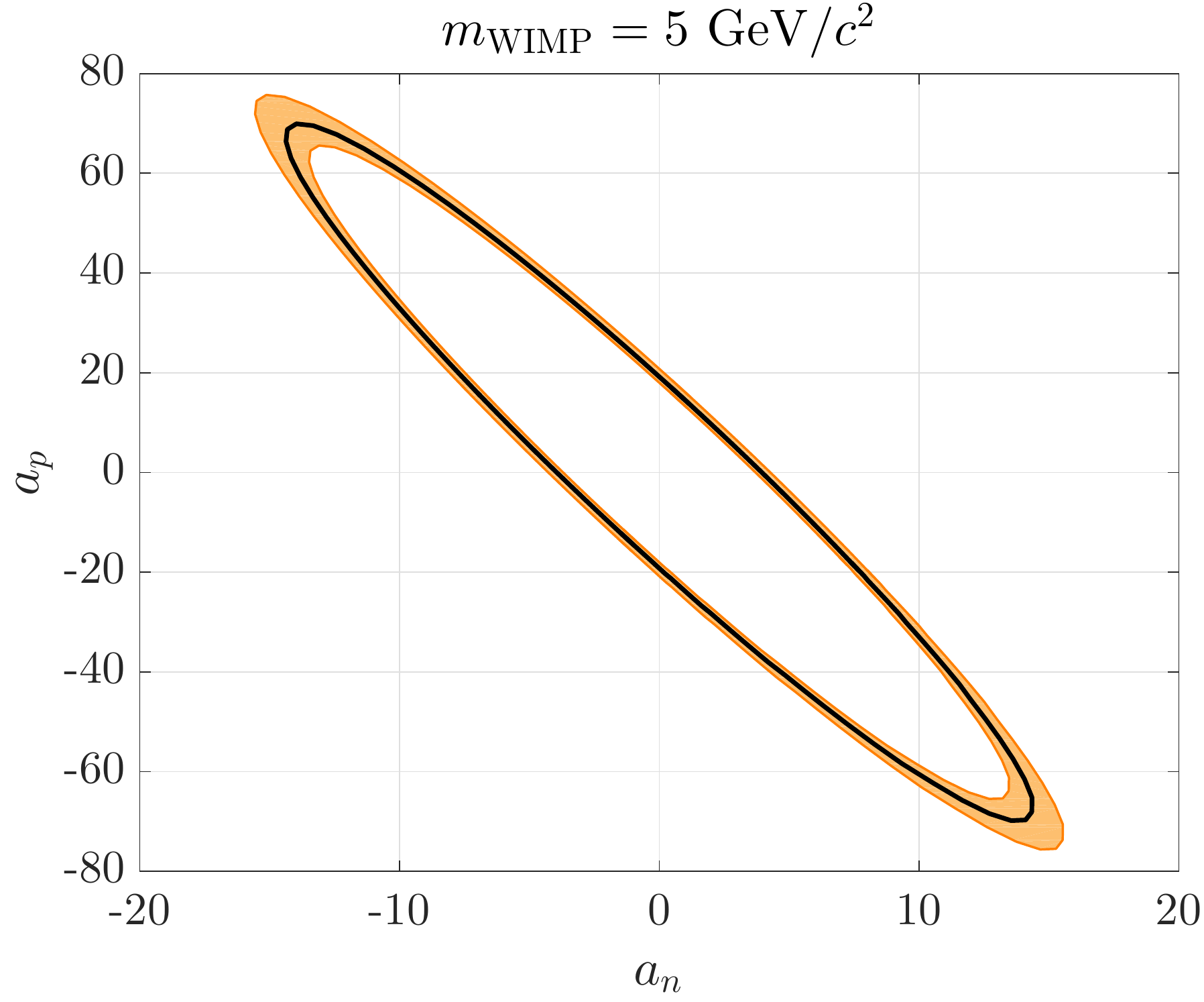}}
    \quad
    \subfloat{
    	\includegraphics[width=0.35\textwidth]{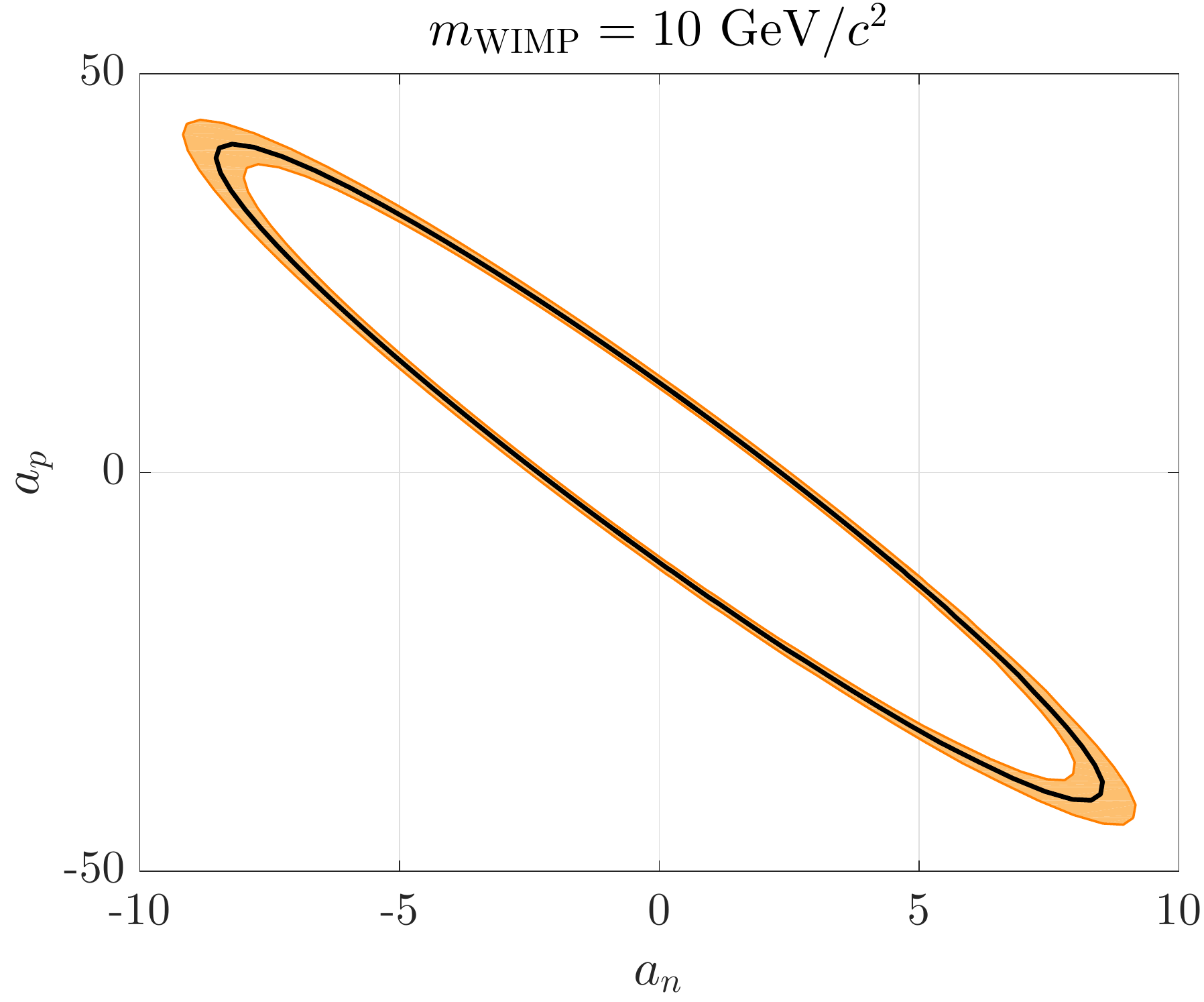}}
    \subfloat{
    	\includegraphics[width=0.35\textwidth]{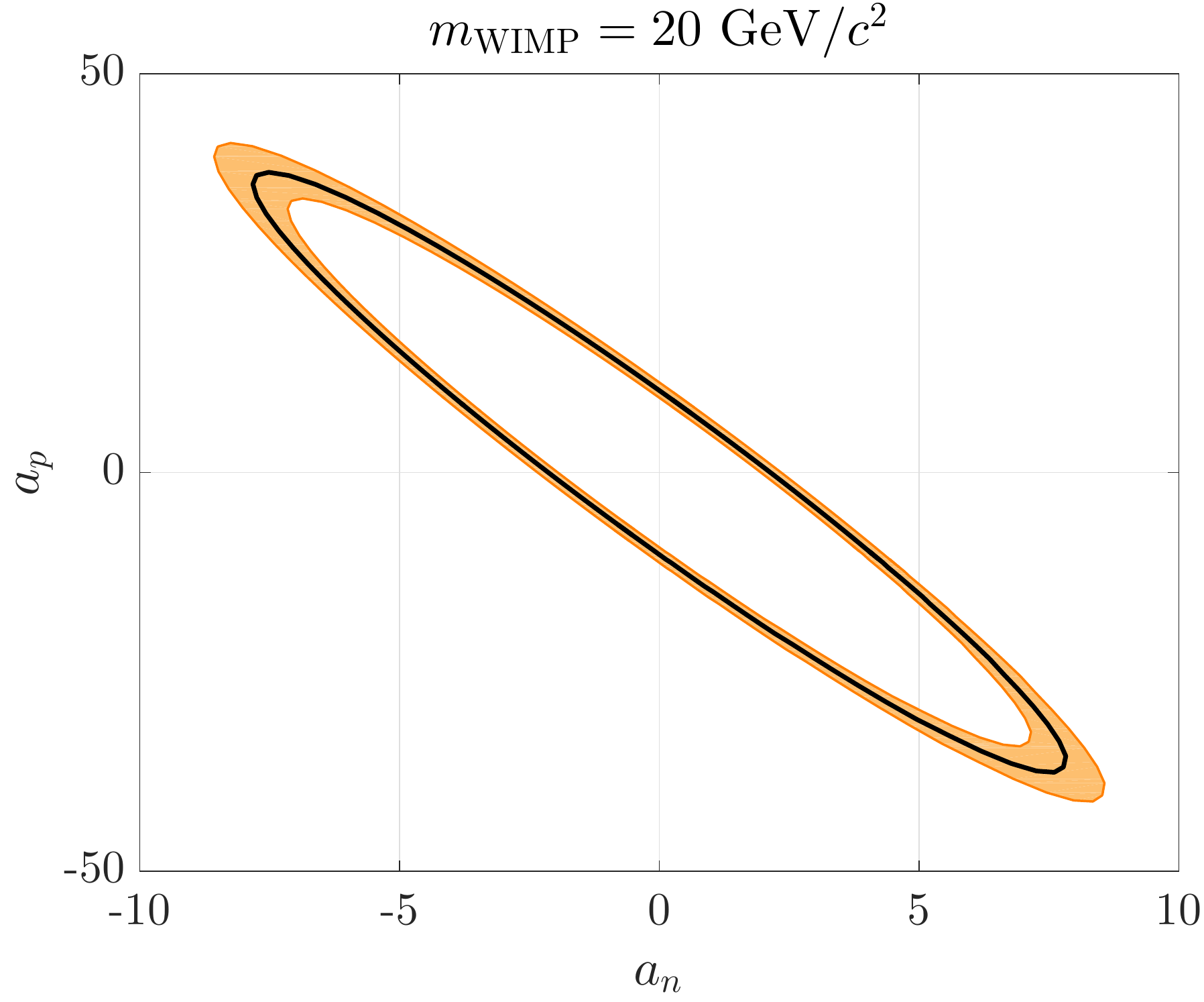}}
    \caption{Median (90\,\% C.L.) upper limit and associated 95\,\% uncertainty (thick black solid curve and orange bands) on the WIMP-nucleon coupling coefficients $a_p$ and $a_n$ from CDMSlite \runTwo for WIMP masses of 2 (top left), 5 (top right), 10 (bottom left), and 20 (bottom right) \gev.  Areas outside the ellipses are excluded for each WIMP mass.}
    \label{fig:SDpolar}
\end{figure*}

Limits are also placed jointly on the coupling coefficients $a_p$ and $a_n$ for four different WIMP masses.  Results in this plane were computed by converting the coefficients to polar coordinates, $a_p=a\sin\theta$ and $a_n=a\cos\theta$, and observing that for a given $\theta$, $S_T{\left(q\right)}\propto a^2$.  The proton- and neutron-only models are recovered for $\theta=\pi/2,\ 0$, respectively.  Values of $\theta$ were scanned, and an upper limit was placed on $a$ for each angle.    Appendix~\ref{app:SDcouplingLimits} discusses different methods for computing these limits and includes justification for the chosen approach.  Limits in the $a_p$ \vs $a_n$ plane are given in \FIG\ref{fig:SDpolar} for $m_{\text{WIMP}}$ of 2, 5, 10, and 20~\gev.  Regions outside of the ellipses are excluded.  The limits were again computed by sampling the analysis uncertainties with the median and 95\,\% intervals for each WIMP mass given in the figure.

\section{Summary and Outlook}
\label{sec:summary}

This paper described in detail the CDMSlite technique for extending dark matter direct detection searches to WIMP masses of ${\sim}$1.5~\gev by achieving analysis thresholds as low as 56~\evee. New analysis techniques were presented and applied to the first two CDMSlite data sets taken with the SuperCDMS Soudan experiment, yielding new limits on spin-dependent interactions and a better understanding of the effects of astrophysical uncertainties on the limits.

There is one more Soudan CDMSlite data set, taken with a different detector, to be analyzed.  Previous studies have indicated that this different detector is less sensitive to \lf noise, and preliminary studies with the new CDMSlite data show a 50\,\% trigger efficiency point as low as 50~\evee.  This data set will be used to develop improved CDMSlite analysis techniques, including: a salting scheme to mitigate analyzer bias, further understanding of the electric-field influence on fiducial volume, and low-energy background modeling to test background subtraction techniques. 

The SuperCDMS Collaboration is also designing a new experiment, SuperCDMS SNOLAB, where the CDMSlite technique will be used in detectors designed specifically for high-voltage operation.  Planned improvements with such detectors include~\cite{Kurinsky2016}: two-sided biasing, which diminishes the reduced bias region of the detector; increasing the surface area coverage of the phonon sensor; operating at higher applied potentials; and fabricating TESs with lower operational temperatures for the phonon read-out. With the latter two improvements, the SuperCDMS Collaboration aims at thresholds ${\lesssim}$10~\evee that will correspondingly provide sensitivity to WIMP masses as low as 400~\mev~\cite{Agnese2017}.

The SuperCDMS Collaboration gratefully acknowledges technical assistance from the staff of the Soudan Underground Laboratory and the Minnesota Department of Natural Resources, as well as the many contributions of David Caldwell, who passed away during the writing of this article. The iZIP detectors were fabricated in the Stanford Nanofabrication Facility, which is a member of the National Nanofabrication Infrastructure Network, sponsored and supported by the NSF.  Part of the research described in this article was conducted under the Ultra Sensitive Nuclear Measurements Initiative and under Contract No. DE-AC05-76RL01830 at Pacific Northwest National Laboratory, which is operated by Battelle for the U.S. Department of Energy.  Funding and support were received from the National Science Foundation, the Department of Energy, Fermilab URA Visiting Scholar Award No. 13-S-04, NSERC Canada, and MultiDark (Spanish MINECO). Fermilab is operated by the Fermi Research Alliance, LLC, under Contract No. De-AC02-07CH11359. SLAC is operated under Contract No. DEAC02-76SF00515 with the United States Department of Energy.

\appendix

\section{Setting limits on spin-dependent coupling coefficients with two-body currents}
\label{app:SDcouplingLimits}

A model-independent method for setting joint limits on the spin-dependent coupling constants $a_p$ and $a_n$ was derived by Tovey \etal in \REF\cite{Tovey2000}.  In that work, the authors derive a simple expression relating the allowed values of the coupling constants, for a given WIMP mass, as
\begin{equation}
	\frac{\pi}{24G_{\text{F}}^2\mu_p^2}\geq\left[\frac{a_p}{\sqrt{\sigma_p^L}}\pm\frac{a_n}{\sqrt{\sigma_{n\vphantom{p}}^L}}\right]^2,
	\label{eq:tovey}
\end{equation}
where $G_{\text{F}}$ is Fermi's constant, $\sigma_{p{/}n}^L$ are the limits on the free-proton/-neutron cross sections for the given WIMP mass (assuming a proton-/neutron-only interaction), the small difference between the WIMP-proton $\mu_p$ and WIMP-neutron $\mu_n$ reduced masses is ignored, and the sign in the brackets is the same as the ratio of nuclear spin-group expectation values $\left\langle S_n\right\rangle/\left\langle S_p\right\rangle$.  This expression is derived from the observation that the allowed total-nucleus cross section $\sigma_0^{\text{SD}}$ must be smaller than the limit set upon it by a given analysis $\sigma_0^L$.   Equation~\ref{eq:tovey} is then found by using the expression for the zero-momentum spin structure function $S_T{\left(0\right)}$ without two-body currents, found by taking $\delta a_1{\left(0\right)}\to0$ in \EQ\ref{eq:zeroMomSpinStruct}.

Including the two-body current contributions to $S_T{\left(0\right)}$ from Klos \etal~\cite{Klos2013} changes this derivation and result.  Starting with $\sigma_0^{\text{SD}}/\sigma_0^L\leq1$ and using \EQ\ref{eq:sigma0SD} for $\sigma_0^{\text{SD}}$ and \EQ\ref{eq:zeroMomSpinStruct} for $S_T{\left(0\right)}$ gives
\begin{multline}
	1\geq\frac{8\left(J+1\right)G_{\text{F}}^2\mu_T^2}{J\pi} \\
	\times\left[\frac{\left|\left(a_0+a'_1\right)\left\langle S_p\right\rangle\right|}{\sqrt{\sigma_0^L}}\pm\frac{\left|\left(a_0-a'_1\right)\left\langle S_n\right\rangle\right|}{\sqrt{\sigma_0^L}}\right]^2,
\end{multline}
where the sign of the $\pm$ is determined by the sign of $\left(a_0-a'_1\right)\left\langle S_n\right\rangle/\left(a_0+a'_1\right)\left\langle S_p\right\rangle$. The limits on the total cross section are not factored out as they are next rewritten in terms of the limits on the free-proton/-neutron cross sections $\sigma_{p{/}n}^L$ in the proton-/neutron-only models, as given by \EQ\ref{eq:sigmaSDnorm}.  In the denominator of the left term, the proton-only model form is used, while the neutron-only form is used under the right term.  The resulting inequality after changing coupling bases to that of the proton and neutron couplings is
\begin{widetext}
	\begin{multline}
		\frac{\pi}{24G_{\text{F}}^2\mu_p^2}\geq\left[\frac{\left|2a_p+\left(a_p-a_n\right)\delta a_1{\left(0\right)}\right|}{\sqrt{\sigma_p^L}}\frac{\left|\left\langle S_p\right\rangle\right|}{\left|\left[2+\delta a_1{\left(0\right)}\right]\left\langle S_p\right\rangle - \delta a_1{\left(0\right)}\left\langle S_n\right\rangle\right|}\right. \\
		\pm\left.\frac{\left|2a_n-\left(a_p-a_n\right)\delta a_1{\left(0\right)}\right|}{\sqrt{\sigma_{n\vphantom{p}}^L}}\frac{\left|\left\langle S_n\right\rangle\right|}{\left|-\delta a_1{\left(0\right)}\left\langle S_p\right\rangle + \left[2+\delta a_1{\left(0\right)}\right]\left\langle S_n\right\rangle\right|}\right]^2.
		\label{eq:2bTovey}
	\end{multline}
\end{widetext}
The simpler \EQ\ref{eq:tovey} is recovered by taking the limit of no two-body currents ($\delta a_1{\left(0\right)}\to0$).  

If proton-/neutron-only limits are computed using the two-body-inclusive spin-structure function, then it is inconsistent to use the simple \EQ\ref{eq:tovey} to compute limits on the coupling constants.  This is particularly important for low-mass \ws as the two-body current has its largest effect for low momentum transfer.

Because of the complexity of \EQ\ref{eq:2bTovey}, the ``polar coordinate'' method for computing coupling constant upper limits was used instead for the current results.  This method transforms coordinates from the Cartesian $\left(a_p,a_n\right)$ to the polar $\left(a,\theta\right)$ as
\begin{align}
	a_p&=a\sin{\theta}	\\
	a_n&=a\cos{\theta}.
\end{align}
In these new coordinates, the momentum-dependent spin-structure function \EQ\ref{eq:spinStructFunc} is
\begin{align}
S_T{\left(q\right)}&=
	\begin{multlined}[t]
		a^2\left[\left(1+\sin{2\theta}\right)S_{00}{\left(q\right)}-\cos{2\theta}S_{10}{\left(q\right)}\right. \\
		\left.+\left(1-2\sin\theta\cos\theta\right)S_{11}{\left(q\right)}\right]
	\end{multlined} \\
	&\equiv a^2f{\left(q,\theta\right)},
\end{align}
where $q$ is the momentum transferred in the collision.  This form of the spin-structure function enters the standard computation by multiplying both sides of \EQ\ref{eq:sigma0SD} by the form factor $F_{\text{SD}}^2=S_T{\left(q\right)}/S_T{\left(0\right)}=a^2f{\left(q,\theta\right)}/S_T{\left(0\right)}$.
The polar-coordinates method is equally valid with or without the inclusion of two-body currents depending upon the functions used for the $S_{ij}$.  The procedure described in \SEC\ref{sec:sdlimits} can then be followed to construct the upper limit curves; \ie, scan over the angle $\theta$ and compute an upper limit on $a^2$ for each angle.

\bibliography{refs}
\bibliographystyle{apsrev4-1-JHEPfix-autoEtAl}

\end{document}